\documentclass[longauth]{aa}
\pdfoutput=1

\usepackage{natbib}
\bibpunct{(}{)}{;}{a}{}{,}

\defcitealias{Pierre2016}{Paper~I}
\defcitealias{Giles2016}{Paper~III}
\defcitealias{Lieu2016}{Paper~IV}
\defcitealias{Pompei2016}{Paper~VII}
\defcitealias{Koulouridis2016}{Paper~XII}
\defcitealias{Lidman2016}{Paper~XIV}

\usepackage[varg]{txfonts}
\usepackage{color}

\usepackage{graphicx}
\usepackage[percent]{overpic}
\usepackage{afterpage}

\usepackage{amsmath}

\LTcapwidth=\textwidth

\usepackage{placeins}

\usepackage{array}
\newcolumntype{C}[1]{>{\centering\let\newline\\\arraybackslash\hspace{0pt}}p{#1}}
\usepackage{booktabs}

\newcommand{\dd}{deg$^{2}$}
\newcommand{\flux}{$\rm erg \, s^{-1} \, cm^{-2}$}
\newcommand{\lum}{$\rm erg \, s^{-1}$}
\newcommand{\BXC}{bright XXL cluster sample}
\newcommand{\Chundred}{XXL-100-GC sample}
\newcommand{\ChundredShort}{XXL-100-GC}
\newcommand{\smspace}{\kern 0.1em}

\newcommand{\ignore}[1]{}
\newcommand{\ClRadMT}{r_{500,\mathrm{MT}}}

\newcommand{\ClRad}{r_{500}}
\newcommand{\ClMassMT}{M_{500,\mathrm{MT}}}
\newcommand{\ClMassWL}{M_{500,\mathrm{WL}}}
\newcommand{\ClMass}{M_{500}}
\newcommand{\Lxxl}{L^{\mathrm{XXL}}_{500,MT}}
\newcommand{\Lkpc}{L^{\mathrm{XXL}}_{300~\mathrm{kpc}}}

\newcommand{\Txxl}{T_{300~\mathrm{kpc}}}
\newcommand{\ClTemp}{T_{500,MT}}
\newcommand\SFunc[2]{P_{#1}\left({\smspace}I{\smspace}|{\smspace}#2{\smspace}\right)}
\newcommand\SFuncShort[1]{P_{#1}\left({\smspace}I{\smspace}\right)}
\newcommand\SFuncXXL{P\left({\smspace}I{\smspace}\right)}

\begin{document}
   \title{
   The XXL Survey
   \thanks{Based on observations obtained with {\it XMM-Newton}, an ESA science mission with
    instruments and contributions directly funded by ESA Member States and NASA.
    Based on observations made with ESO Telescopes at the La Silla and Paranal Observatories
    under programme ID 089.A-0666 and LP191.A-0268}\fnmsep\thanks{The Master Catalogue 
    is available at the CDS via anonymous ftp to {\tt cdsarc.u-strasbg.fr} ({\tt 130.79.128.5}) 
    or via {\tt http://cdsarc.u-strasbg.fr/viz-bin/qcat?J/A+A/vol/page}}}  
   \subtitle{II. The bright cluster sample: catalogue and luminosity function}
   %\saythanks

   \author{
     F.~Pacaud\inst{\ref{AIfA}}
     \and N.~Clerc\inst{\ref{MPE}}
     \and P.~A. Giles\inst{\ref{Bristol}}
     \and C.~Adami\inst{\ref{LAM}}
     \and T.~Sadibekova\inst{\ref{SAp}}
     \and M.~Pierre\inst{\ref{SAp}}
     \and B.~J. Maughan\inst{\ref{Bristol}}
     \and M.~Lieu\inst{\ref{Bhm}}
     \and J.~.P.~Le~F\`evre\inst{\ref{SEDI}}
 	 \and S.~Alis\inst{\ref{IstanUniv},\ref{OCA}}
 	 \and B.~Altieri\inst{\ref{ESAC}}
     \and F.~Ardila\inst{\ref{UFlorida}}
     \and I. Baldry\inst{\ref{Liverpool}}
     \and C.~Benoist\inst{\ref{OCA}}
 	 \and M.~Birkinshaw\inst{\ref{Bristol}}
 	 \and L.~Chiappetti\inst{\ref{Milan}}
 	 \and J.~D\'emocl\`es\inst{\ref{Bhm},\ref{SAp}}
 	 \and D.~Eckert\inst{\ref{Geneve}}
 	 \and A.~E.~Evrard\inst{\ref{UMich}}
 	 \and L.~Faccioli\inst{\ref{SAp}}
 	 \and F.~Gastaldello\inst{\ref{Milan}}
 	 \and L.~Guennou\inst{\ref{KwaZulu}}
 	 \and C.~Horellou\inst{\ref{Onsala}}
 	 \and A.~Iovino\inst{\ref{Brera}}
 	 \and E.~Koulouridis\inst{\ref{Athens},\ref{SAp}}
 	 \and V.~Le Brun\inst{\ref{LAM}}
 	 \and C.~Lidman\inst{\ref{AAO}}
 	 \and J.~Liske\inst{\ref{ESOG}}
 	 \and S.~Maurogordato\inst{\ref{OCA}}
     \and F.~Menanteau\inst{\ref{Illinois}}
 	 \and M.~Owers\inst{\ref{Macquarie},\ref{AAO}}
 	 \and B.~Poggianti\inst{\ref{Padova}}
     \and D.~Pomar\`ede\inst{\ref{SEDI}}
 	 \and E.~Pompei\inst{\ref{ESO}}
     \and T.~J.~Ponman\inst{\ref{Bhm}}
     \and D.~Rapetti\inst{\ref{Copenh},\ref{LMU}}
     \and T.~H.~Reiprich\inst{\ref{AIfA}}
     \and G.~P.~Smith\inst{\ref{Bhm}}
     \and R. Tuffs\inst{\ref{Heidelberg}}
 	 \and P.~Valageas\inst{\ref{IPhT}}
 	 \and I.~Valtchanov\inst{\ref{ESAC}}
 	 \and J.~P.~Willis\inst{\ref{UVic}}
 	 \and F.~Ziparo\inst{\ref{Bhm}}
 	 }

   \institute{
     	Argelander-Institut f\"ur Astronomie,
     	University of Bonn,
     	Auf dem H\"ugel 71,
     	D-53121 Bonn,
     	Germany\\
	\email{fpacaud@astro.uni-bonn.de}\label{AIfA}
    \and
    	Max-Planck-Institut f\"ur Extraterrestrische Physik,
    	Giessenbachstra{\ss}e,
    	D-85748 Garching bei M\"unchen,
    	Germany\label{MPE}
    \and
    	H.H. Wills Physics Laboratory,
    	University of Bristol,
    	Tyndall Avenue,
    	Bristol, BS8 1TL,
    	UK\label{Bristol}
    \and
    	Aix-Marseille Universit\'e, CNRS, 
        LAM (Laboratoire d'Astrophysique de Marseille) UMR~7326
    	13388, Marseille,
    	France\label{LAM}
    \and
    	Service d’Astrophysique AIM,
    	CEA/DSM/IRFU/SAp,
    	CEA Saclay,
    	F-91191 Gif sur Yvette
    	France\label{SAp}
  	\and
  		School of Physics and Astronomy,
  		University of Birmingham,
  		Birmingham B15 2TT,
  		UK\label{Bhm}
  	\and
  		Service d'\'Electronique des D\'etecteurs et d'Informatique,
  		CEA/DSM/IRFU/SEDI,
  		CEA Saclay,
  		F-91191 Gif-sur-Yvette,
  		France\label{SEDI}
    \and
        Department of Astronomy and Space Sciences,
        Faculty of Science,
        Istanbul University,
        34119 Istanbul,
        Turkey\label{IstanUniv}
    \and
        Laboratoire Lagrange, UMR 7293,
    	Universit\'e de Nice Sophia Antipolis, CNRS,
    	Observatoire de la C\^ote d’Azur,
    	F-06304 Nice,
    	France\label{OCA}
  	\and
  	    European Space Astronomy Centre,
  	    ESA, P.O. Box 78,
  	    E-28691 Villanueva de la Ca$\mathrm{\tilde{n}}$ada,
  	    Madrid, Spain\label{ESAC}
  	\and
  		Department of Astronomy,
  		University of Florida,
  		Gainesville,
  		FL 32611, USA\label{UFlorida}
  	\and
  		INAF, IASF Milano,
  		via Bassini 15,
  		I-20133 Milano,
  		Italy\label{Milan}
  	\and
  		Department of Astronomy, University of Geneva,
  		ch. d'\'Ecogia 16,
  		CH-1290, Versoix,
  		Switzerland\label{Geneve}
    \and
    	Departments of Physics and Astronomy, and Michigan Center for Theoretical Physics,
    	University of Michigan,
    	Ann Arbor, MI 48109
    	USA\label{UMich}
  	\and
  		Astrophysics and Cosmology Research Unit,
  		University of KwaZulu-Natal,
  		Durban, 4041,
  		South Africa\label{KwaZulu}
  	\and
  		Australian Astronomical Observatory,
  		PO BOX 915,
  		North Ryde, 1670,
  		Australia\label{AAO}
    \and
    	Department of Earth and Space Sciences,
    	Onsala Space Observatory,
    	Chalmers University of Technology,
    	SE-439 92 Onsala,
    	Sweden\label{Onsala}
    \and
		INAF - Osservatorio Astronomico di Brera,
		Via Brera 28, I-20122 Milano
		\& via E. Bianchi 46, I-23807 Merate,
		Italy\label{Brera}
    \and
    	Institute for Astronomy \& Astrophysics, Space Applications \& Remote Sensing,
    	National Observatory of Athens,
    	GR-15236 Palaia Penteli, Athens,
    	Greece\label{Athens}
    \and
		National Center for Supercomputing Applications and Department of Astronomy
		University of Illinois at Urbana-Champaign,
		Urbana, IL 61801,
		USA\label{Illinois}
  	\and
  		Department of Physics and Astronomy,
  		Macquarie University,
  		Sydney NSW 2109,
  		Australia\label{Macquarie}
  	\and
  		INAF - Osservatorio Astronomico di Padova,
  		Vicolo dell'Osservatorio, 5,
  		I-35122, Padova,
  		Italy\label{Padova}
  	\and
  		European Southern Observatory,
  		Alonso de Cordova 3107,
  		Casilla 19001, Vitacura,
  		Santiago de Chile,
  		Chile\label{ESO}
  	\and
  	    Dark Cosmology Centre, Niels Bohr Institute,
  	    University of Copenhagen,
  	    Juliane Maries Vej 30,
  	    DK-2100 Copenhagen,
  	    Denmark\label{Copenh}
  	\and
        Faculty of Physics,
        Ludwig-Maximilians University,
        Scheinerstrasse 1,
        D-81679 Munich, Germany\label{LMU}
  	\and
  		Institut de Physique Théorique,
  		CEA/DSM/IPhT and CNRS/URA\,2306,
  		F-91191 Gif-sur-Yvette, Cédex,
  		France\label{IPhT}
  	\and
  		Department of Physics and Astronomy,
  		University of Victoria,
  		3800 Finnerty Road,
  		Victoria, BC,
  		Canada\label{UVic}
  	\and
  		 Astrophysics Research Institute,
  		 Liverpool John Moores University, IC2,
  		 Liverpool Science Park, 146 Brownlow Hill,
  		 Liverpool L3 5RF, UK\label{Liverpool}
  	\and
  		ESO,
		Karl-Schwarzschild-Str. 2
		D-85748 Garching bei M\"unchen
		Germany  \label{ESOG}
	\and
		Astrophysics Department
		Max Planck Institut f\"ur Kernphysik
		Saupfercheckweg 1
		D-69117 Heidelberg
		Germany	 \label{Heidelberg}
  }

   \date{Received 3 July 2015 / Accepted 25 November 2015}

  \abstract
  % context heading (optional)
  {The XXL Survey is the largest survey carried out by the {\it XMM-Newton} satellite and covers
  a total area of 50 square degrees distributed over two fields. It primarily  aims
  at investigating the large-scale structures of the Universe using the distribution of galaxy
  clusters and active galactic nuclei as tracers of the matter distribution. The survey
  will ultimately uncover several hundreds of galaxy clusters out to a redshift of $\sim2$
  at a sensitivity of $\sim10^{-14}\rm\,erg\,s^{-1}cm^{-2}$ in the [0.5-2] keV band.}
  % aims heading (mandatory)
  {This article presents the XXL bright cluster sample, a subsample of 100 galaxy clusters
  selected from the full XXL catalogue by setting a lower limit of
  $3\times10^{-14}\rm\,erg\,s^{-1}\,cm^{-2}$ on the source flux within a 1$^\prime$ aperture.}
  % methods heading (mandatory)
  {The selection function was estimated using a mixture of Monte Carlo simulations
  and analytical recipes that closely reproduce the source selection process.
  An extensive spectroscopic follow-up provided redshifts for 97 of the 100 clusters.
  We derived accurate X-ray parameters for all the sources. Scaling relations were
  self-consistently derived from the same sample in other publications of the series.
  On this basis, we study the number density, luminosity function, and spatial distribution
  of the sample.}
  % results heading (mandatory)
  {The bright cluster sample consists of systems with masses between $M_{500}=7\times10^{13}$
  and $3\times10^{14}\,M_\odot$, mostly located between z=0.1 and 0.5.
  The observed sky density of clusters is slightly below the predictions from the WMAP9
  model, and significantly below the prediction from the {\it Planck} 2015 cosmology.
  In general, within the current uncertainties of the cluster mass calibration, models with
  higher values of $\sigma_8$ and/or $\Omega_M$ appear more difficult to accommodate.
  We provide tight constraints on the cluster differential luminosity function and find no
  hint of evolution out to $z\sim1$.
  We also find strong evidence for the presence of large-scale structures in the XXL bright
  cluster sample and identify five new superclusters.}
  % conclusions heading (optional)
  {}

\titlerunning{The bright XXL cluster sample}
\authorrunning{F. Pacaud et al.}
\maketitle

\section{Introduction}
%______________________________________________________________

In the current paradigm of cosmological formation of structures, the  evolution of dark
matter halos results from the competition between gravity and expansion.
Following General Relativity, the expansion rate of the universe is governed by
its energy content. As the largest bound and virialised entities in the universe,
galaxy clusters hold traces of the expansion history.
Therefore, provided we are able to estimate their masses reliably, the cluster population
constitutes an ideal cosmological `object' through which to test dark energy models that describe
the acceleration of the expansion at late cosmological times \citep[e.g.][]{Weinberg2013}.

Constraining cosmological models requires large samples of galaxy clusters spanning a wide
range of masses and redshifts.
The principle is to capture the evolution of the halo mass function and the halo
spatial distribution across cosmic times.
While cluster surveys can be carried out at different wavelengths
\citep[e.g]{Vikhlinin2009b,Mantz2010a,Rozo2010,Sehgal2011,Benson2013,
Planck2014XX}, the X-ray band offers special advantages: (i) X-ray cluster properties can
be easily analytically, and self-consistently, predicted by ab initio models and (ii)
catalogues are affected to a lesser extent by projection effects. Moreover, medium-depth
surveys routinely allow the detection of  massive clusters  out to $z\sim 1.5$ as well as
the systematic inventory of the group-size population at intermediate redshifts.
Given that cluster masses are not directly observable quantities, we usually rely on
scaling relations based on proxies such as the X-ray temperature or luminosity to
determine the cosmologically important halo mass distribution. For the most massive
objects, weak-lensing techniques provide independent mass estimates.
Because the formation of clusters is essentially gravity driven, it is generally assumed
that cluster properties are self-similar as a function of mass, size, and cosmic time
\citep{Kaiser1986}. This hypothesis is, however, still a matter of debate, especially for
the low-mass objects and needs to be definitively assessed or amended before self-consistent
cosmological analyses can be undertaken.
In addition, it has been shown \citep{Stanek2006,Pacaud2007,Nord2008} that the determination
of the scaling relations requires detailed knowledge of the selection function for the
samples in question; a quantity, which is, of course, also mandatory for the final-stage
cosmological analysis \citep{Sahlen2009,Mantz2010a,Mantz2015}.
The ab initio formalism describing the evolution of the  X-ray properties of clusters is
also, for the determination of the selection function, an advantage over the other wavelengths.
This advantage still holds when performing calculation of the selection function by means
of hydrodynamical simulations.

The systematic search for X-ray clusters has a long history, starting from the early times
of X-ray astronomy with the pioneering  HEAO-1 X-ray observatory \citep{Piccinotti1982}
and the subsequent Einstein Medium Sensitivity Survey \citep{Gioia1990a,Henry1992}.
A significant step was achieved by the ROSAT All-Sky Survey (RASS) \citep{Truemper1993},
which provided a wealth of galaxy cluster samples at various depths and completeness
levels: REFLEX-I, II \citep{Boehringer2001,Boehringer2014}, NORAS \citep{Boehringer2000},
the ROSAT-NEP \citep{Henry2006}, and MACS \citep{Ebeling2001} to cite only a few of them.
At the same time, it was quickly realised that the characterisation  of cluster samples
strongly benefits from  multiwavelength observations  (identification, spectroscopic
measurements, and extended calibration of mass-observable relations).
With the advent of {\it XMM-Newton} and {\it Chandra}, a great effort was devoted to small- and
medium-area surveys: COSMOS \citep{Scoville2007, Finoguenov2007b}, XMM-LSS \citep{Pierre2007},
and XMM-BCS \citep{Suhada2012}, complemented with serendipitous searches: XMM-XCS
\citep{Romer2001, Mehrtens2012}, X-CLASS \citep{Clerc2012b, Sadibekova2014}, and XDCP
\citep{Fassbender2011}.

The XXL Survey was designed to bridge the gap between these deep and/or
narrow surveys and the RASS.
Its ultimate science goal is to provide independent and self-sufficient cosmological
constraints, in particular on the dark energy equation of state, before
the start of missions covering very large fractions of the sky, such as Spectrum-Roentgen-Gamma
(SRG), which will carry the eROSITA instrument, and {\it Euclid}.

The survey concept is presented by \citet[hereafter Paper I]{Pierre2016}.
It consists of two independent  25~deg$^2$ fields observed with {\it XMM-Newton} at a
depth of 10~ks leading to the detection of clusters down to $\sim 10^{-14}$~ergs\,s$^{-1}$cm$^{-2}$
soft-band flux.
Two main aspects received special attention during the design of the survey.
The first is the ability to construct a sample containing a few hundred moderately massive
clusters, with very well understood selection effects, hence suitable for cosmological studies.
While the most massive clusters in the universe have been well identified out to a redshift
of about unity, complete samples in the $10^{13}-10^{14}M_\odot$ range are still lacking.
The advantage of having a unique extended connected area was also put forward for studies
of the large-scale structures.
The second aspect is the need for an associated comprehensive multiwavelength follow-up
programme with the most advanced ground-based and space observatories. This ensures both
internal cross-checks of the selection function and, when possible, a self-consistent
determination of the scaling relations for the cluster sample. In addition, the uniform XXL
data set enables the systematic assessment of the massive cluster density between redshifts
1 and 2.

The present article is part of the first XXL release based on the data collected during
the XMM AO10-11 observing periods.
The descriptions of the available X-ray data, their initial processing and quality checks
are presented in \citetalias{Pierre2016} together with a thorough description of the project's aims and methods.
While some 450 cluster candidates have been inventoried in the XXL Survey to date, we focus
here on the 100 brightest objects (hereafter the {\em \BXC}\ or the \Chundred).
This complete sample contains the highest signal-to-noise objects and allows a number of
in-depth analyses.
We perform various statistical studies of the cluster catalogue in a cosmological context.
Complementary results from the same sample are presented in other XXL papers of the first
series, in particular: the soft-band luminosity versus temperature scaling relation
\citep[hereafter Paper III]{Giles2016} and the quantification of cluster masses using weak
lensing \citep[hereafter Paper IV]{Lieu2016}.

The paper is organised as follows:
Section~\ref{sect:Dproc} presents the main steps of the data processing and the requirements
imposed on the data quality.
Section~\ref{sect:Ssel} describes the extraction of the XXL cluster catalogue and the
selection of the \BXC\ from the initial list of XXL extended source  candidates.
The resulting sample, including spectroscopic redshifts and X-ray luminosities, is
presented in Sect.~\ref{sect:cat}.
Section~\ref{sect:SelFunc} describes the derivation of the selection function.
The luminosity function, including evolution modelling,  is then compared with
cosmological models in Sect.~\ref{sect:Cosmo} and we conclude by a discussion
in Sect.~\ref{sect:Discuss}.

Unless otherwise stated, the results presented in this paper rely on the cosmological
parameters measured by \cite{Hinshaw2013}, based on the final, nine-year cosmic microwave
background (CMB) observations of the Wilkinson Microwave Anisotropy Probe satellite (WMAP),
combined with a set of baryon acoustic oscillation (BAO) measurements, and with constraints
on $H_0$ from Cepheids and type Ia supernovae: ($H_0 = 69.7\,\mathrm{km\,s^{-1}Mpc^{-1}}$,
$\Omega_m=0.282$, $\Omega_\Lambda=0.718$, $\sigma_8= 0.817$, $\Omega_b = 0.0461$,
$n_{s}=0.9646$ ).

\begin{figure*}
\begin{center}
	\resizebox{\hsize}{!}{\includegraphics[height=6.9cm,viewport=25 8 520 320,clip]{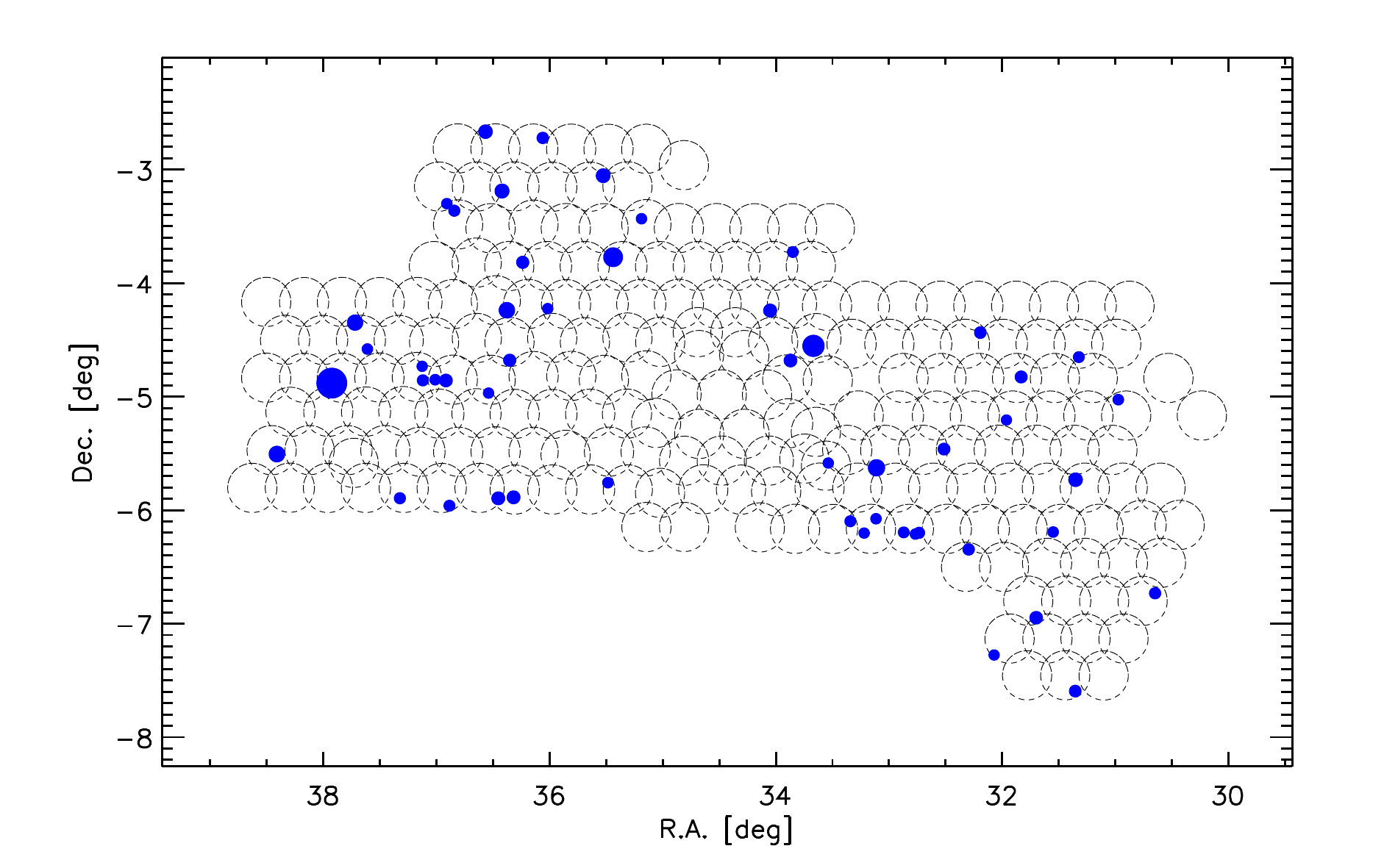}
	                      \includegraphics[height=6.9cm,viewport=0 10 345 318,clip]{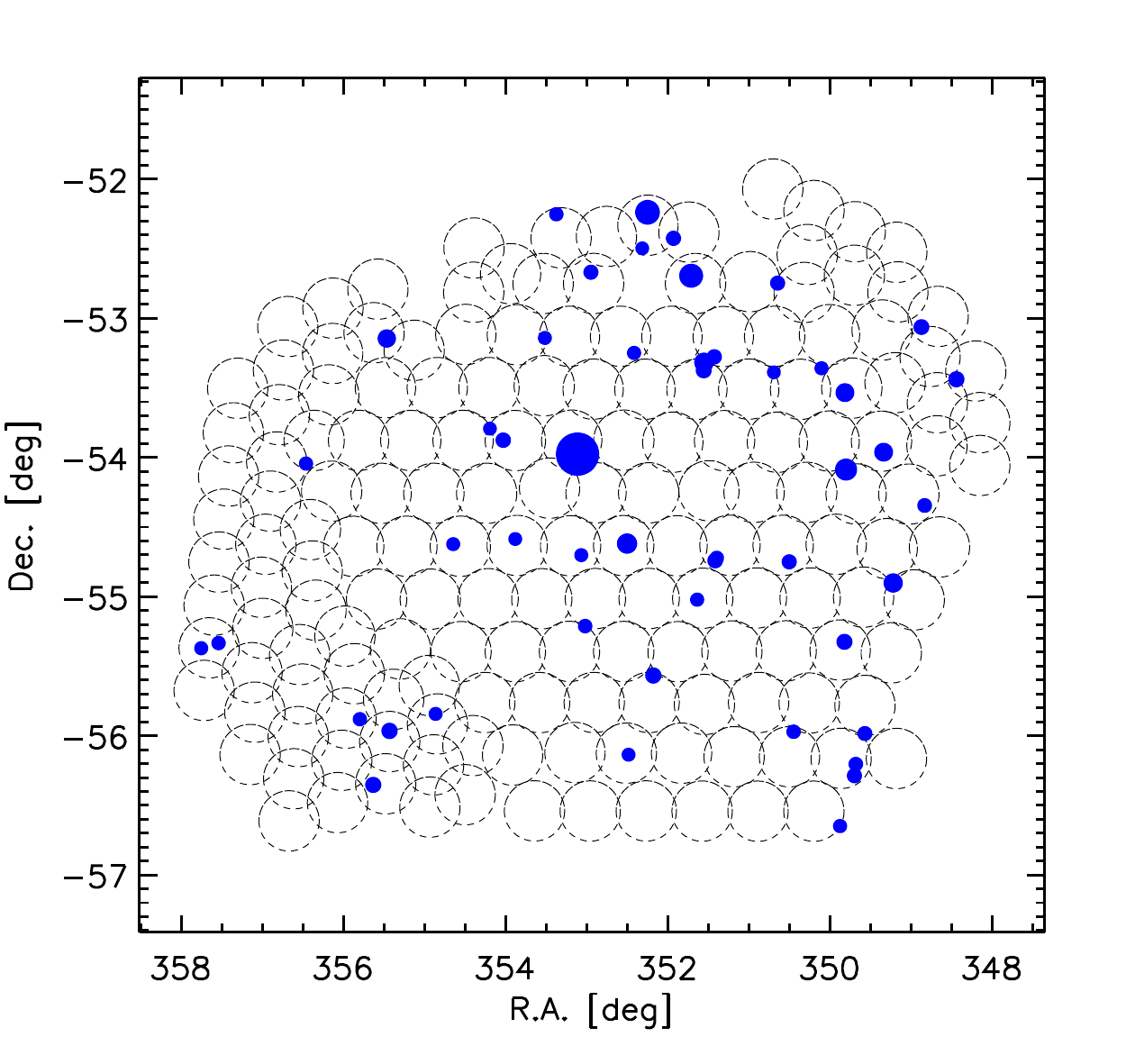}}
	\caption{
    \label{fig:pointings} Sky distribution of the XXL pointings that passed the selection
    criteria for the construction of the cluster sample.
    The dashed circles represent the area used for cluster detection on each pointing,
    i.e. a 13$^\prime$ radius around the average optical axis.
	The location of the members of the \BXC\ is shown with blue dots, whose size is
	proportional to their $60^{\prime\prime}$ aperture flux.
    The $\sim$ 1 \dd\ area located around RA=34.5 and Dec =-5, is the Subaru Deep Survey,
    whose position was selected for its lack of bright X-ray sources.}
\end{center}
\end{figure*}

\section{Data processing}
\label{sect:Dproc}

The full XXL X-ray data set consists of 2.7\,Ms of XMM observations from a dedicated AO10-AO11
Very Large Program (VLP), combined with more than 3\,Ms of earlier archival observations.
After splitting up XMM mosaic mode observations into the relevant individual exposures,
542 XMM pointings are available in total.
These are spread over 417 distinct sky positions and in total cover 50.9\,\dd\ of the
extragalactic sky.
We refer the reader to \citetalias{Pierre2016} for a complete description of the survey layout and design.

The raw data were re-processed  using recent calibration information and
{\tt XMM-SAS} version 10.0.2. Periods of high radiation level or particle contamination
were identified and filtered out from broadband ([0.3-10] keV) light curves, as proposed
by \cite{Kuntz2008}.
Following \cite{Pratt2002}, a histogram of the light curve values was fitted with
a Poisson law for each camera and the time intervals falling outside the $\pm3\sigma$
range were excluded.

Images, exposure maps and detector masks in several bands were then generated from the
filtered event-lists and telescope attitude using 2.5\arcsec pixels and were processed with
the XMM-LSS/XXL source detection pipeline, {\sc Xamin} \citep{Pacaud2006,Clerc2012b},
which we briefly summarise below.
The raw images were filtered in wavelet space using the method described by \cite{Starck1998},
which combines a rigorous treatment of the Poisson noise in low intensity signal with an
iterative image reconstruction process that prevents flux loss.  We then ran {\sc SExtractor}
\citep{Bertin1996} with a low detection threshold on the filtered images to extract a
conservative source catalogue and generate source masks.
Finally, a dedicated XMM maximum likelihood fitting procedure was used to determine
morphological parameters and flux for the source candidates, together with likelihood
ratios to assess the detection and source extent probabilities.

Given that a number of sky positions have been observed several times, usually because
of high flaring rates, we generated pseudo pointings by co-adding the images and exposure maps
of all suitable (possibly truncated) pointings at a given location; for the great
majority, there is only one relevant pointing per position.
The decision as to which pointings should enter a given stack was taken by optimising
the average signal-to-noise achieved in a 1\arcmin aperture.
The \BXC\ was defined based on the {\sc Xamin} catalogues of 412 stacks.
Indeed, 5 of the 417 sky positions correspond to pointed follow-up observations of
XMM-LSS/XXL clusters which fully overlap genuine survey pointings.
These were excluded from the source extraction process to avoid selection biases.

Despite the overall good homogeneity of the XXL data set, we observe variations in the
quality of the different stacks - this is particularly noticeable in the areas covered by
the AO10-AO11 VLP since most of the sky positions have been observed only once.
We ran extensive quality checks in order to filter out observations that are not suitable
for the subsequent derivation of the selection function.
These checks are discussed in detail in \citetalias{Pierre2016}.
This results in two limits: stacks with less than 3\,ks of clean observing time (18 of them)
were not considered useful owing to their strong initial contamination, poor depth, and the
difficulty to assess the background properties; high-background pointings
($>$4.5\,ct/s/pointing in the $[$0.5-2$]$~keV band) were also rejected since the modelling
of the selection function would become uncertain for low surface brightness sources.
This results in a subselection of 386 stacks.
Considering only the innermost 13\arcmin\ of each pointing, the survey covers a total
geometric area of 46.6\,\dd\ at high galactic latitudes (24.4\,\dd\ in the north and 22.2\,\dd\
in the south).
The selected pointings and the corresponding (geometric) sky coverage are shown in
Fig.~\ref{fig:pointings}. The figure also shows the lower density of pointings in the
southern field resulting from the larger spacings used for the preliminary XMM-BCS Survey
\citep{Suhada2012}. This, combined with the smaller total sky area, results in a lower sky
coverage for the southern field.

\section{Sample selection}
\label{sect:Ssel}
\subsection{The full XXL cluster sample}
\label{sect:XXLfull}
Extended source candidates are selected from the {\sc Xamin} maximum likelihood  outputs
in the {\tt extent / extent\_likelihood / detection\_likelihood} 3D space
\citep{Pacaud2006}. By default, all sources with measured {\tt extent} greater than
5\arcsec\ and {\tt extent\_likelihood} greater than 15 are considered as extended.
Among these, we differentiate a high significance sample -- the C1 class with an
{\tt extent\_likelihood} greater than 33 and a {\tt detection\_likelihood} greater than 32 --
from the remaining candidates, which we term the C2 class. The C1 sample has proven to be mostly free of
contamination by spurious detections or blended point sources both from simulation
\citep{Pacaud2006,Clerc2012b} and observations  \citep{Pacaud2007,Clerc2014}.
In comparison, the C2 population is about 50\% contaminated \citep{Pierre2006,Adami2011}.

For typical XMM cluster extents -- a core radius of 20\arcsec\ -- and nominal XXL observing
conditions, the C1 and C2 classes correspond to flux limits of $\sim2\times10^{-14}$\,\flux
and $\sim8\times10^{-15}$\,\flux, respectively.
However, strictly speaking, the {\sc Xamin} selection is a non-straightforward function of
both cluster flux and extent, as shown by \cite{Pacaud2006} and \cite{Clerc2012b}.
These aspects are discussed in more detail in Sect.~\ref{sect:SelFunc}.

Cluster substructures and multiple detections on neighbouring pointings have been
cross-identified with a 2\arcmin search radius and controlled by a human moderator before
removal. In doing so, we prioritised detections with the lowest off-axis angle.

%______________________________________________________________

\subsection{Identifying the hundred brightest XXL clusters}

The initial inventory of the XXL source population indicates that the cluster cosmology sample
will  ultimately contain approximately 500 clusters of galaxies.
The spectroscopic validation of such a large sample of cluster candidates
is a long  process, especially tedious at high redshift.
As of April 2015 (two years after the completion of the XMM observations), spectroscopic or
photometric redshifts have been determined for $\sim$ 85\% of the $\sim$ 450 X-ray
cluster detections to date.

For the first set of scientific studies, we focus on a complete subsample
of the brightest XXL clusters so as to enable statistical studies of cluster physics
for the group-mass range at intermediate redshifts (i.e. for the dominant population
of the survey) along with a preliminary cosmological analysis. This sample will allow
the most detailed X-ray analyses for the XXL cluster population and is intended to stand
as a benchmark for all future studies.

We decided to set the size of the \BXC\ to 100 objects and opted for a flux-limit
selection in a fixed angular aperture.
Such a model-independent procedure ensures that the selection is easily reproducible by
alternative processings.
The goal was to select a homogeneous set of clusters with at least 100 X-ray counts
in a high signal-to-noise area, taking into account the sensitivity drop of about 1/2 due
to the vignetting at the edge of the XMM field of view. It also minimises the impact of the
initial C2 selection since we only consider the most prominent objects in the survey in
terms of surface brightness. This was confirmed a posteriori by the fact that most of the
clusters in the final sample are classified as C1.

This high-flux selection effectively limits the bright sample to redshift of $z \sim 1$,
while 1-2 clusters per \dd\ are expected beyond redshift of unity in XXL \citep{Pierre2011}.
These clusters will not be used in the initial XXL cosmological studies: current
uncertainties on cluster evolution have a critical impact on the modelling
of the selection function of these objects and of their physical properties.
Rather, the distant cluster sample is undergoing dedicated multiwavelength studies
that will characterise their properties more accurately and will permit to develop
procedures that lead to reliable mass estimates.

\begin{figure*}
   \begin{center}
      \vspace{-0.3cm}
      \resizebox{\hsize}{!}{\fbox{\includegraphics[height=6.5cm]{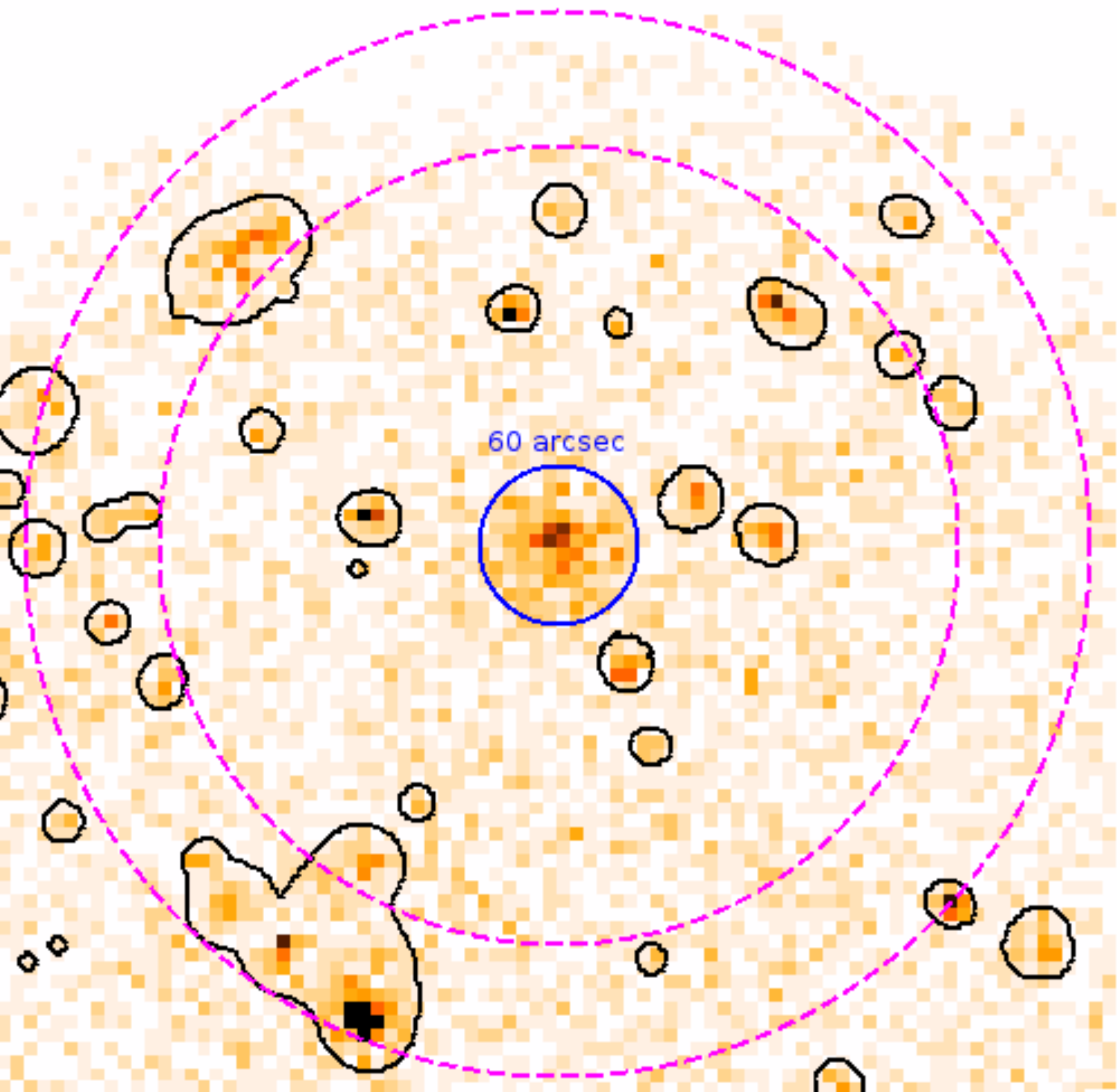}}
                            \includegraphics[height=6.5cm,viewport=10 10 440 293,clip]{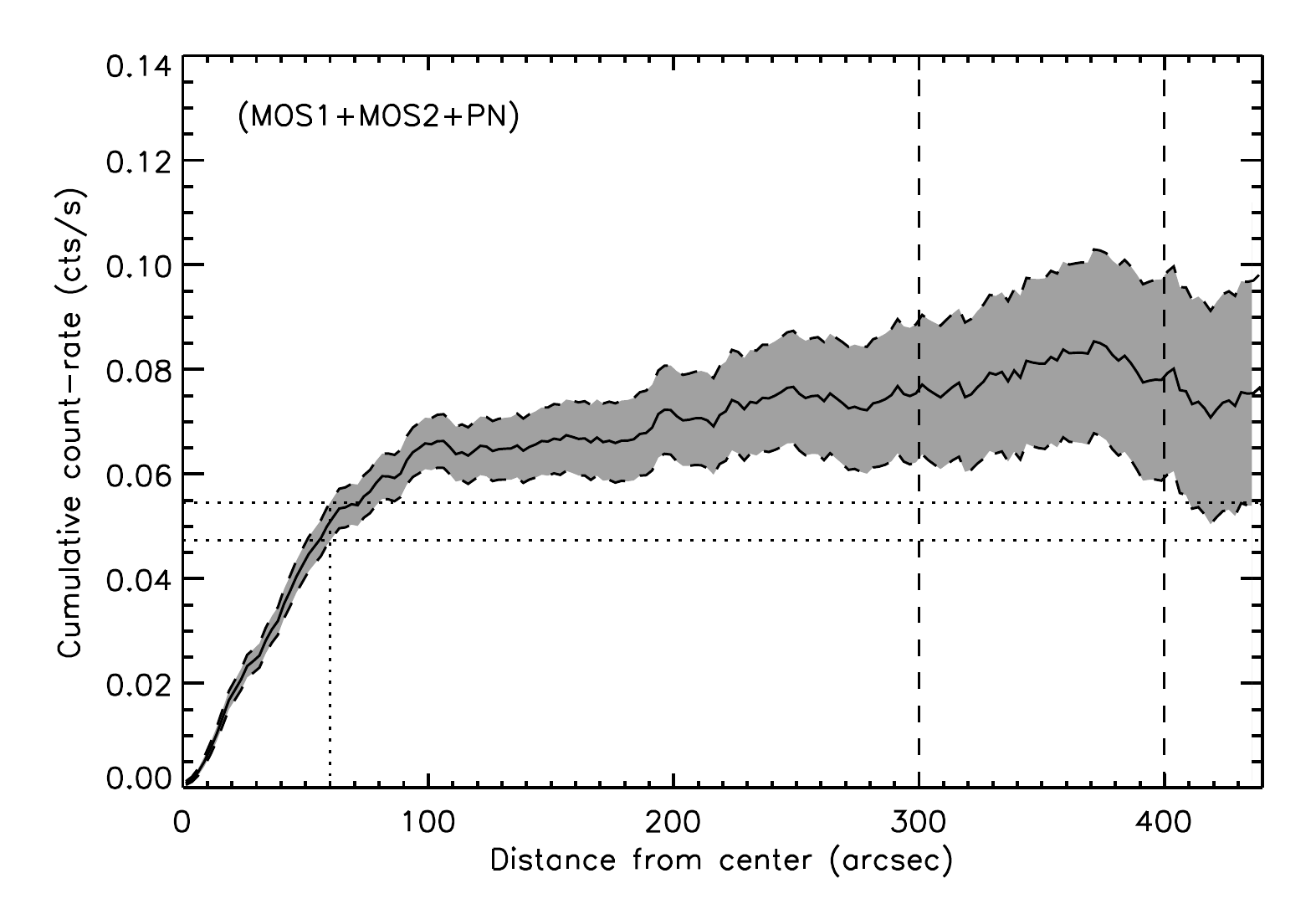}}
      \vspace{0.1cm}
      \caption{\label{GCA}Growth curve analysis of galaxy cluster XLSSC~010 at $z=0.33$.
        The 60\arcsec aperture flux of this cluster is close to the median value in the \BXC.
		{\it Left}: Source masking is shown by the black lines.
		The two dashed pink circles delimit the annulus used for the background estimate.
		The cluster selection was made from the integrated [0.5-2] keV count rate measured
		within a radius of 60\arcsec (blue line).
		{\it Right}: Growth curve of XLSSC~010. The grey area represents the 1-$\sigma$
		confidence interval.
	    The dotted lines show the 60\arcsec extraction radius (vertical line) and the corresponding
	    1-$\sigma$ interval for the aperture count rate (horizontal lines).
	    The dashed vertical lines indicate the radial range within which the background was estimated.
        \vspace{-0.2cm}}  
  \end{center}
\end{figure*}

To select the brightest sources, we start from the cleaned C1+2 source list (i.e. the union
of the C1 and C2 clusters), for which the {\sc Xamin} pipeline provides  angular core radius
($\theta_{c}$) and total flux estimates from a fit with a $\beta = 2/3$ surface brightness
model \citep{Cavaliere1976}.
This enables us to derive approximate count rates within a given radius.
A preliminary step was to select the 184 brightest clusters from the current
XXL C1+2 catalogue.
This pre-selection was based on pipeline total count rates in the [0.5-2] keV band,
integrated out to a fixed angular radius of 60\arcsec; this radius was chosen to encompass
a significant fraction of the cluster emission at all distances (a cluster
core radius of 180 kpc ranges from 100 to  23\arcsec for $0.1<z<1$). Only clusters with a
total pipeline count rate above 0.025~cts/s in the aperture were retained.
Subsequently, the aperture photometry of every cluster from this primary selection
was re-evaluated using a growth curve analysis (GCA) as described by \citet{Clerc2012b}.
In particular, nearby-source masking, cluster centring and the local background estimate
were interactively optimised; one example is shown in Fig.~\ref{GCA}.
This finally allowed us to select the 100 brightest XXL clusters within a 60\arcsec flux aperture.
In the process, we assume that there is a one-to-one correspondence between count rate and
flux and so we ignore the temperature dependence of the energy-to-flux conversion factor
(ECF). This is necessary since some sources were not observed by one of the detectors,
either because of a misalignment of the MOS and PN detectors or because some of the MOS1
CCDs are switched off in recent observations. The flux selection can therefore be considered as a count rate limit,
in which average ECFs were used to extrapolate the information for missing detectors.
The assumption of a fixed ECF has little impact in practice, since the
cluster selection operates in the narrow [0.5-2] keV band where the XMM effective area curve
shows little variation over typical cluster spectra.
By iteratively comparing the 60\arcsec count rates from the pipeline output and from the
growth curve analysis, we estimate that the sample defined in this way is more than
99\% complete, for the initial pre-selection of 184 clusters.
However, uncertainties on count rate measurements from the growth curve analysis
(typically 15 \% at the faint end of the sample) render the selection limit of the faint
objects somewhat fuzzy.
Given the count rate distribution of the pre-selection, this concerns about 20 clusters just
above and below the selection threshold which are randomly reshuffled.
Therefore, a model of the selection function must include the measurement errors arising from
the growth curve analysis.

The flux limit of the final catalogue is 3$\times 10^{-14}$ \flux\ in a $60''$ radius aperture;
this is equivalent to a total MOS1+MOS2+PN count rate of 0.0332 cts/s in $[$0.5-2$]$ keV,
with the assumed ECF of $1.107\times 10^{12}\, \mathrm{cts\,erg^{-1} cm^{-2}}$\footnote{All
conversion from luminosities to count rates or fluxes in this paper use an average $n_H$
column density over the two XXL fields of $1.8\times10^{20}\,\mathrm{cm^{-2}}$.}.
%______________________________________________________________

\section{The catalogue}
\label{sect:cat}
%______________________________________________________________

The \BXC\ consists of 96 C1 and 4 C2 clusters, almost equally distributed between
the two XXL fields: 51 in XXL-N and 49 in XXL-S. The clusters and their main parameters
are presented in Table~\ref{BXCTab}.
In addition, seven additional clusters above the flux limit were identified but not included
in the sample: five that were detected in the bad quality stacks and two that did not pass
the pipeline C1+C2 selection criteria. These clusters are listed in Table~\ref{SuplTab} for
completeness, but they will not be used in the scientific analysis presented here.
Some members of the \BXC\ are already known from previous works:  22 of them are in
the initial XMM-LSS area \citep{Clerc2014}, 13 in the initial BCS X-ray
sample \citep{Suhada2012}, and 12 from other projects (including the ROSAT All-Sky Survey
and optical surveys).

\subsection{Cluster redshift validation}

The determination and validation of the cluster redshifts is an important part of the XXL
Survey effort necessary to reach the full survey scientific potential.
Obtaining spectroscopic redshift for several hundred clusters out to a redshift of 1 and
beyond is nevertheless a long-term process involving numerous observing facilities,
hence requiring a dedicated management and substantial manpower. \\
In the first step, we used the CFHTLS
T0007\footnote{{\tt http://www.cfht.hawaii.edu/Science/CFHTLS/T0007/}}  and BCS
\citep{Menanteau2010} photometric redshift catalogues to assign preliminary cluster
redshifts; the procedure looks for a galaxy overdensity in redshift-space around the
X-ray centroid.
Subsequent spectroscopy has shown that the CFHTLS and BCS estimates have errors $<0.08$ and
$0.12$, respectively, for 99 \% of the clusters; these precisions are valid up to
$z \sim 1$.\\

\begin{figure}[t!]
 \begin{center}
 \includegraphics[width=7.5cm,viewport=23 8 415 295,clip]{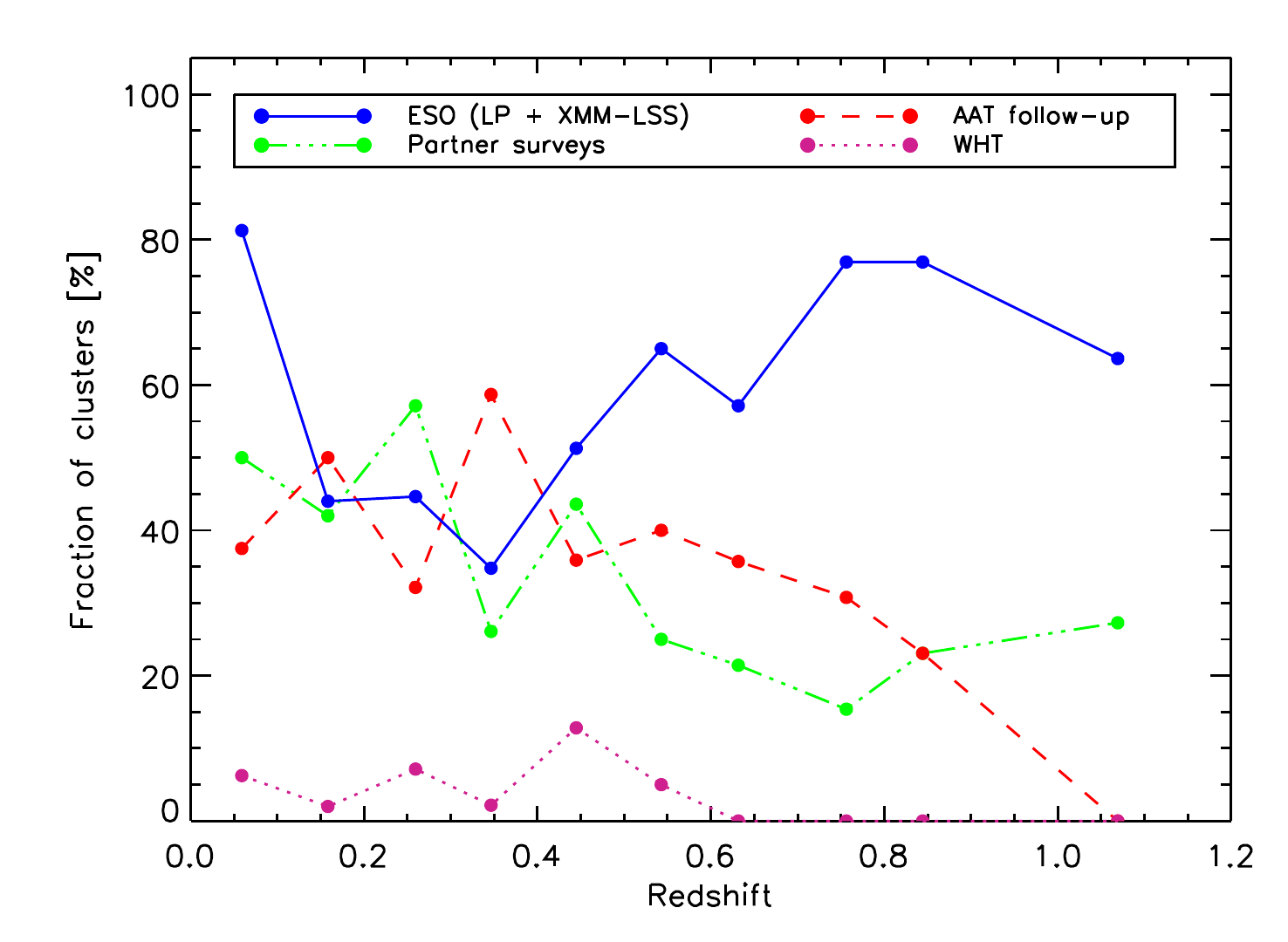}
 \caption{Origin of the \Chundred\  spectroscopic redshifts in different redshift slices.
 The available spectroscopic data are split into four categories: (i) the follow-up data obtained
 by the dedicated XXL ESO Large Program (LP, PI: C. Adami) using a combination of NTT/EFOSC2
 and VLT/FORS2, together with previous data obtained by the XMM-LSS team with the same instruments,
 (ii) the XXL follow-up program based on the  the Anglo-Australian Telescope (AAT, PI: C. Lidman),
 (iii) the XXL William Herschel Telescope (WHT) follow-up program (PI: B. Poggianti), and (iv)
 redshifts obtained through partner spectroscopic surveys, including GAMA \citep{Driver2011},
 VIPERS \citep{Garilli2014} and the VVDS \citep{LeFevre2013}. The observations and data reduction for the redshifts
 obtained at the WHT and AAT are respectively the subject of XXL \citetalias{Koulouridis2016}
 and \citetalias{Lidman2016}.
 For each category, the plot shows the proportion of clusters with galaxy redshifts coming from
 the corresponding project.\label{Zorigin}}
\vspace{-0.3cm}
\end{center}
\end{figure}

As described in \citetalias{Pierre2016}, a comprehensive follow-up programme has been tailored in order to
obtain spectroscopic redshifts (and, possibly, velocity dispersions) for all XXL clusters
and for a significant and representative fraction of the active galactic nucleus (AGN)
population.
The enterprise involves already available redshifts from external programmes, like
VIPERS \citep{Garilli2014} or GAMA \citep{Driver2011}, as well as from dedicated XXL follow-up
campaigns at ESO (Large Programme), Anglo-Australian Telescope \citep[hereafter Paper XIV]{Lidman2016},
William Herschel Telescope \citep[hereafter Paper XII]{Koulouridis2016}, and others.
We refer to our dedicated website\footnote{{\tt http://xxlmultiwave.pbworks.com}}, for detailed
and regularly updated information on the XXL follow-up activities.
All available spectra in both areas along with redshift values, measurement errors and
quality flags are stored in the CESAM database in
Marseille\footnote{{\tt http://cesam.lam.fr/xmm-lss/}}, from which the spectroscopic
identification of the X-ray sources is performed.
For clusters of galaxies, we follow here the procedure developed for the XMM-LSS survey
\citep{Adami2011}.
This task is conducted mostly in interactive mode and undergoes independent checks by at
least two moderators.
We assume that a cluster is spectroscopically confirmed either i) when there
are three concordant redshifts within a radius of 500 kpc from the X-ray centroid or
ii) when at least the cD galaxy has a spectroscopic redshift, taking into account the
preliminary information provided by the photometric redshift analysis.
For the vast majority of the \BXC , the spectroscopic identification was straightforward;
in some cases, where for example a cluster is undergoing a merger event or two velocity
groups are superposed along the line of sight, a human decision was necessary;
these very few cases are documented in Appendix~\ref{append:ClusterProp}.
The origin of the redshifts of the \BXC\ is shown in Fig.~\ref{Zorigin}.
At the time of paper submission, all clusters but three have been spectroscopically confirmed;
the remaining ones are scheduled for the next ESO observing periods (Large Programme by
C. Adami).
For these few objects we still use the photometric redshift estimates for the first series
of XXL papers (see Table~\ref{BXCTab}).

\begin{figure}[t!]
\begin{center}
\includegraphics[width=8.5cm,viewport=5 10 412 486,clip]{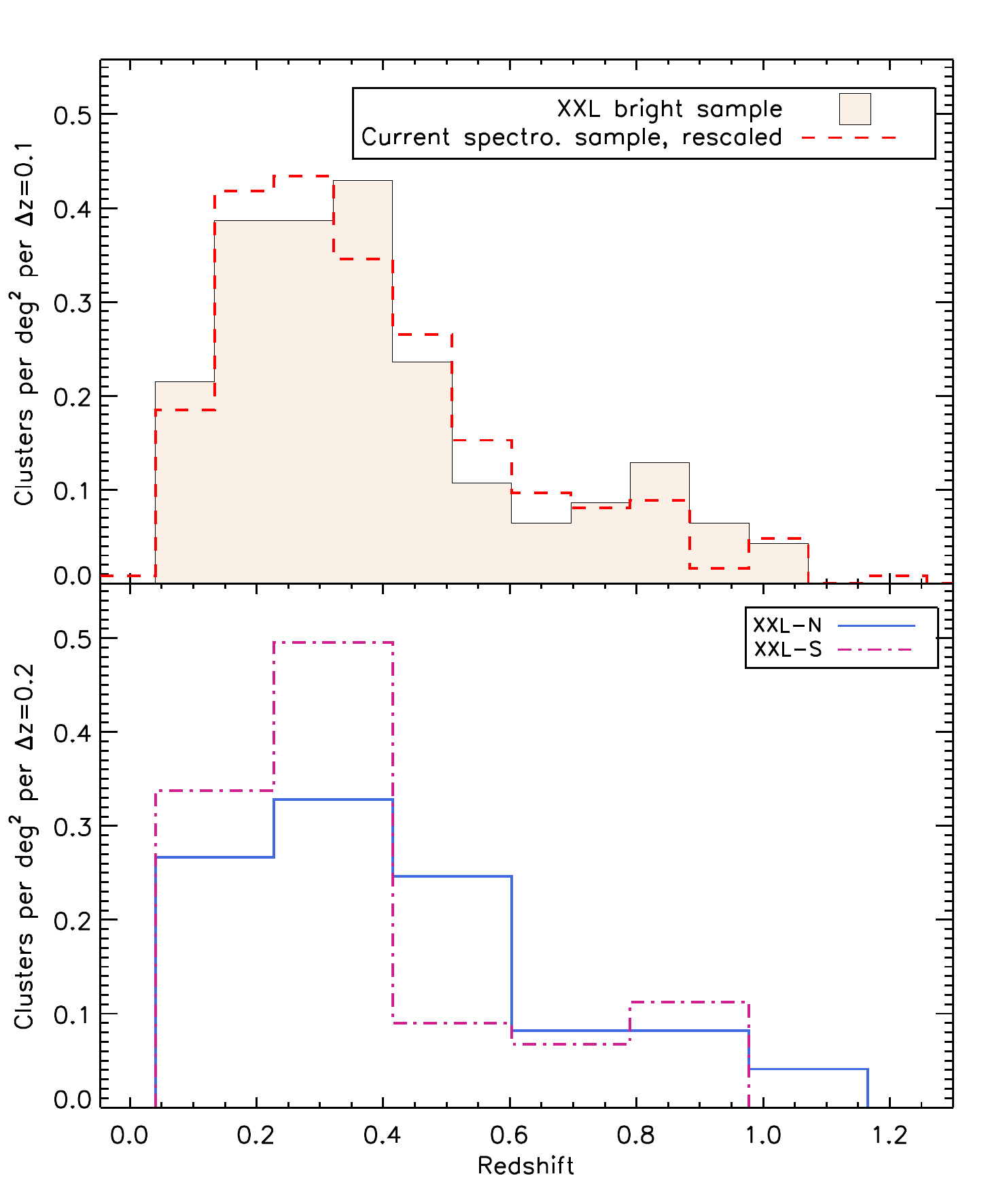}
\caption{Redshift distribution of the \BXC .
   Top: The bright sample compared with the distribution of 267 spectroscopically
   confirmed XXL clusters (to date). The latter has been rescaled to the same total number.
   Bottom: North/south fields separately (the bin width was doubled to obtain
   a shot noise comparable to that of the complete histogram).
 \label{Zdist}}
\end{center}
\end{figure}

The redshift distribution of the \BXC\ is shown in Fig.~\ref{Zdist}.
It is very similar, within the statistical noise, to the distribution
of the total set of currently confirmed XXL clusters, which is $\sim$2.5
times larger. This probably results from our choice of a fixed angular aperture for
the flux cut, which favours the inclusion of the more compact high redshift clusters
despite their lower average total flux.

The cluster density is similar in the two XXL fields.
Despite the slightly larger area and deeper exposure in the north (see
\citetalias{Pierre2016} and Sect.~\ref{sect:Dproc}), we observe marginally higher
numbers in the southern field.
The deviations are, however, always within the expectations from shot noise.
Interestingly, the redshift distribution of both the \BXC\ and the currently confirmed
XXL clusters do not seem to reproduce the deficit of clusters at redshifts of  $0.4<z<0.6$
previously reported by \cite{Clerc2014}, who studied a subsample of the full XXL
C1 catalogues. Either this lack of clusters was simply the result of cosmic variance,
or it results from a deficit of low-mass systems below the flux limit imposed in the
present work.

\vspace{-0.15cm}
\subsection{Cluster X-ray properties}
The clusters underwent a detailed X-ray spectroscopic analysis, which is extensively
described in \citetalias{Giles2016}.
Briefly, XMM spectra were extracted for all the available detectors in a fixed 300~kpc
circular aperture around the pipeline position. The background was modelled, for each
XMM instrument, using an annulus encompassing the same off-axis range as the source
aperture. When the source was too close to the centre, a surrounding annulus was used
instead.
The spectra were fitted with a thermal emission model including absorption
by Galactic neutral hydrogen \citep{Kalberla2005}. Since the metallicity could only
be constrained for a few systems, we fixed it to 0.3$Z_\odot$ for all clusters to
ensure self-consistency.
In addition to the spectroscopically weighted temperature, $\Txxl$, the rest
frame $[$0.5-2$]$ keV luminosity $\Lkpc$ in the extraction region was derived
as part of the fit and corrected for the masked areas.
To obtain physically meaningful luminosities, we estimated $\ClRadMT$ and
$\ClMassMT$\footnote{The radius within which the average total mass density of the cluster
equals 500 times the critical density estimated at the cluster redshift, and the total
mass within this radius.} from the best fit temperatures and the average
$\ClMassWL-\Txxl$ scaling relation derived in \citetalias{Lieu2016} from a weak lensing analysis
of the \Chundred\ clusters overlapping the CFHTLenS area \citep{Erben2013} combined with
clusters from the Canadian Cluster Comparison Project \citep{Hoekstra2015}.
The parameters of this scaling relation are given in Table~\ref{tab:ScalPar}.
The values of $\Lkpc$ were then extrapolated to $\ClRadMT$ using, for consistency,
the same $\beta$-model as for the selection function modelling (see Sect.~\ref{sect:SelFunc}).

\begin{figure}[h!]
   \resizebox{\hsize}{!}{\includegraphics[width=9cm,viewport=18 10 415 295,clip]{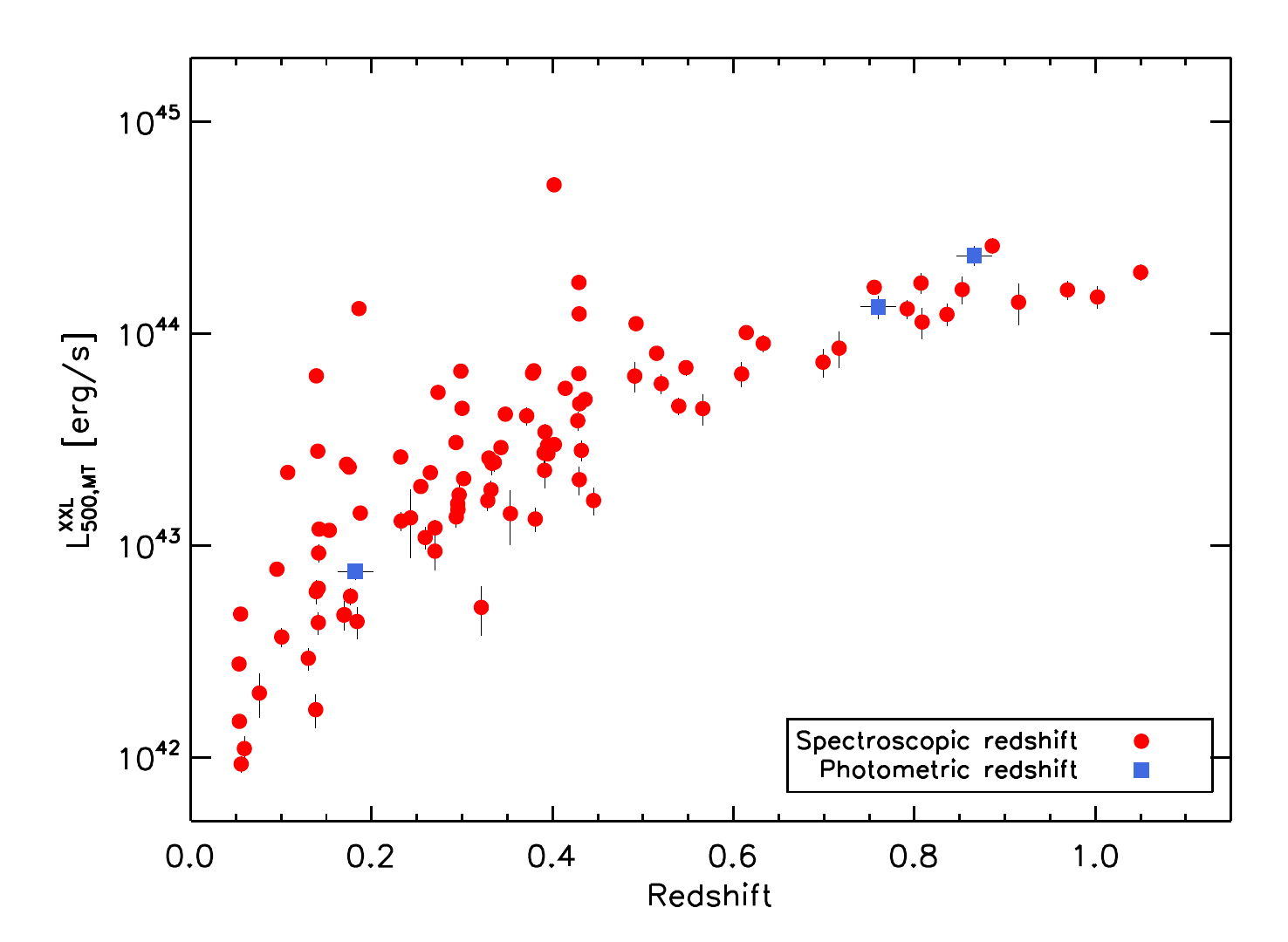}}
   \caption{Distribution of the \BXC\ in the $\Lxxl$ vs redshift plane.
   The luminosity is measured in the [0.5-2] keV band and integrated to $\ClRadMT$.
   The data show a clear Malmquist bias due to the flux selection combined
   with volume effects that prevent us from detecting massive (and thus luminous)
   clusters at low redshift.\label{fig:LzDist}}
   \vspace{-0.1cm}
\end{figure}

\begin{figure}[t!]
   \resizebox{\hsize}{!}{\includegraphics[width=9cm,viewport=18 10 415 295,clip]{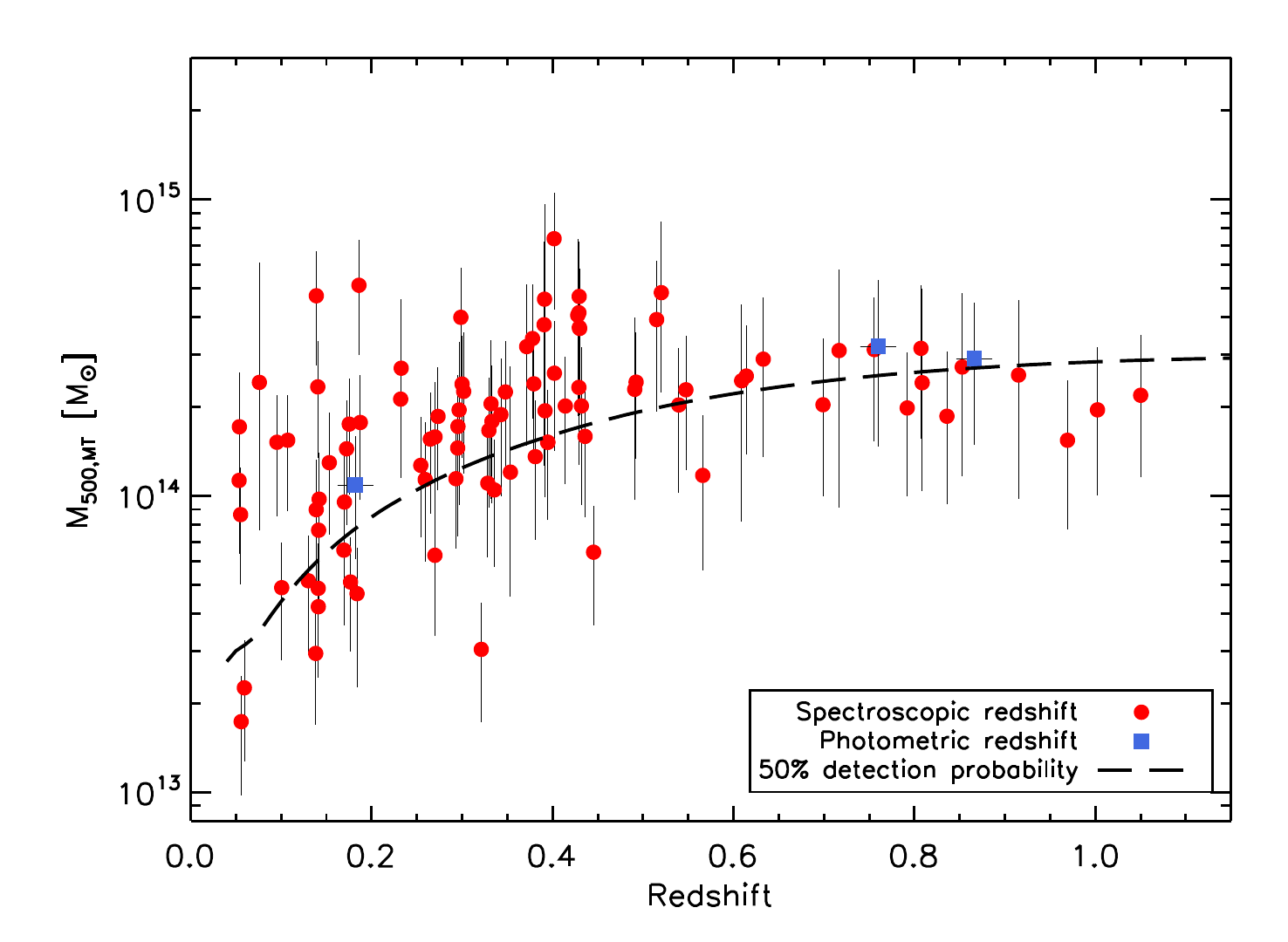}}
   \caption{Distribution of the \BXC\ in the $M_{500,MT}$ vs redshift plane.
   Although many low mass systems are detected at low redshift ($\sim20\%$ of the sample
   with $M_{500,MT}<\times10^{14}\,M_\odot$), the high redshift subsample mostly contains
   similar mass clusters. The dashed line shows the 50\% completeness limit in the WMAP9
   cosmology from the model described in Sect.~\ref{sect:PhysMod}.\label{fig:MzDist}}
\end{figure}

The resulting cluster luminosities and masses within $\ClRadMT$, $(\Lxxl,\ClMassMT)$ are
listed in Tables~\ref{BXCTab}-\ref{SuplTab}. The quoted errors on cluster masses combine
the uncertainty on the cluster temperatures with the intrinsic scatter of the
$\ClMassWL-\Txxl$ relation but do not include the uncertainty on the scaling relation
parameters (again for consistency with the selection model). However, the corresponding
errors on $\ClRadMT$ are not propagated to $\Lxxl$ (but the effect is small).

Figure~\ref{fig:LzDist} shows the distribution of the \BXC\ in the $(z, L_X)$ plane,
revealing the usual Malmquist bias of flux/magnitude-limited samples.
The figure also shows the lack of massive clusters at low redshift resulting from volume
effects, implying that the sample shows a strong mass versus redshift correlation. This
correlation increases the degeneracy between the redshift evolution and the other parameters
in the self-consistent scaling relation analysis based on the XXL clusters alone. Such errors
propagate straight into the cosmological analysis of the sample, as we will show later.
We note, however, that the correlation will become milder with the full XXL cluster sample
as we add lower luminosity clusters at all redshifts.

The mass distribution of the sample is illustrated in Fig.~\ref{fig:MzDist}.~Surprisingly,
the high redshift part of the sample ($\simeq$$40\%$~of the total) covers a narrow range in
cluster masses ($M_{500,MT}\!=\!2$$-$$3\times10^{14}\,M_\odot$). We point out that this is not a direct
consequence of picking up clusters with similar luminosities (close to the detection limit)
at high $z$  since the $\ClMassMT$ estimates are derived from the  measured temperatures
rather than luminosities.
Instead, it likely results from a subtle balance between the evolution of the
cluster number density, the evolution of scaling laws and the apparent sizes of clusters.

In Fig.~\ref{fig:MzComp}, we compare the distribution of the \ChundredShort\ clusters in mass
and redshift with samples extracted from different surveys. The \Chundred\ probes lower
masses than most previous ICM-based cluster surveys. The 400d survey \citep{Burenin2007}
stands as an example of projects that rely on deep archival ROSAT pointings.
For consistency, we estimated the cluster masses from their measured temperature combined
with the XXL $\ClMassWL-\Txxl$ scaling relation of \citetalias{Lieu2016}.
ROSAT surveys do not exhibit the flat mass
selection observed at high redshift in the \Chundred. We also show the largest samples selected
from the Sunyaev-Zel'dovich effect (SZE) with the South Pole Telescope \citep[][SPT, ]{Bleem2015}
and the {\it Planck} satellite  \citep{Planck2015XXVII}. For these data, we relied on the masses
derived from scaling relations between $M_{500}$ and the SZE signal by each of the teams.
A few XXL clusters correspond to SZE detected massive systems. They are listed in the
individual notes of Appendix~\ref{append:ClusterProp} and reveal a good agreement between
the $\ClMassWL-\Txxl$ based masses obtained in this work and estimates based on the SZE.

Clearly, the \Chundred\ probes lower masses than most previous ICM-based cluster
samples, including those selected from deep ROSAT pointed observations. The full XXL sample
will go even deeper and will uncover a new, little-studied, high redshift, and low mass
cluster population.

\begin{figure*}
\begin{center}
  \resizebox{\hsize}{!}{\includegraphics[height=7cm]{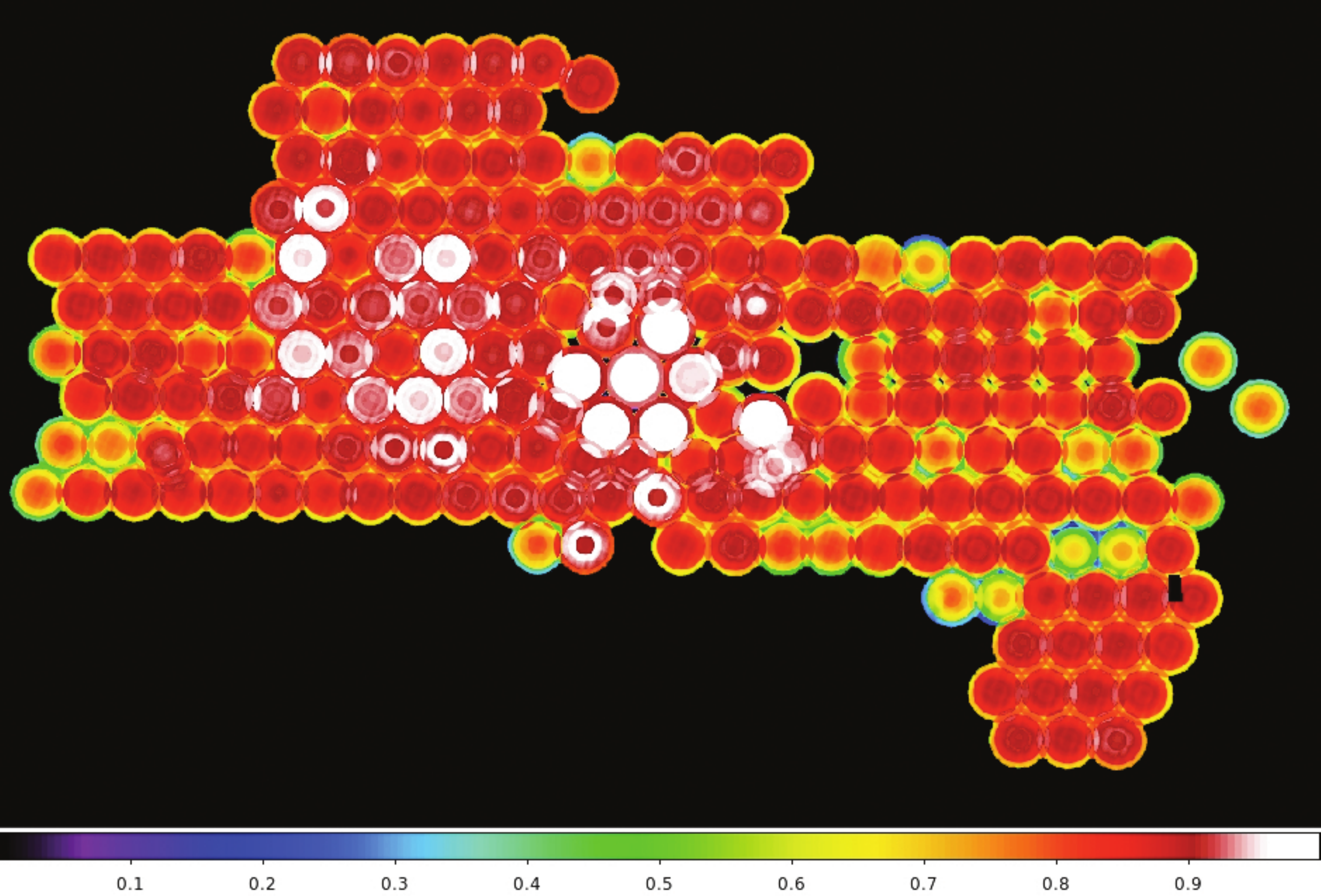}
  \includegraphics[height=7cm]{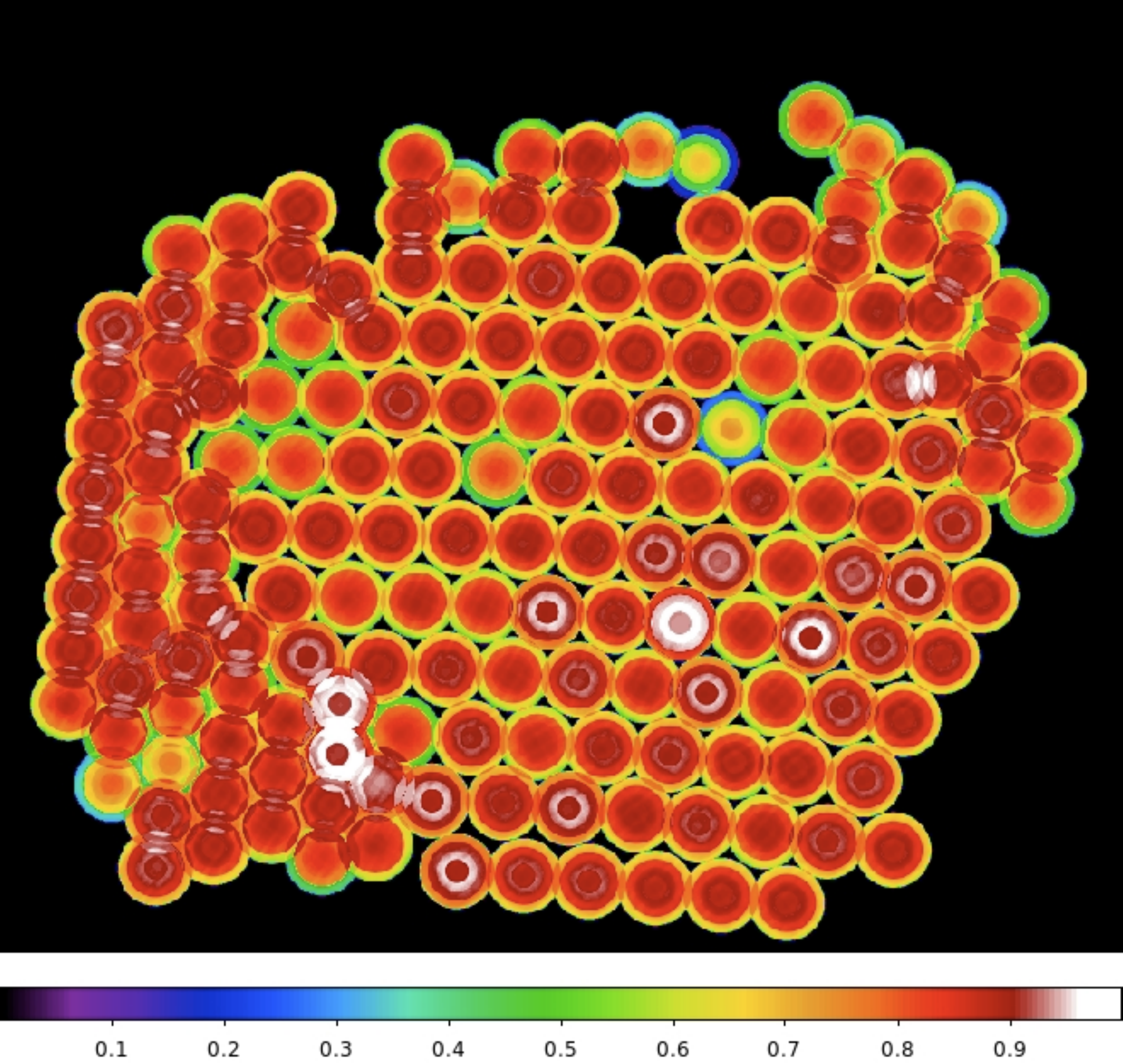}}
  \caption{Detection efficiency map for a cluster with a core radius $r_c$=20\arcsec and a
  total XMM count-rate (integrated to infinity) of $CR_\infty$=0.055~cts/s.
  The maps include the effect of the initial C1+2 selection, the subsequent aperture flux cut
  with its measurement error, and the combination of overlapping pointings.\label{fig:2Dproba}}
  \vspace{-0.2cm}
\end{center}
\end{figure*}

\section{Selection function}
\label{sect:SelFunc}

The \BXC$\,$ relies on the initial C1+2 selection combined with a later flux cut.
Given the high subsequent~flux cut, one would expect the C1+2 incompleteness to have
little impact on the final selection function. However, there is still a residual effect,
which causes very extended sources to be detected less efficiently or compact sources
to be mistaken for point sources. In addition, the flux measurement errors of the~GCA
($\sim$15\% near the threshold) play an important role in the selection process and depend
on the local exposure time (and~therefore the pointing on which the source was detected).
Finally, the estimate of the completeness in regions where several pointings overlap
must exactly follow the order and manner in which the two selection steps are applied
to the source candidates.

In this section we describe a solution to each of these problems and discuss the
resulting \Chundred\ selection function.

%______________________________________________________________
\subsection{Pipeline selection function}

The C1+2 selection has been studied in depth in previous works, using simulations
of $\beta$-model clusters (e.g. \citealt{Pacaud2006}).
In this work we follow the methodolgy of \citet{Clerc2012b}, where clusters are simulated
in synthetic observations with variable exposure time, $t_{exp}$, and particle background
level, adding AGNs as a spatially uncorrelated point-source distribution that follows
the flux distribution of \citet{Moretti2002}. For the background, we take as nominal values the
average sky emission and XMM instrumental background measured by \citet{Read2003}, and allow
for a variable particle background level by applying a rescaling factor, $b$, to the
instrumental background. For each set of $\beta$-model parameters, the C1+2 selection
function~was then estimated in four off-axis angle bins (0$-$4\arcmin, 4$-$7\arcmin, 7$-$10\arcmin\
and 10-13\arcmin) and for several values of the exposure time and background level in the
range $3\,\mathrm{ks}<t_{exp}<40\,\mathrm{ks}$ and $0.25<b<4$.

In practice, evaluating the average exposure time of a given pointing is straightforward,
but this is not as trivial for the background parameter. To estimate $b$, we use a
least-square matching procedure that compares the background measured by the {\sc Xamin}
pipeline at the position of each detected source to the one derived at a similar off-axis
angles in simulations with different values of $b$.
We then interpolate the selection functions to the proper $b$ and $t_{exp}$ for each
pointing.

Although the $\beta$-model cluster profiles are a function of both $r_c$ and $\beta$, we showed
in \cite{Pacaud2007} that the C1+2 selection mostly depends on the width of the profile (e.g.
FWHM) and the flux in the central parts. Since all the analysis in the following relies on
the flux within 60\arcsec or $\ClRadMT$ and ignores the contribution from the cluster
outskirts, the impact of $\beta$ on the selection is dominated by the degeneracy between
$\beta$ and $r_c$. We therefore fix $\beta$ to a canonical value of 2/3 in the remaining.
For each value of $b$ and $t_{exp}$, the C1+2 detection probability is estimated for core
radii in the range 10\arcsec$<r_c<$100\arcsec and total XMM count rates (summed over the
three detectors and integrated to infinite radius) spanning 0.0025~ct/s$<CR_\infty<$0.5~ct/s.

From here on, we denote the pipeline selection function in pointing $p$, i.e. the
probability of inclusion in the sample, as $\SFunc{C1+2,p}{CR_\infty,r_c,RA,Dec}$, where
the dependence on $t_{exp}$ and $b$ is encapsulated in the pointing under consideration and
the off-axis angle is implicitly given by the sky position.

%______________________________________________________________
\subsection{Modelling the flux cut}

While a simple limit on the aperture flux would be easy to implement, it would not
accurately reproduce the \ChundredShort\ selection process since the GCA photometry is affected by
noise. A proper understanding of the count rate measurement errors is therefore
necessary.

\begin{figure}
    \begin{center}        
        \includegraphics[width=8.3cm,viewport=20 7 415 325,clip]{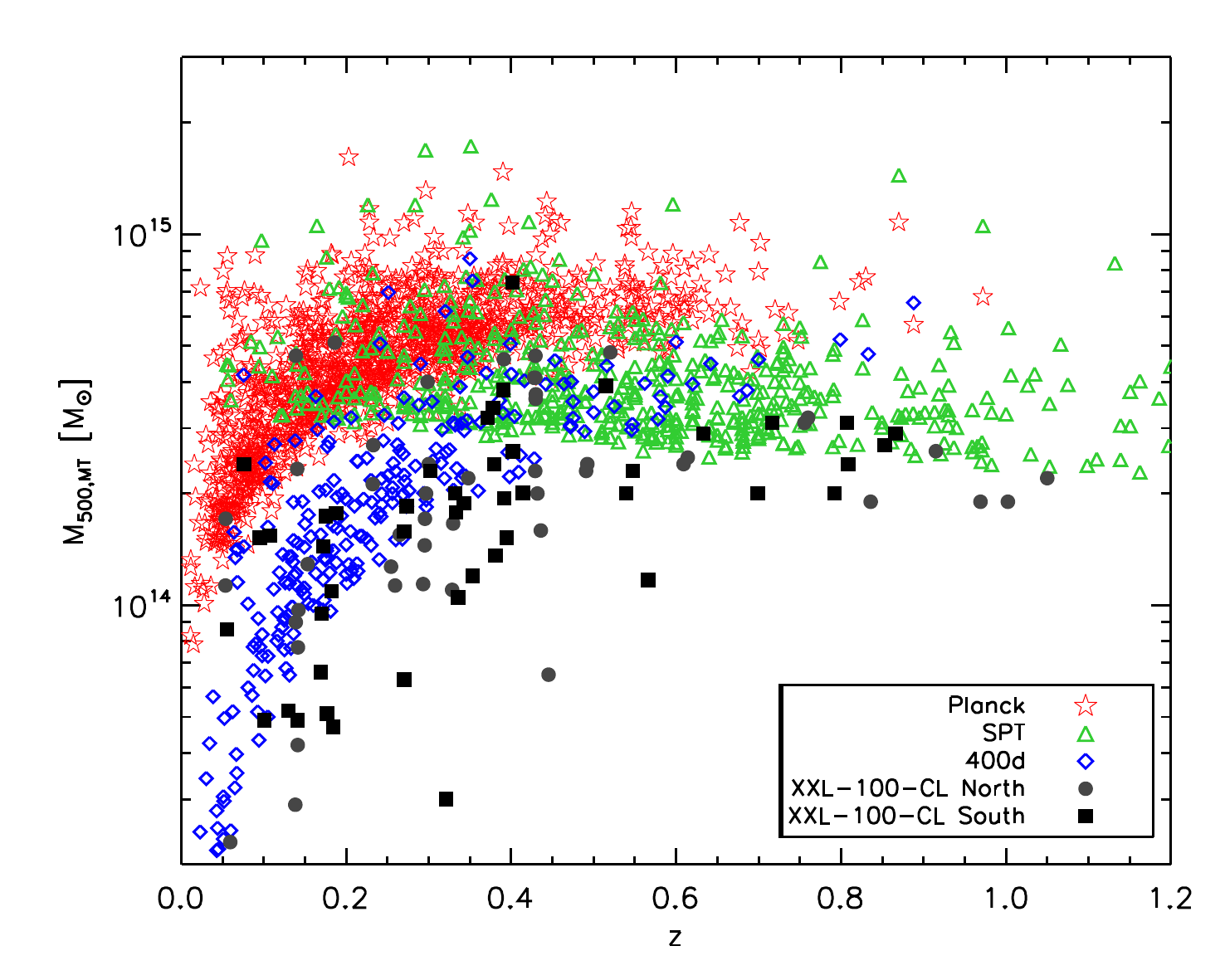}
	\caption{Comparison of the mass and redshift distribution of the \BXC\ with other
	cluster samples. We detect on average lower-mass / higher-redshift clusters than
	current Sunyaev-Zel'dovich cluster surveys \citep{Bleem2015,Planck2015XXVII}, as well
	as the $400\,\mathrm{deg^2}$ ROSAT survey based on deep archival pointings.
	At high redshift, the selection of the \Chundred\ is also much closer to being mass-limited
	than for ROSAT-based surveys such as the 400d.\label{fig:MzComp}}
   \vspace{-0.3cm} 
   \end{center}
\end{figure}

Actually, a simple Poisson approximation (which ignores the subtle effects of background
estimation, correction for masked areas and varying exposure) proved sufficient to reproduce
the measured errors for the 100 observed clusters. In this model, we generate a counts image
by distributing the count rate uniformly within the 60\arcsec\ aperture, multipling by the
exposure map and adding a modelled background. The Poisson error on the count rate is simply
obtained from the total counts in the aperture and a global rescaling by the average
exposure time in the aperture. This provides us with maps of the GCA measurement errors
for different source fluxes and positions. Since these error maps are tailor-made for each
pointing, they naturally account for gaps and missing MOS CCDs in addition to the local
exposure time and background level.

As we always have more than 50 photons in the 60" aperture, the Poisson error on the
aperture counts is close to a Gaussian distribution. For simplicity,
we assume that the same holds for all other sources of uncertainty (e.g. background
estimation, correction for varying exposure in the aperture). In this case,
the probability of exceeding the count rate cut $CR_{cut}$ for a source with a true
aperture flux $CR_{60}$ given the local measurement error
$\sigma_m\left(CR_{60},RA,Dec\right)$ is
\vspace{-0.1cm}
\begin{multline}
   \SFunc{cut}{CR_{60},RA,Dec} = \\
   \frac{1}{2} \left(1 + {\rm erf}\left[\frac{CR_{60}-CR_{cut}}{\sigma_m\left(CR_{60},RA,Dec\right)\sqrt{2}}\right]\right).
\end{multline}
\vspace{-0.2cm}

\noindent Including the pipeline C1+2 incompleteness, the exact selection~of the \BXC\ becomes,
for a single pointing~$p_i$,
\vspace{-0.2cm}
\begin{multline}
   \SFunc{p_i}{CR_\infty,r_c,RA,Dec} = \frac{\SFunc{C1+2,\ p_i}{CR_{\infty},r_c,RA,Dec}}{2} \\
	\times\ \left(1 + {\rm erf}\left[\frac{\epsilon_{60}(r_c) CR_\infty-CR_{cut}}{\sigma_m\left(\epsilon_{60}(r_c) CR_\infty,RA,Dec\right)\sqrt{2}}\right]\right),
    \label{eq:Sel1Point}
\end{multline}
\vspace{-0.2cm}

\noindent where $CR_\infty$ is the total $\beta$-model flux (integrated to infinity) and
$\epsilon_{60}(r_c)=\left(1 - \left[1+(60"/r_c)^2\right]^{1.5 - 3\beta}\right)$ is the
fraction of this flux included in the 1$^\prime$ aperture.

%______________________________________________________________
\subsection{Pointing overlaps}
Modelling the effect of pointing overlaps would be straightfoward if the \Chundred\ was
just the union of independent cluster catalogues from different pointings.
The probability of not detecting a source on any pointing would simply be the product
of the non-detection probabilities on all available pointings.
This is, for instance, the case for the completeness of the full C1+2 sample provided by the
XXL detection pipeline,
\begin{equation}
   \SFuncShort{C1+2} = 1 - \prod_{i=1}^{N} \left[1 - \SFuncShort{C1+2,p_i}\right],
   \label{eq:SimpleCombi}
\end{equation}
where we temporarily dropped the dependence on $CR_\infty$, $r_c$ and position ($RA$,
$Dec$) for brevity.

Such a simple prescription does not apply to the \BXC, for which the whole selection process
was not repeated independently over each pointing.
Instead, we first assembled the union of C1+2 detections on the different pointings and
then ran the GCA only once per cluster on the good pointing where it was detected with
the lowest off-axis angle.
In this case, the combined selection function reads
\begin{equation}
   \SFuncXXL =  \sum_{i=1}^{N} \left( \SFuncShort{p_i} \prod_{j<i} \left[1 - \SFuncShort{C1+2,p_j}\right]\right),
   \label{eq:SelFunc}
\end{equation}
where the $N$ pointings in the summation and product must be ordered by the off-axis angle for
the position under consideration. This effectively accounts for the flux cut on a given
pointing $p_i$ only if the cluster has not been detected in any pointing $p_j$ with a lower
off-axis angle ($j<i$).

For each value of $CR_\infty$ and $r_c$, we generated a 2D completeness map
based on Eq.~(\ref{eq:SelFunc}). Examples of such maps for a typical cluster
count rate and core radius are shown in Fig.~\ref{fig:2Dproba}. In these maps, the
spatial variation of the detection efficiency mainly reflects the local exposure time.
This is a result of the stringent selection on background level imposed for the construction
of the sample, while we use a wide range of exposures from 3\,ks in the shallowest dedicated
XXL observations to 80\,ks in the Subaru/XMM Deep Survey \citep{Ueda2008}, which is part of the northern field.

\subsection{Sky coverage}
\begin{figure}
\begin{center}
  \resizebox{\hsize}{!}{\includegraphics[width=8.5cm,viewport=20 0 420 340,clip]{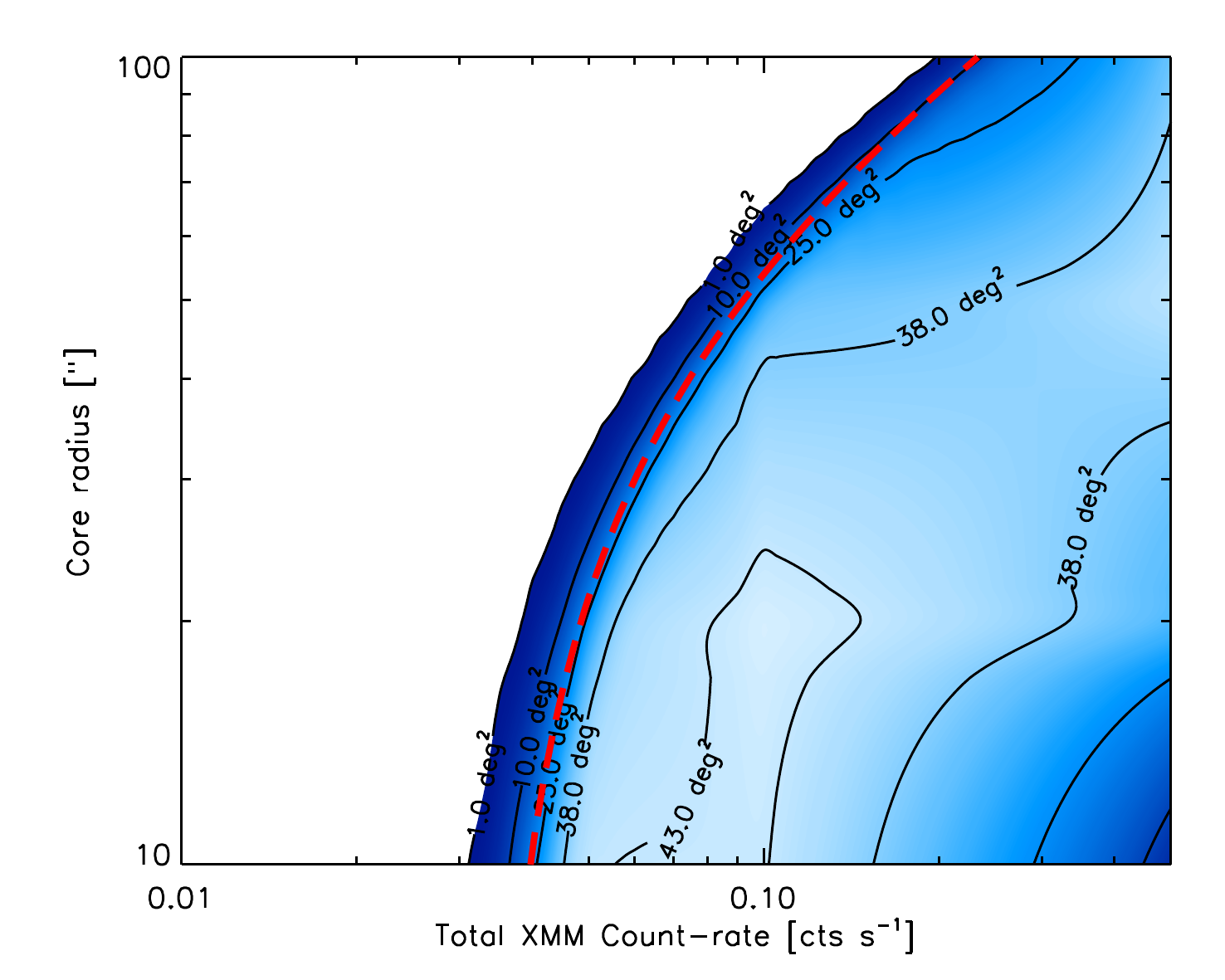}}
  \caption{Sky coverage of the \Chundred, displayed as contours of constant sky coverage
   in the source parameter space - cluster core radius for a $\beta$-model with $beta=2/3$
   vs total XMM count rate, summed over the three imaging cameras and integrated to infinite
   radius. For comparison, the dashed red curve shows the aperture flux cut
   of $CR_{60}=0.0332$~cts/s used for the subselection of the \BXC.\label{fig:SelFunc2D}}
\end{center}
\end{figure}

For most applications, the full 2D selection function is not required.
The sky coverage of the sample is straightforwardly derived from the sensitivity maps as
\begin{equation}
   \Omega_S\left(CR_\infty,r_c\right) = \int \SFunc{}{CR_\infty,r_c,RA,Dec} d\Omega.
\end{equation}
The resulting selection function for the \Chundred\ is shown in Fig.~\ref{fig:SelFunc2D}.
This figure captures the impact of each step of the selection process: the thick
dashed curve corresponds to the fixed 60\arcsec flux cut of $3\times10^{-14}$\flux, while the
transition between this line and the 1\dd\ area curve shows the spread around the cut due
to measurement errors in the GCA. Finally, the modulations at large core radii and for
count rates exceeding the flux cut reflect the initial C1+C2 selection function.

\begin{figure*}
\begin{center}
  \resizebox{\hsize}{!}{\hbox{\includegraphics[height=7.5cm, viewport=25 6 420 325,clip]{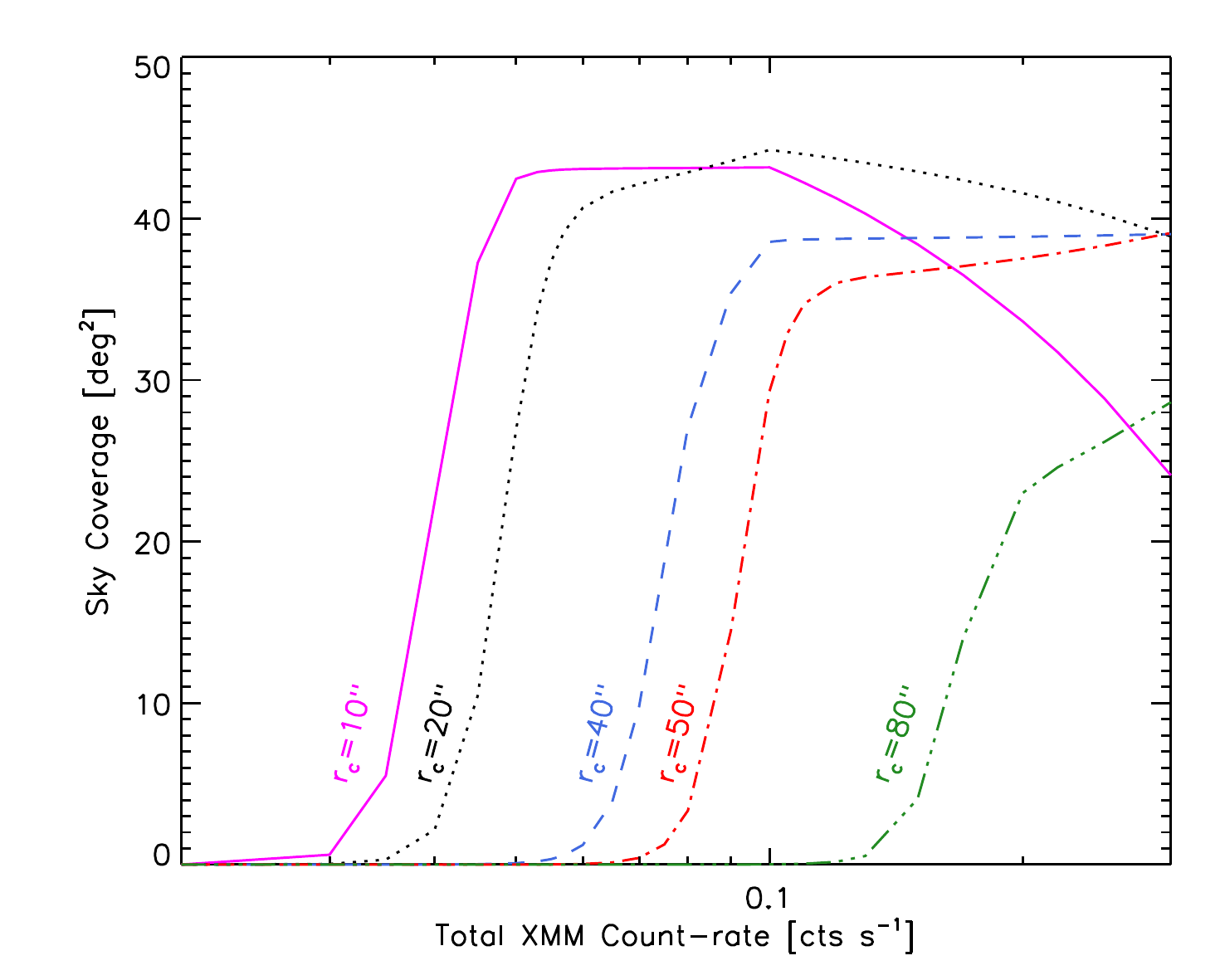}
       \includegraphics[height=7.5cm, viewport=25 6 415 325,clip]{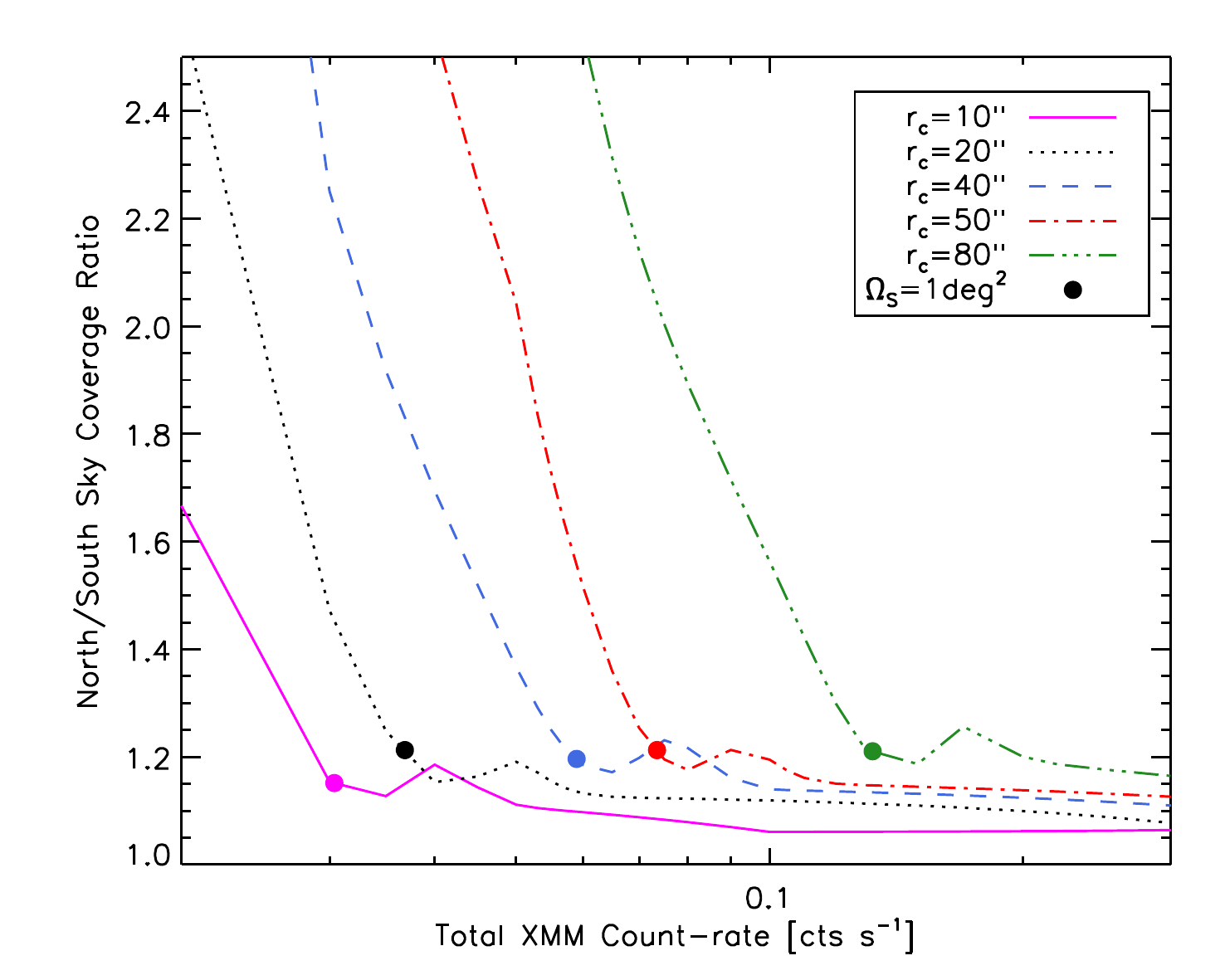}}}
  \caption{Sky coverage of the \Chundred, displayed as a function of the total XMM count
      rate (i.e. including the three imaging cameras and integrated to infinite radius).
    {\it Left}: Total sky coverage for different cluster core radii. These curves are
      the slices through the two-parameter selection function displayed in
      Fig.~\ref{fig:SelFunc2D}.
    {\it Right}: Ratio of sky coverage between the northern and southern fields.
      For sufficiently large count-rates, the sky coverage of the northern field is
      consistently 10-15\% larger than in the south, while the difference is more pronounced
      when the total sky coverage falls below $1\,\mathrm{deg}^2$.\label{fig:SelFunc1D}}
\end{center}
\end{figure*}

A more accurate visualisation of the selection function is provided in the left panel of
Fig.~\ref{fig:SelFunc1D} which shows slices of the sky coverage at constant core radius,
as a function of count rate.
The faint-end part of each curve for $10$\arcsec$<r_c<50$\arcsec results from the \Chundred\ flux
cut broadened by measurement uncertainties, while additional pipeline selection alters the
detection efficiency at larger radii and fluxes.

The relative contribution of the two fields to the total sky coverage is shown in the right
panel of Fig.~\ref{fig:SelFunc1D}. For the faintest sources, the sky coverage is much larger
in the north due to the existence of a few deep areas like the Subaru/XMM deep Survey
\citep{Ueda2008}. At larger count rates, however, the sky coverage is 10 to 15\% larger in
the northern field as expected from the geometry of the survey.

\section{Cosmological analysis}
\label{sect:Cosmo}

\begin{table*}
\begin{center}
\caption{Reference scaling relations used in the analysis of the $[$0.5-2$]$ keV luminosity
function and cosmological modelling.\label{tab:ScalPar}}
 \begin{tabular}{lcccccc}
 \hline\hline\\[-1.5ex]
 Relation & $X_0$ & $\rm A_{XT}$ & $\rm B_{XT}$ & $\rm \gamma$ & $\rm \sigma_{XT}$ & Reference\\[2pt]
 \hline\\[-1.5ex]
 $\ClMassWL - \Txxl$ & $\mathrm{2\times10^{14}}\ M_\odot$ & 1.16 & 1.67 & $-$1 & $-$ & \citetalias{Lieu2016}\\[4pt]
 $\Lxxl - \Txxl$ & $\mathrm{3\times10^{43}\,erg\,s^{-1}}$ & 0.71 & 2.63 & 1.64 & 0.47 & \citetalias{Giles2016}\\[4pt]
 \hline
 \end{tabular}
 \vspace{-0.3cm}
\end{center}
\tablefoot{For a parameter X, the average scaling with temperature is parametrised as
$X/X_0=A_{XT}\,(T/3keV)^{B_{XT}}\,E(z)^{\gamma_{XT}}$ and the dispersion around the relation
follows a log-normal distribution with parameter $\sigma_{XT}$.}
\end{table*}

In this section we analyze the distributions of the \Chundred\ in both luminosity
and position on the sky, and compare these distributions with the predictions of
the baseline WMAP9 cosmological model, with masses converted to X-ray luminosities via the scaling
relations from papers III and IV.

\subsection{Physical modelling of the cluster population}
\label{sect:PhysMod}

In all that follows, the cluster mass function is obtained from the \cite{Tinker2008}
mass function combined with the high accuracy approximation of the total matter transfer
function provided by \cite{Eisenstein1999}, including the effect of BAOs.
The cluster mass function is then converted to a temperature function using the average
weak lensing $\ClMassWL$-$\Txxl$ scaling relation measured in the companion \citetalias{Lieu2016},
without including the scatter. Finally, we make use of the soft band $\Lxxl$-$\Txxl$
scaling relation obtained in \citetalias{Giles2016} to model the cluster number per unit volume,
$n$, in X-ray parameter space as:
\begin{equation}
  \frac{\mathrm{d}n\left(L,T,z\right)}{ \mathrm{d}L\,\mathrm{d}T} =
    \frac{\mathrm{d}n}{\mathrm{d}M\,\mathrm{d}z}\left(\hat{M}\right)\frac{\mathrm{d}\hat{M}(T,z)}{\mathrm{d}T} \mathcal{LN}\left[L\,|\,\hat{L}\left(T,z\right),\sigma_{LT}\right],
    \label{eq:dnOverdLdT}
\end{equation}
where $\hat{M}(T,z)$ is the average $\ClMassWL$-$\Txxl$ scaling relation,
$\mathcal{LN}(x|\hat{x},\sigma)$ is a log-normal distribution with Gaussian parameters
$\ln\left(\hat{x}\right)$ and $\sigma$, $\hat{L}$ that results from the average $\Lxxl$-$\Txxl$
scaling relation and $\sigma_{LT}$ is the scatter around this relation.
For reference, the parameters of both scaling relations are provided in Table~\ref{tab:ScalPar}.
The presence of cool cores with different strengths in a large fraction of the cluster population
generates a positive correlation in the scattering of $\Lxxl$ and $\Txxl$ at a given mass,
the amount of which remains poorly quantified. Combining the dispersions around both scaling relations
without including this correlation would probably overestimate the total scatter. We therefore
decided to model only the major contribution to the total dispersion, $\sigma_{LT}$.

In order to apply the \Chundred\ selection function, $(\Lxxl,\Txxl,z)$ needs to be converted
to an XMM count rate. For this, we use the on-axis XMM response, an APEC model with metal
abundances set to 0.3 of the solar values, and the luminosity distance in the WMAP9
cosmology.
In this way, we assume that the difference between $\ClTemp$ and $\Txxl$ is negligible
and that the result can be identified with the count rate within the projected $\ClRad$.
This is justified because most of the cluster flux originates from within
300~kpc and because we always work in the soft band which minimises the impact of the
temperature on the K-correction.
As the scatter between mass and $\Txxl$ is larger than with $\ClTemp$ owing to the
impact of cluster cores, identifying the two values provides us with another justification
to neglect the scatter of the $\ClMassWL$-$\Txxl$ scaling relation in the modelling of the
cluster population.

Finally, a very important assumption is the value of the core radius. Motivated by
previous studies of the cluster $\beta$-model parameters (e.g. \citealt{Helsdon2000},
\citealt{Henning2009}, or \citealt{Alshino2010}), the core radius is assumed to scale linearly
with the cluster's physical size as $r_c = x_{500}\times \ClRad$ with $x_{500}=0.15$ for
$\beta=2/3$. For consistency, the same surface brightness model was also used in \citetalias{Giles2016}
to extrapolate the observed luminosity within $300~kpc$ to $\ClRad$ and to account for the
selection function in the scaling relation fits. In the modelling, it also serves to
extrapolate the XMM count-rate within $\ClRad$ to infinity, as required by the definition
of the \ChundredShort\ sky coverage.
We discuss further the impact of the choice of $x_{500}$ in Sect.~\ref{sect:DiscussDens}.

\begin{figure*}
  \begin{center}
    {\includegraphics[width=8.5cm,viewport=30 10 410 325,clip]{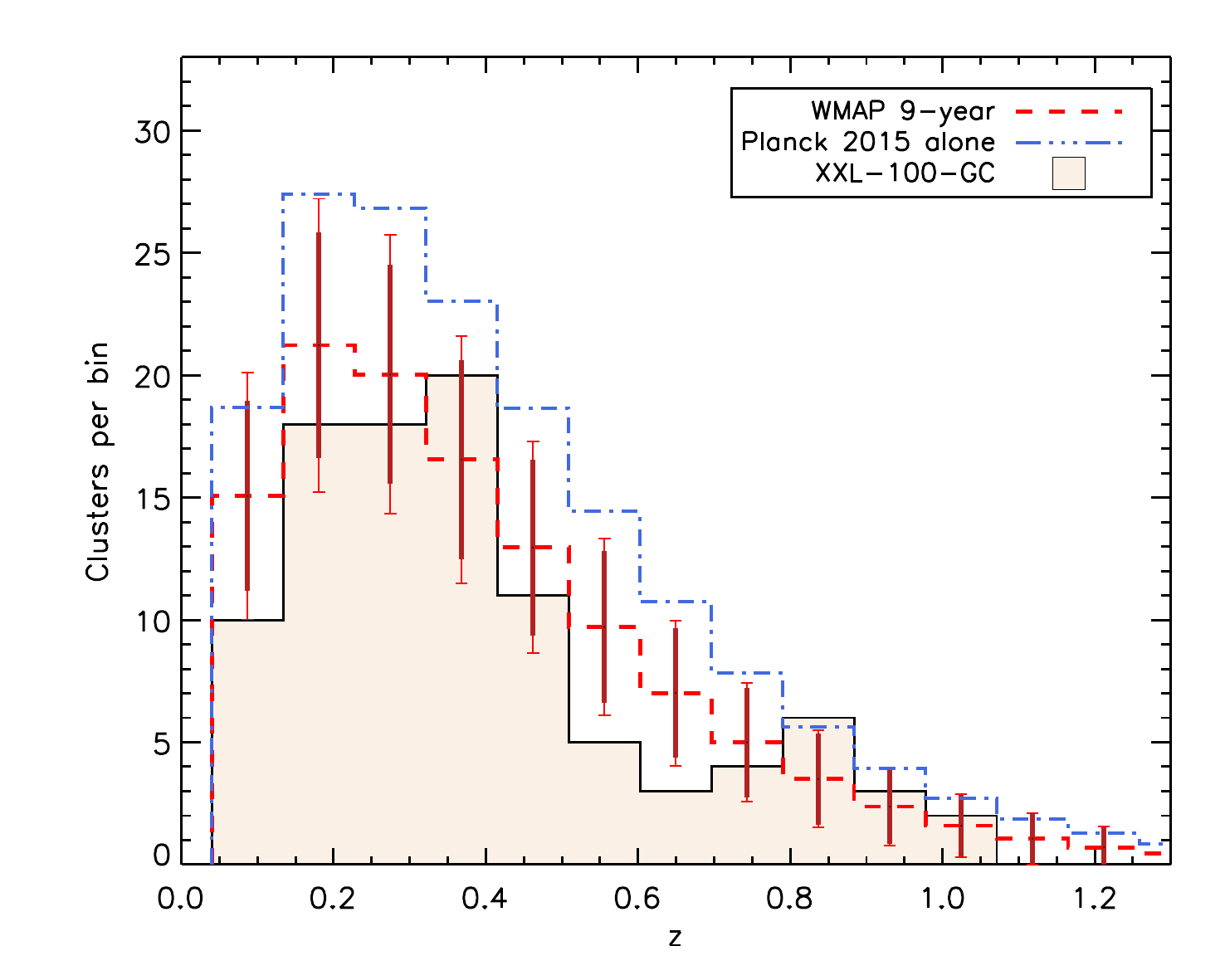}\hspace{1cm}\includegraphics[width=8.5cm,viewport=30 10 410 325,clip]{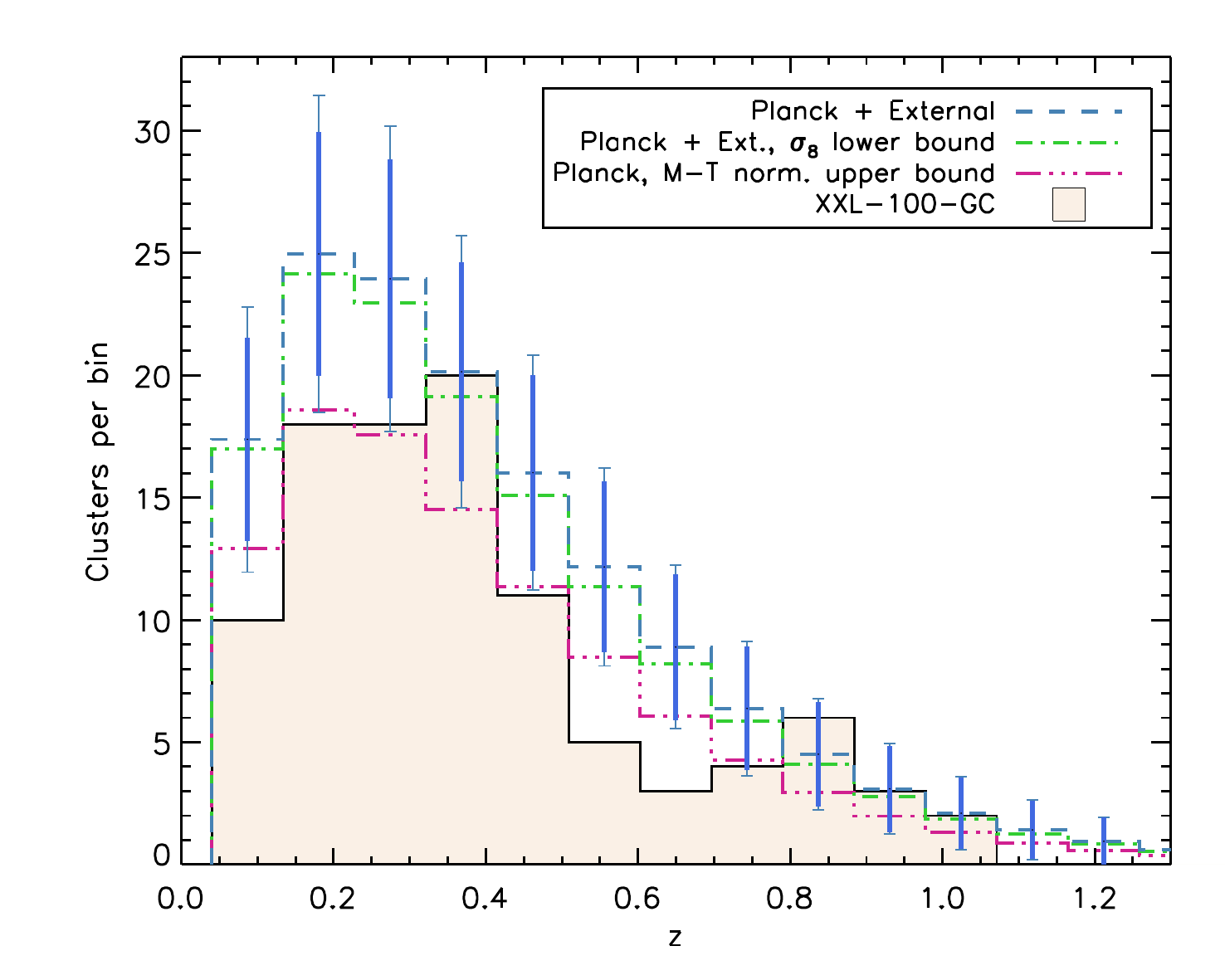}}
   \caption{Redshift distribution of the \Chundred\ (filled histogram) compared with
    different model expectations. By default, the model predictions are based on the mass
    and temperature scaling relations of \citetalias{Giles2016} and \citetalias{Lieu2016},
    and assume a $\beta$-model with $\beta=2/3$ and $x_{500}=0.15$.
    {\it Left}: The fiducial WMAP9 cosmology (red dashed line) compared with the
    {\it Planck} 2015 cosmological parameters obtained only from the CMB data (blue dot-dashed).
    {\it Right}: Other models derived
    from \citet{Planck2015XIII}. The {\it Planck}+External set of cosmological parameters includes
    additional BAO and H0 constraints (blue dashed). The green dot-dashed line is the same,
    but fixing $\sigma_8$ to the $1\sigma$ lower bound allowed by the {\it Planck}+External data set.
    The purple triple dot-dashed line uses the {\it Planck}-only parameters, but the normalisation
    of the $M_{500,WL}-T_{300kpc}$ scaling relation has been increased to its $1\sigma$ upper bound.
    The error bars (shown only for the WMAP9 and {\it Planck}+External cosmologies) include
    both the shot noise (thick part) and the cosmic variance.\label{fig:dndzComp}}
 \vspace{-0.3cm} 
 \end{center}
\end{figure*}

This model predicts a total of 117 clusters in the fiducial WMAP9 cosmology, slightly more
than the actual 100 detections. The significance of this deficit in clusters
is less than $1.5\,\sigma$ accounting for Poisson noise alone, and falls to $1\,\sigma$
including the additional sample variance arising from cosmic density fluctuations,
which we compute following the method of \cite{Valageas2011}.
The expected redshift distribution is also in reasonable agreement with the observed one
(see the left panel of Fig.~\ref{fig:dndzComp}).
As a useful cross-check of the modelling assumptions, we directly computed the selection
function of the \Chundred\ in terms of mass and redshift for this fiducial cosmology, set of
scaling relations, and surface brightness profile. The redshift-dependent mass limit
corresponding to a 50\% detection probability is shown in Fig.~\ref{fig:MzDist}.
It is consistent with the mass selection as reflected by the distribution of the cluster
data points, although it is a bit high at redshift $z>0.6$. The lack of massive
clusters at high redshift combined with the steepness of the mass function suffices to
explain the numerous clusters detected just below the 50\% completeness limit.
We therefore conclude that the number density of the \Chundred\ is fully consistent with
the fiducial WMAP9 cosmology.

In comparison, the set of cosmological parameters recently determined by the
\cite{Planck2015XIII} from the analysis of the CMB power spectrum measurements of the
Planck satellite \citep{Planck2011I} significantly overpredicts the density of the
\Chundred\ with a total of 165 clusters based on the scaling relations of \citetalias{Giles2016}
and \citetalias{Lieu2016}\footnote{These scaling relations were measured assuming the WMAP9
cosmology. Their use is justified here since the distance scales between the two cosmologies only vary by a few
percent over the considered redshift range and the fitting procedures do not rely on the
normalisation of the mass function, as explained in Appendix~\ref{append:Like}.}.
This results in great part from a larger value of $\sigma_8=0.831$, but also from the
decrease in the Hubble parameter ($H_0=67.27\,\mathrm{km\,s^{-1}Mpc^{-1}}$) and the increase
in total matter density ($\Omega_m=0.3156$), which alters both the survey volume (+$\sim$5\%)
and the mass function (+25$-$35\% depending on mass and redshift).
The relative effects of changes in $\sigma_8$ or the background geometry can be distinguished
by considering the modelled cluster population for a third set of cosmological parameters
obtained in \cite{Planck2015XIII} from the combination of their CMB measurement with other
cosmological tracers, which we term {\it Planck}+External cosmology. It has essentially
the same geometry as the baseline {\it Planck} CMB fit and mostly differs in the value of
$\sigma_8$, which is $0.8159$ and therefore comparable to the WMAP9 estimate.
Despite the lower $\sigma_8$, this model still predicts 143 clusters and outnumbers
the observed \ChundredShort\ cluster density at all redshifts (see right panel of
Fig.~\ref{fig:dndzComp}).

The observed mismatch between the {\it Planck} CMB results and the late-time tracers of matter
fluctuations is well known and was reported by the {\it Planck} collaboration itself using cluster
samples selected with the Sunyaev-Zel'dovich effect (\citealt{Planck2014XX}; \citealt{Planck2015XXIV}).
To investigate the significance of the mismatch with the \BXC, we considered
two altered models based on the~{\it Planck} 2015 cosmology (which we also show  in the right panel
of Fig.~\ref{fig:dndzComp}). In the first, the {\it Planck}+External set of parameters is
assumed with $\sigma_8$ fixed to the allowed $1\,\sigma$ lower bound of $0.8073$.
This only mildly decreases the predicted number of clusters to 136, implying that the
uncertainty on the cosmological parameters derived by Planck cannot explain the discrepancy.
In the second model, we stick to the cosmological parameters obtained from the {\it Planck} CMB
dataset alone, but increase the normalisation of the XXL $\ClMassMT-\Txxl$ relation of
\citetalias{Lieu2016} to its allowed $1\,\sigma$ upper bound.
This alteration in the normalisation at 1~keV also incorporates part of the degeneracy
with the slope of the scaling relation, since the median temperature of our sample is closer
to 3~keV. Therefore, it serves to approximates a $1\sigma$ deviation in the 2D
parameter space. This change results in an almost perfect match with 102 predicted clusters.
From these considerations, we conclude that the tension between the {\it Planck}
2015 cosmology and the \Chundred\ cannot yet be established with a strong significance.
Further analysis of the XXL cluster population might result in a better agreement.

\subsection{Luminosity function}
\label{sec:LumFunc}

With the prescriptions presented above for the cluster surface brightness profile
and scaling relations, we can model the selection effects in the \Chundred\ for
population studies.

\begin{figure*}
  \begin{center}
    \resizebox{\hsize}{!}{\includegraphics[height=6cm,viewport=15 2 415 328,clip]{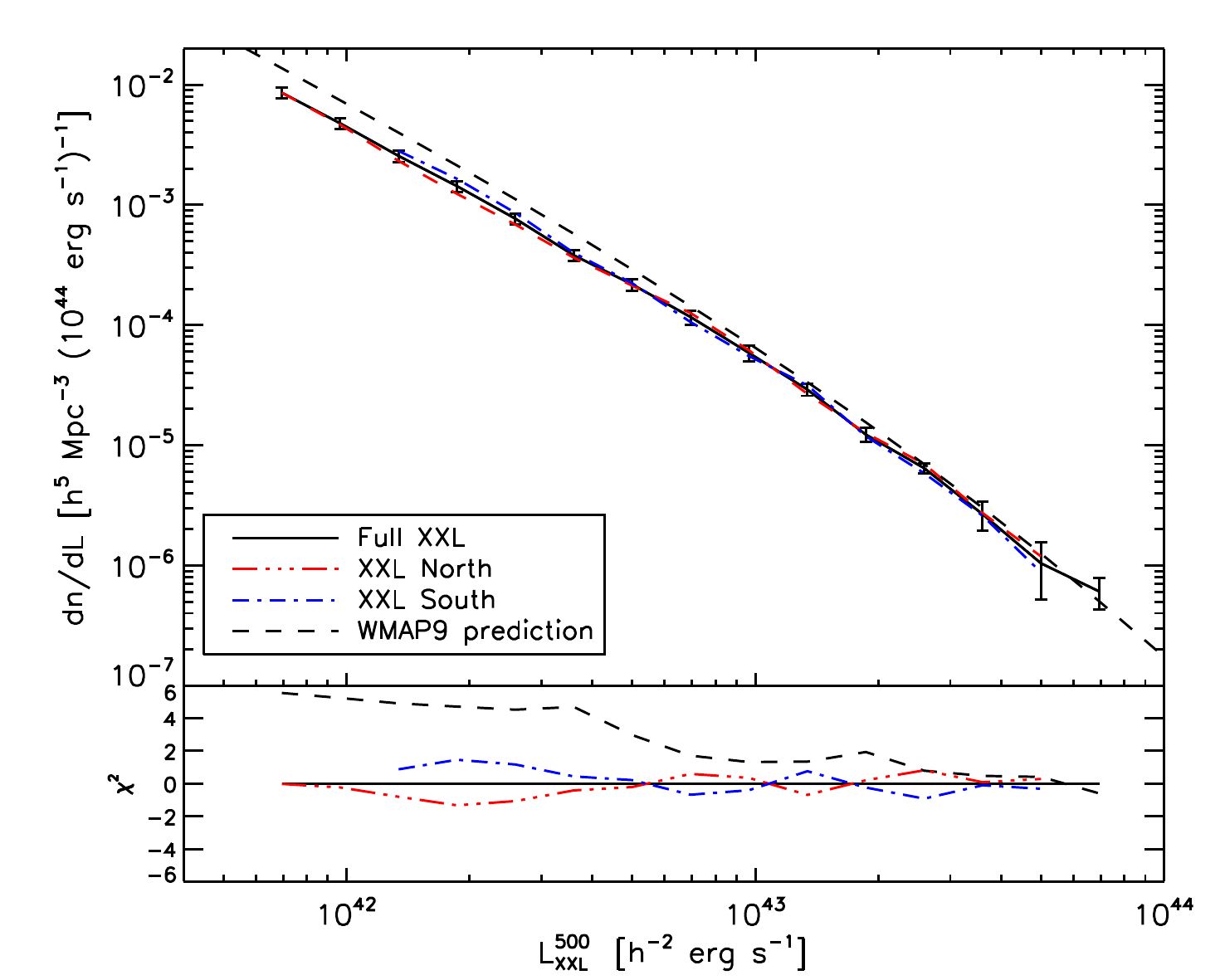}
                          \includegraphics[height=6cm,viewport=15 2 415 328,clip]{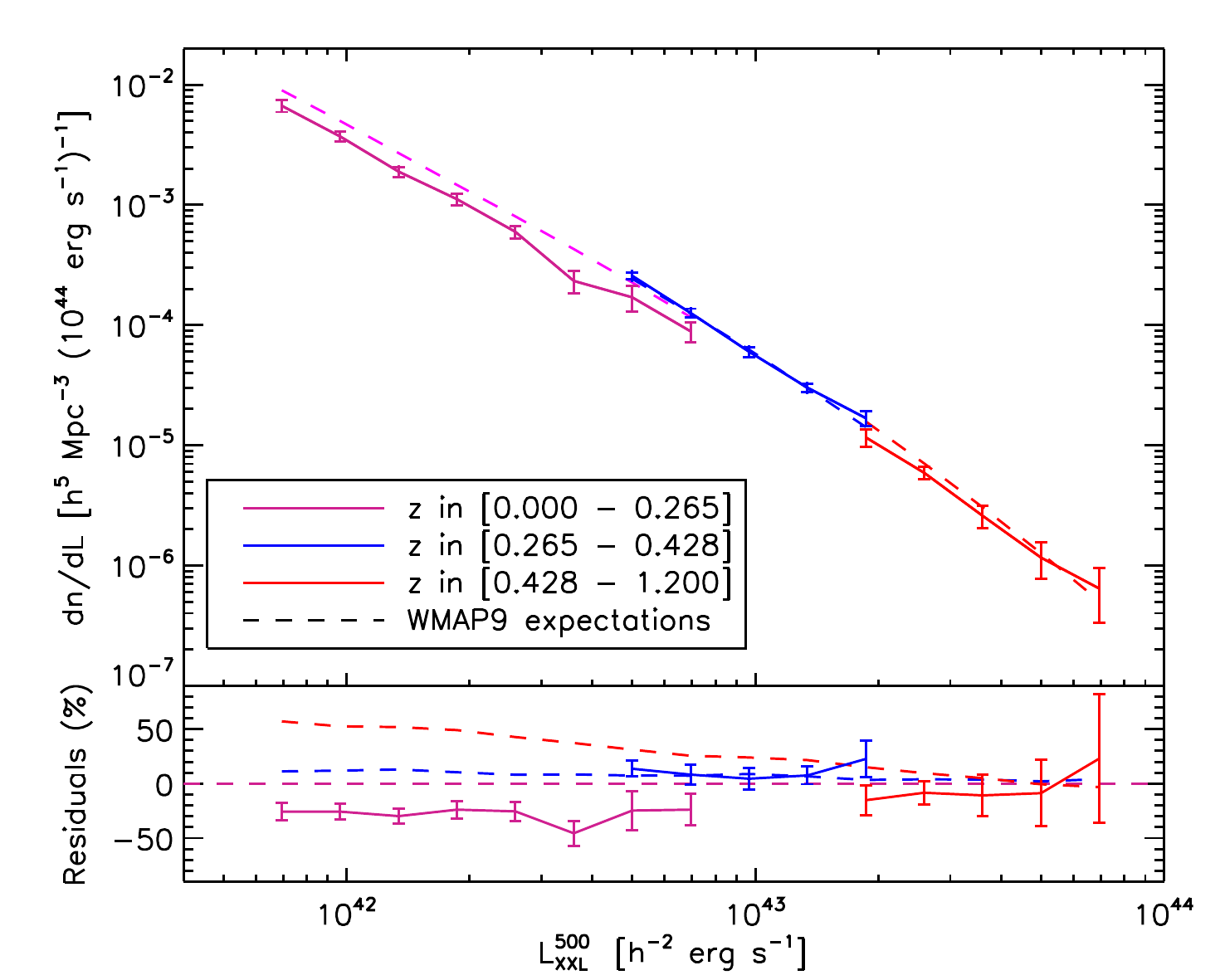}}
   \caption{Differential luminosity function of the \Chundred\ in the WMAP9 cosmology
    measured with the cumulative estimator defined by Eq.~(\ref{eq:CumulCorr}). The
    effective volume correction derives from the \ChundredShort\ scaling relations
    provided in Table~\ref{tab:ScalPar}. The dashed lines show the predictions of the
    luminosity function in the WMAP9 cosmological model for the same redshift bins
    as the measurements, and using the same colour code.
    {\it Left}: Differential luminosity function averaged over the full redshift range of
    $[$0-1.2$]$ covered by the \Chundred, and for the northern/southern field separately.
    The $\chi^2$ plot shows how the deviation of each subfield from the complete analysis
    compares with the combined error bars.
    {\it Right}: Differential luminosity function of the \BXC\ in three redshift bins.
    The lower plot shows the residuals with respect to the low-redshift WMAP9 prediction.
    \label{fig:LumFunc}}
  \end{center}
\end{figure*}

The most direct statistic widely used to characterise the X-ray cluster population is
its luminosity function, which is simply obtained by counting clusters in luminosity bins
and correcting for the effective volume probed by the survey. Several methods exist for
calculating this correction. In this section, we use the following extimator,
\vspace{-0.1cm}
\begin{equation}
    \frac{\mathrm{d}n}{\mathrm{d}L}(L) = \frac{\mathrm{d\ }}{\mathrm{d}L}\left[\frac{N_{>L}}{V_{>L}}\right],
   \label{eq:CumulCorr}
\end{equation}
which we term the cumulative estimator of the differential luminosity function.
Here $N_{>L}$ is the total number of clusters with $\Lxxl>L$  in the redshift slice
under consideration, while $V_{>L}$ is the average survey volume for clusters with such
luminosities in the same redshift range (defined in the same way as in Eq.~(\ref{eq:CumulFunc})).
In practice, we thus compute the cumulative luminosity function ($N_{>L}/V_{>L}$) and differentiate
it numerically.
A discussion of other commonly adopted estimators and how they compare with the method
adopted in this paper is provided in Appendix~\ref{append:LFcomp}. Briefly, the cumulative estimator enables us to
estimate the luminosity function for many luminosity values from a reasonably wide luminosity
range around each point. This smooths out the shot noise in the estimation and provides tighter
error bars at the expense of introducing correlation between the data points.

The errors on the luminosity function were computed from a thousand Monte Carlo
simulations and include the effect of shot noise as well as the uncertainties on the measured
luminosities and $\Lxxl-\Txxl$ scaling relation. The number of clusters in each simulation
follows a Poisson distribution with a mean of 100. We then generate a bootstrap sample from
the cluster catalogue, randomise their luminosities and recompute the luminosity function
with a randomised set of scaling relation parameters, derived from the $\Lxxl-\Txxl$
covariance matrices obtained in \citetalias{Giles2016}.

The global \Chundred\ luminosity function, averaged over the whole redshift range, is shown
in Fig.~\ref{fig:LumFunc} (left panel), together with expectations for  the WMAP9 cosmology
obtained by integrating Eq.~(\ref{eq:dnOverdLdT}) over temperature and redshift.
Assuming the \ChundredShort\ set of scaling relations, the agreement between the measurements
and the predictions  is excellent at intermediate to high luminosities, but we observe a
significant deficit of low luminosity clusters in the observed sample ($\sim4\sigma$, reduced
to $2-3\sigma$ when accounting for cosmic variance, which our error models do not include).
This is consistent with the 15\% higher total cluster counts predicted by the model.
The large redshift span and the $z$ versus $\Lxxl$ degeneracy that characterises the \Chundred\
however complicates the interpretation of the observed cluster deficit.
We observe no hint of deviations between the measurements obtained from the two XXL fields
taken separately, although we note that the luminosity function of the northern field reaches
fainter luminosities than could be achieved in the south, and that the deficit is more
pronounced there.

To proceed further and investigate a possible evolution of the luminosity function, we
divided the sample into three mutually exclusive redshift bins ($0<z<0.265$, $0.265<z<0.428$,
and $0.428<z<1.2$) containing a similar number of \ChundredShort\ clusters (34, 33, and 33,
respectively).
The resulting luminosity functions are shown in the right panel of Fig.~\ref{fig:LumFunc}.
Clearly, the lack of low luminosity clusters apparent in the global luminosity function
stems from a slight deficit of low redshift clusters at almost all luminosities. This was
already visible, in a more compact way, in the redshift distributions of Fig.~\ref{fig:dndzComp}
where the WMAP9 cosmology predicts more clusters in all bins below $z\sim0.3$.
Within the error bars, we see no hint of evolution between the three redshift bins. This
result was first reported by \citet{Rosati1998} for $z\la0.8$ although, ironically, not in
the same cosmology (a flat universe without cosmological constant) and confirmed by a few
other studies (see \citealt{Rosati2002} for a review).
Within the luminosity range probed by the \Chundred, this is fully consistent with the
predictions of the WMAP9 model, although a strong positive evolution is expected for faint,
high redshift clusters, a regime that has not yet been explored by current surveys.

\setlength{\defaultaddspace}{-0.05cm}
\begin{table*}
\begin{center}
\caption{Tabulated values of the \Chundred\ differential luminosity function.\label{tab:LfuncValues}}
{\small
\begin{tabular}{ccccccccc}
\hline\hline
                                         &  \multicolumn{2}{c}{Full $z$ range}        &  \multicolumn{2}{c}{$0.0 < z < 0.265$}      & \multicolumn{2}{c}{$0.265 < z < 0.468$}     & \multicolumn{2}{c}{$0.468 < z < 1.2$} \rule{0pt}{2.6ex}\\[-0.05cm]\cmidrule(l{.5cm}r{.5cm}){2-3}\cmidrule(l{.5cm}r{.5cm}){4-5}\cmidrule(l{.5cm}r{.5cm}){6-7}\cmidrule(l{.5cm}r{.5cm}){8-9}\addlinespace
$L^{500}_{XXL}$                 &  $dn/dL$                 &  $\Delta(dn/dL)$ &  $dn/dL$                &  $\Delta(dn/dL)$  &  $dn/dL$                &  $\Delta(dn/dL)$  &  $dn/dL$                &  $\Delta(dn/dL)$           \\[2pt]
[$10^{42}\,h^{-2}\mathrm{erg\,s^{-1}}$]  &  [LF unit]$^\dagger$   &        \%        &  [LF unit]$^\dagger$  &        \%         &  [LF unit]$^\dagger$  &        \%         &  [LF unit]$^\dagger$  &        \%                     \\
\hline
 0.69 & $8.58 \times 10^{-3}$ &   10.8 & $6.67 \times 10^{-3}$ &   11.1 &         $-$		&    $-$ &         $-$  	 &   $-$  \rule{0pt}{2.6ex}\\
 0.97 & $4.77 \times 10^{-3}$ &   10.6 & $3.71 \times 10^{-3}$ &    9.6 &         $-$		&    $-$ &         $-$  	 &   $-$  \\
 1.34 & $2.56 \times 10^{-3}$ &   11.4 & $1.89 \times 10^{-3}$ &    9.3 &         $-$		&    $-$ &         $-$  	 &   $-$  \\
 1.86 & $1.43 \times 10^{-3}$ &   10.3 & $1.11 \times 10^{-3}$ &   10.6 &         $-$		&    $-$ &         $-$  	 &   $-$  \\
 2.59 & $7.69 \times 10^{-4}$ &   10.0 & $5.97 \times 10^{-4}$ &   11.5 &         $-$		&    $-$ &         $-$  	 &   $-$  \\
 3.60 & $3.81 \times 10^{-4}$ &   10.8 & $2.33 \times 10^{-4}$ &   21.1 &         $-$		&    $-$ &         $-$  	 &   $-$  \\
 5.00 & $2.18 \times 10^{-4}$ &   11.2 & $1.70 \times 10^{-4}$ &   23.7 & $2.57 \times 10^{-4}$ &    6.3 &         $-$  	 &   $-$  \\
 6.95 & $1.17 \times 10^{-4}$ &   13.5 & $8.88 \times 10^{-5}$ &   19.2 & $1.26 \times 10^{-4}$ &    8.5 &         $-$  	 &   $-$  \\
 9.65 & $5.86 \times 10^{-5}$ &   14.8 &         $-$	       &   $-$  & $6.01 \times 10^{-5}$ &    9.6 &	   $-$           &   $-$  \\
 13.4 & $2.91 \times 10^{-5}$ &   11.3 &         $-$	       &   $-$  & $3.02 \times 10^{-5}$ &    7.5 &	   $-$           &   $-$  \\
 18.6 & $1.23 \times 10^{-5}$ &   13.6 &         $-$	       &   $-$  & $1.68 \times 10^{-5}$ &   13.6 & $1.16 \times 10^{-5}$ &   16.1 \\
 25.9 & $6.45 \times 10^{-6}$ &   10.1 &         $-$	       &   $-$  &	  $-$		&    $-$ & $5.90 \times 10^{-6}$ &   11.5 \\
 36.0 & $2.67 \times 10^{-6}$ &   27.5 &         $-$	       &   $-$  &	  $-$		&    $-$ & $2.59 \times 10^{-6}$ &   21.4 \\
 50.0 & $1.04 \times 10^{-6}$ &   50.0 &         $-$	       &   $-$  &	  $-$		&    $-$ & $1.16 \times 10^{-6}$ &   33.5 \\
 69.5 & $6.07 \times 10^{-7}$ &   28.8 &         $-$	       &   $-$  &	  $-$		&    $-$ & $6.42 \times 10^{-7}$ &   48.1 \\
\hline
\end{tabular} }
\tablefoot{Because of the luminosity vs redshift degeneracy in the \Chundred, only a limited
range of luminosities is available for each redshift slice. A graphical display of these values is provided in Fig.~\ref{fig:LumFunc}.
$^{(\dagger)}$: all luminosity function values in this table are in units of [\,$h^5\,\mathrm{Mpc^{-3}\,(10^{44}\,erg\,s^{-1})^{-1}}$\,].}
\vspace{-0.2cm}
\end{center}
\end{table*}

\begin{table*}
\begin{center}
\caption{Tabulated values of the \Chundred\ cumulative luminosity function.\label{tab:CumLfuncValues}}
{\small \begin{tabular}{ccccccccc}
\hline\hline
                                &  \multicolumn{2}{c}{Full $z$ range}        &  \multicolumn{2}{c}{$0.0 < z < 0.265$}      & \multicolumn{2}{c}{$0.265 < z < 0.468$}     & \multicolumn{2}{c}{$0.468 < z < 1.2$} \rule{0pt}{2.6ex}\\[-0.05cm]\cmidrule(l{.5cm}r{.5cm}){2-3}\cmidrule(l{.5cm}r{.5cm}){4-5}\cmidrule(l{.5cm}r{.5cm}){6-7}\cmidrule(l{.5cm}r{.5cm}){8-9}\addlinespace
$L^{500}_{XXL}$                 &  $n(>L)$                &  $\Delta[n(>L)]$ &  $n(>L)$                &  $\Delta[n(>L)]$  &  $n(>L)$                &  $\Delta[n(>L)]$  &  $n(>L)$                &  $\Delta[n(>L)]$              \\[2pt]
[$10^{42}\,h^{-2}\mathrm{erg\,s^{-1}}$]  &  [$h^3\mathrm{\,Mpc^{-3}}$]   &        \%        &  [$h^3\mathrm{\,Mpc^{-3}}$]  &        \%         &  [$h^3\mathrm{\,Mpc^{-3}}$]  &        \%         &  [$h^3\mathrm{\,Mpc^{-3}}$]  &        \%                     \\
\hline
 0.69 & $6.63 \times 10^{-5}$ &   10.8 & $5.13 \times 10^{-5}$ &    9.2 &         $-$           &    $-$ &	   $-$         &     $-$  \rule{0pt}{2.6ex}\\
 0.97 & $4.90 \times 10^{-5}$ &   11.2 & $3.76 \times 10^{-5}$ &    9.1 &         $-$           &    $-$ &	   $-$         &     $-$  \\
 1.34 & $3.60 \times 10^{-5}$ &   11.0 & $2.78 \times 10^{-5}$ &    9.4 &         $-$           &    $-$ &	   $-$         &     $-$  \\
 1.86 & $2.64 \times 10^{-5}$ &   11.1 & $2.09 \times 10^{-5}$ &   10.2 &         $-$           &    $-$ &	   $-$         &     $-$  \\
 2.59 & $1.85 \times 10^{-5}$ &   11.9 & $1.41 \times 10^{-5}$ &   12.9 &         $-$           &    $-$ &	   $-$         &     $-$  \\
 3.60 & $1.33 \times 10^{-5}$ &   12.7 & $1.07 \times 10^{-5}$ &   13.3 &         $-$           &    $-$ &	   $-$         &     $-$  \\
 5.00 & $9.47 \times 10^{-6}$ &   13.0 & $8.59 \times 10^{-6}$ &   12.6 & $1.03 \times 10^{-5}$ &    6.5 &         $-$         &     $-$  \\
 6.95 & $6.17 \times 10^{-6}$ &   14.9 & $5.12 \times 10^{-6}$ &   22.3 & $6.75 \times 10^{-6}$ &    7.4 &         $-$         &     $-$  \\
 9.65 & $4.14 \times 10^{-6}$ &   12.9 &          $-$          &   $-$  & $4.57 \times 10^{-6}$ &    7.3 &         $-$         &     $-$  \\
13.4 & $2.44 \times 10^{-6}$ &   15.9 &           $-$          &   $-$  & $2.93 \times 10^{-6}$ &   12.3 &         $-$         &     $-$  \\
18.6 & $1.57 \times 10^{-6}$ &   16.5 &           $-$          &   $-$  & $1.91 \times 10^{-6}$ &   14.7 & $1.53 \times 10^{-6}$ &   14.9 \\
25.9 & $9.35 \times 10^{-7}$ &   22.5 &           $-$          &   $-$  &	  $-$	        &    $-$ & $9.34 \times 10^{-7}$ &   15.0 \\
36.0 & $4.74 \times 10^{-7}$ &   38.8 &           $-$          &   $-$  &	  $-$           &    $-$ & $5.27 \times 10^{-7}$ &   23.9 \\
50.0 & $3.03 \times 10^{-7}$ &   27.1 &           $-$          &   $-$  &         $-$	        &    $-$ & $3.21 \times 10^{-7}$ &   42.2 \\
69.5 & $1.31 \times 10^{-7}$ &   20.6 &           $-$          &   $-$  &         $-$	        &    $-$ & $1.45 \times 10^{-7}$ &   20.8 \\
\hline
\end{tabular} }
\tablefoot{Because of the luminosity vs redshift degeneracy in the \Chundred, only a limited
range of luminosities is available for each redshift slice.}
\end{center}
\vspace{-0.3cm}
\end{table*}

A few studies have also reported a negative evolution of the cluster soft-band luminosity
function at high redshift, for instance \citealt{Mullis2004} (for clusters at
$1.2\,<\,\Lxxl\,<\,4.9\times10^{44}\,\mathrm{erg\,s^{-1}}$ and $0.6<z<0.8$) 
and  \citealt{Koens2013} (for $0.9\,<\,\Lxxl\,<\,12\times10^{44}\,\mathrm{erg\,s^{-1}}$
and $0.6<z<1.1$).
We cannot support these findings, but this does not imply any tension since these studies
typically involve more luminous clusters than the bulk of the \ChundredShort\ population.

Tabulated values of the \ChundredShort\ differential and cumulative luminosity functions
are given for all considered redshift intervals in Tables~\ref{tab:LfuncValues} and
\ref{tab:CumLfuncValues}.

\subsection{Constraints from the luminosity distribution}
\label{sec:LumConst}
So far, we have assumed that cluster scaling relations are perfectly known from
\citetalias{Giles2016} and \citetalias{Lieu2016}. In this section, we use the cluster
luminosity function to put independent constraints on the parameters of the $\Lxxl-\Txxl$ distribution.
This provides a useful consistency check of the cluster luminosity function analysis 
presented in Sect.~\ref{sec:LumFunc} since we do not use the temperature information, but only the
cluster number density\footnote{The cluster number density was ignored in the analysis of
\citetalias{Giles2016}, as explained in Appendix~\ref{append:Like}}.
To this purpose, we go back to the raw distribution of the sample in the $(\Lxxl,z)$
plane and compare it with different models using the likelihood function described
in Appendix~\ref{append:Like}, a variation over the likelihood derived by \citet{Mantz2010a}
that also accounts for cosmic variance.
For full consistency, we excluded the cluster XLSSC~504 from the fit, as it was also
excluded from the $\Lxxl-\Txxl$ analysis of \citetalias{Giles2016} owing to its
suspected AGN contamination and abnormally large temperature. To account for this modification,
we included an additional completeness factor of 0.99 in the analysis.

The results are reported in Table~\ref{tab:LT} for a few different parametrisations of the
$\Lxxl-\Txxl$ relation.
We first considered a global change in the normalisation of the scaling relation, which
yielded slightly lower values than in \citetalias{Giles2016}. This was to be expected since the fiducial
model slightly overpredicts the number of \Chundred\ clusters. Freeing the evolution
parameter slightly changes the balance between normalisation and evolution but the deviations
are insignificant. Finally, fitting also for the slope of the scaling relation, indicates
that the slightly lower density of observed \Chundred\ clusters compared to the fiducial
model is better accounted for by a steepening of the scaling relation rather than a
change in normalisation.

In all cases, the fitted  parameters fall within the statistical errors of the measurements
presented in \citetalias{Giles2016}.
Without the temperature information, the data do not allow further constraints on the scatter.

\begin{table}
   \begin{center}
   \caption{Indirect constraints on the $\Lxxl-\Txxl$ relation parameters obtained by fitting the $(L,z)$ number density.\label{tab:LT}}
   \vspace{-0.1cm}
   \renewcommand{\baselinestretch}{1.3}\selectfont
   \begin{tabular}{lccc}
	\hline\hline
 	  Free parameters   &    $A_{LT}$   & $B_{LT}$  & $\gamma_{LT}$  \\
	\hline
	  Reference         &   0.71                     &          2.63            &  1.64                  \\
	  Norm. only        &   $0.63^{+0.07}_{-0.06}$   &          2.63            &  1.64                  \\
	  Norm.+evol        &   $0.64^{+0.11}_{-0.10}$   &          2.63            &  $1.62^{+0.44}_{-0.46}$  \\
	  Norm.+pow+evol    &   $0.72^{+0.20}_{-0.15}$   &  $2.79^{+0.24}_{-0.22}$  &  $1.26^{+0.67}_{-0.76}$  \\[2pt]
	\hline
   \end{tabular}
   \renewcommand{\baselinestretch}{1.0}\selectfont
   \tablefoot{The scaling relation parameters are defined in the same way as in Table~\ref{tab:ScalPar}.
   The error bars include Poisson errors as well as the total cosmic variance predicted for the
   \Chundred\ using the formalism of \citet{Valageas2011}.}
   \end{center}
\end{table}

\begin{figure}
    \resizebox{\hsize}{!}{\includegraphics[width=8cm,viewport=20 7 420 325,clip]{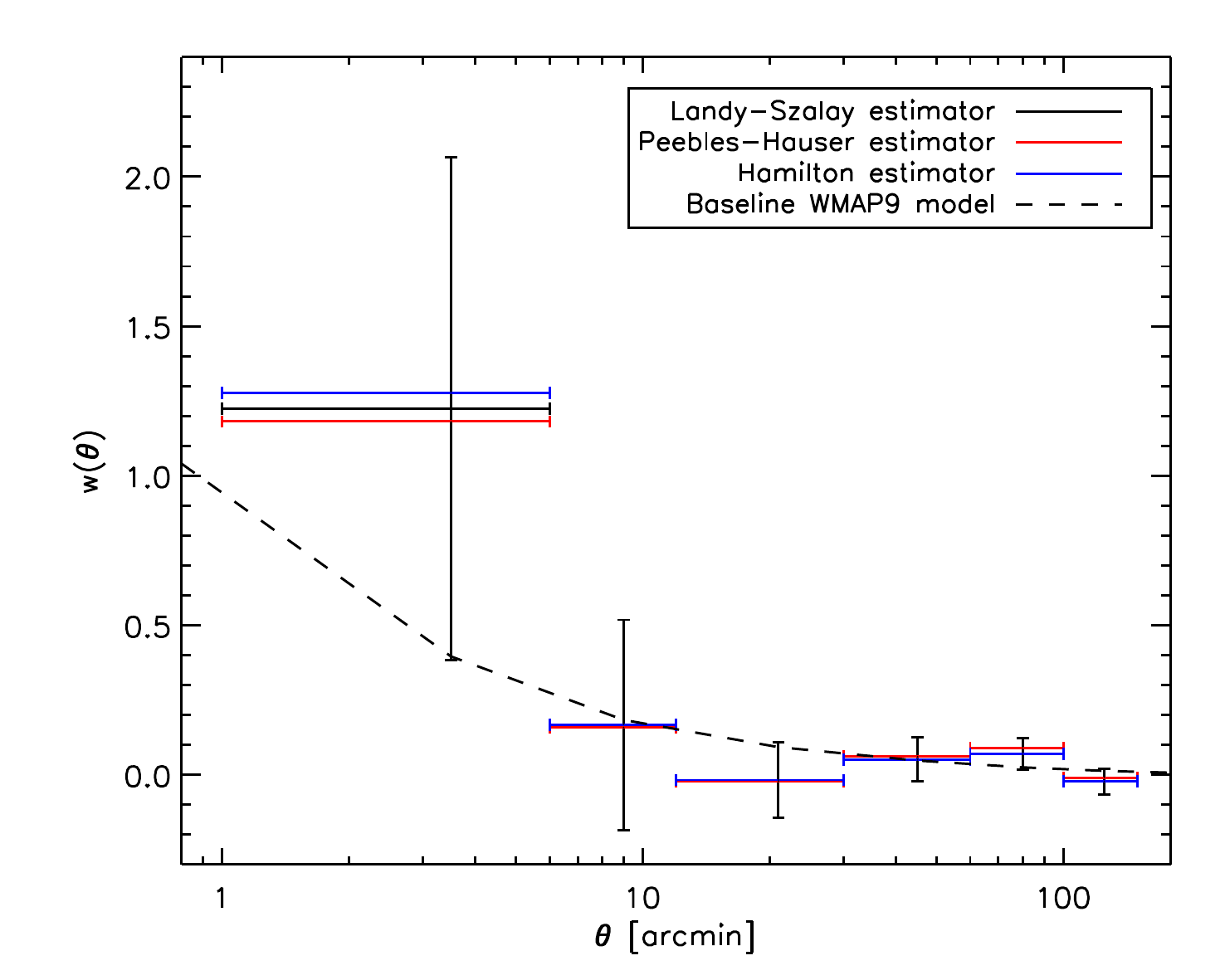}}
   \caption{Angular two-point correlation function of the \BXC. The different levels
     of correlation come from different estimators. The scatter between these estimators
     is well within the Poisson uncertainties, shown by the vertical error bars. The dashed
     line shows the expected correlation in the WMAP9 cosmology with the set of scaling
     relations obtained in \citetalias{Giles2016} and \citetalias{Lieu2016}.
    \label{fig:CorrFunc}}\vspace{-0.1cm}
\end{figure}

\subsection{Spatial distribution}
\label{sec:SpatDist}

The distribution of the \ChundredShort\ clusters over the sky (Fig.~\ref{fig:pointings})
shows hints of clustering. To quantify this visual impression,
we estimated the two-point angular correlation function (ACF) of the sample.
First, we generated random, unclustered catalogs using the 2D selection function maps
(see Fig.~\ref{fig:2Dproba}).
Each cluster was characterised by its pipeline core radius and its GCA count rate in the
60" aperture and we generated 100 dedicated selection function maps by interpolating over
the count rates and core radii used to estimate the sky coverage.
Random catalogues with 10000 members were finally obtained by simulating 100 realisations
of each cluster's position.
From these, we estimated the ACF using the \cite{Landy1993}, \cite{Hamilton1993}
and natural \citep{Peebles1974} estimators.
All give closely comparable results and show a positive correlation for
scales lower than 6\arcmin\ (see Fig.~\ref{fig:CorrFunc}). This signal is slightly
higher than the expectations from the fiducial WMAP9 model (also shown in
Fig.~\ref{fig:CorrFunc}), but the deviations are compatible within the error bars.

With this small sample and sky projection, the overall significance of this measurement is
not high enough yet to derive any useful constraints.
To proceed further, we ran a friend-of-friend (FoF) algorithm in the 3D physical space
and investigated the presence of large-scale structures among the \ChundredShort\ clusters.
Previous studies to identify superclusters based on optical cluster samples used linking
lengths in the range of 30 to 50\,Mpc, for instance 24~h$^{-1}$Mpc for \citet{Einasto2001} and
20 to 40~h$^{-1}$Mpc (redshift dependent) for \citet{Chow-Martinez2014}.
We therefore opted for a linking length of 35\,Mpc. This value matches the average linking
length used by \citet{Chon2013}, who considered several percolation thresholds applied to
the REFLEX-II X-ray cluster sample \citep{Boehringer2014}.

To qualify as a superstructure, a FoF detection had to contain at least three
connected \ChundredShort\ clusters, and among these at least one pair had to have
a separation of less than 10\,Mpc, which corresponds to $\sim10$ times the virial radius of a cluster of mass
$\ClMass=2\times10^{14}$ at redshift $z=0.3$.  This second criterion  ensures that the
detected superclusters are gravitationally bound.

With this procedure, we identified five structures.
We summarise here their main properties:

\begin{itemize}
\item \underline{XLSSC-a} is a very nearby ($z\sim0.05$) association of four groups spread
over the whole XXL-North field. All four groups belonging to the \Chundred\ -- XLSSC~011,
XLSSC~052, XLSSC~054, and XLSSC~062 -- have luminosities in the range
$1-3\times10^{42}~\mathrm{erg\,s^{-1}}$.
\item \underline{XLSSC-b} shows two subcomponents located at a redshift of $z\sim0.14$.
The eastern part consists of four \ChundredShort\ clusters (including the merger XLSSC~050), while
the western part is concentrated around one of the three Abell clusters located in the northern
XXL field (XLSSC~060, also known as Abell~329).
\item \underline{XLSSC-c} is located at $z=0.17$ and is the only superstructure that we found in the
southern field. Centred on the double system XLSSC~535/536 (see Appendix~\ref{append:ClusterProp}
for more details), it consists of six clusters, all from the \Chundred.
\item \underline{XLSSC-d} consists of four \ChundredShort\ clusters plus three other XXL C1+2 clusters
located at redshift $z=0.29$.
\item \underline{XLSSC-e} is located at $z=0.43$ and consists of six clusters, three of
which belong to the \Chundred. This structure is exceptionally compact over the sky; all
the clusters reside in the same XMM pointing. It is. however, much more elongated along
the line of sight. This structure is the subject of \citet[hereafter Paper VII]{Pompei2016}.
\end{itemize}

\begin{figure*}
   \begin{center}
    \resizebox{\hsize}{!}{{\fbox{
    \begin{overpic}[height=8cm,tics=10,trim=20 0 10 20,clip]{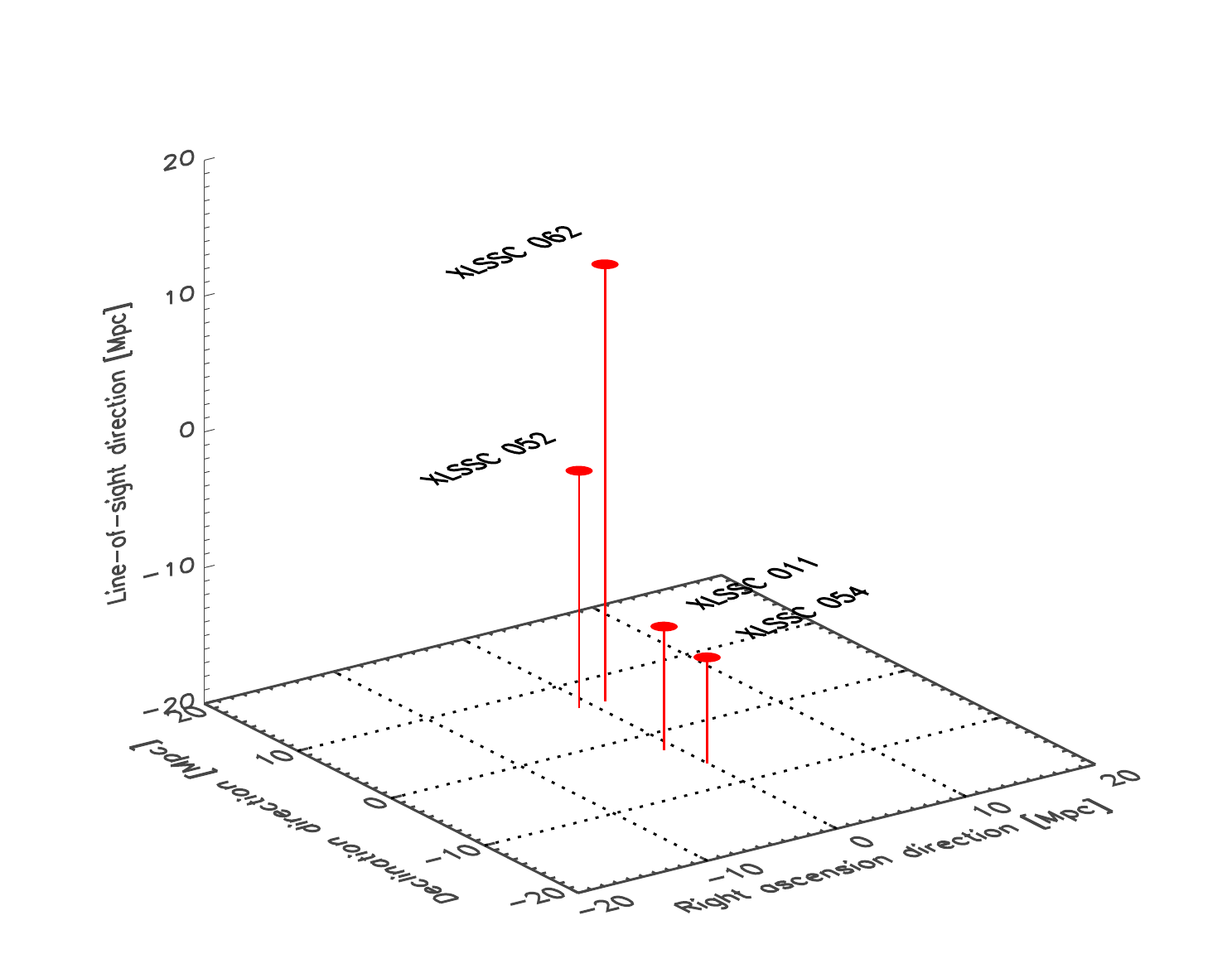}
      \put (78,73) {\large {\bf (XLSSC-a)}}
      \put (81,68.5) {\large {\bf z=0.05}}
    \end{overpic}}
    \fbox{\begin{overpic}[height=8cm,tics=10,trim=10 0 20 20,clip]{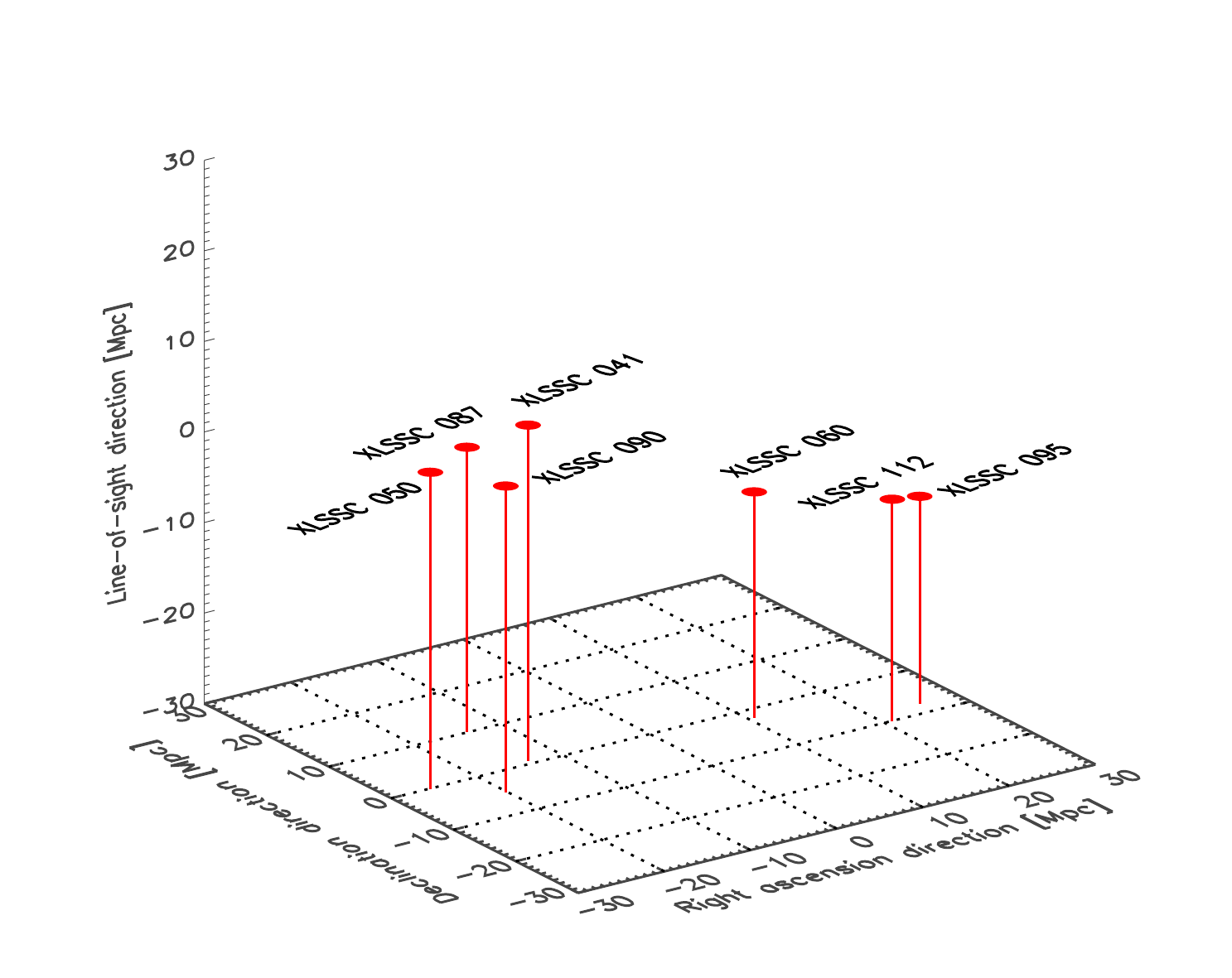}
      \put (78,73) {\large {\bf (XLSSC-b)}}
      \put (81,68.5) {\large {\bf z=0.14}}
    \end{overpic}}}}
    \resizebox{\hsize}{!}{{\fbox{
    \begin{overpic}[height=8cm,tics=10,trim=20 0 10 20,clip]{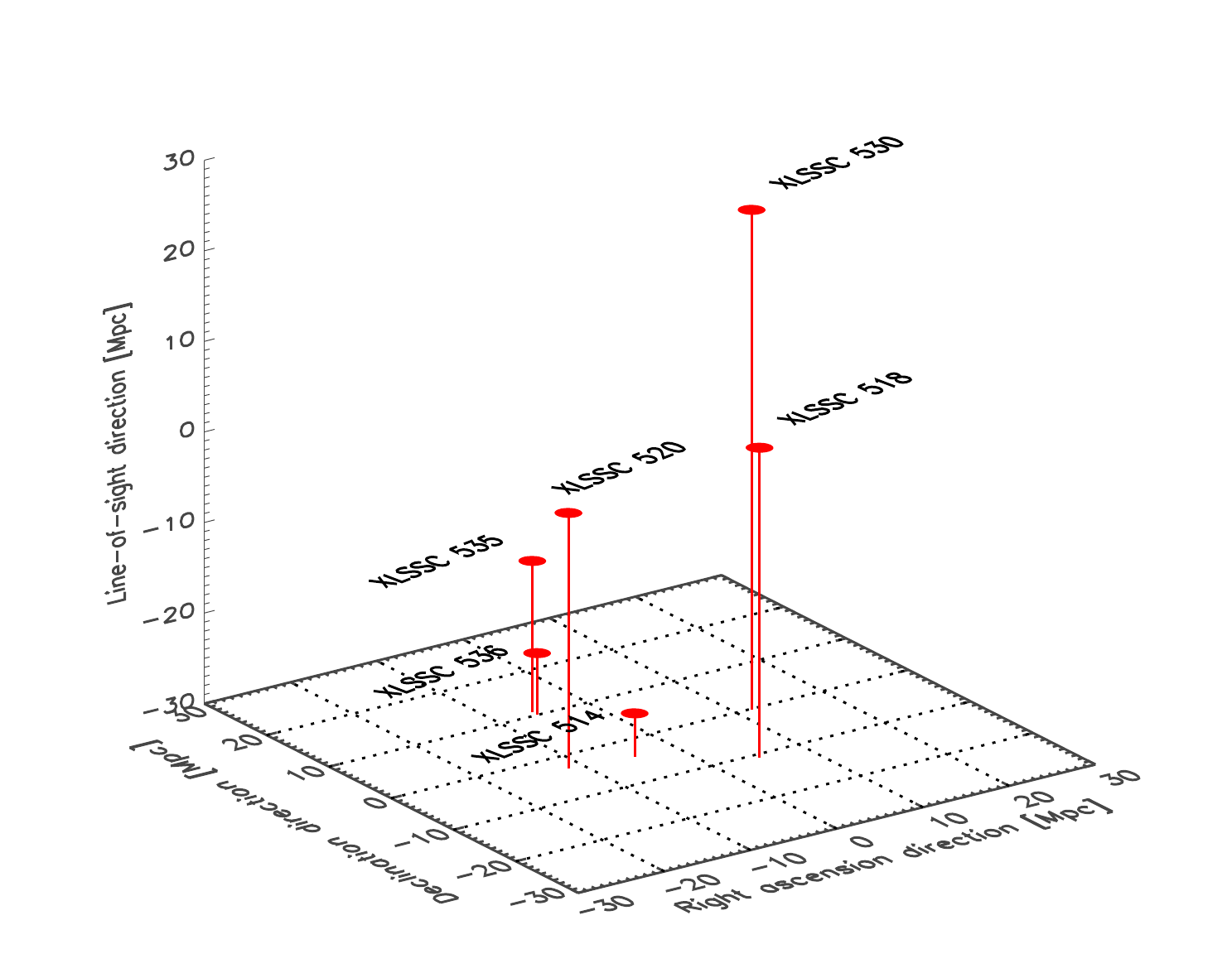}
      \put (78,73) {\large {\bf (XLSSC-c)}}
      \put (81,68.5) {\large {\bf z=0.17}}
    \end{overpic}}
    \fbox{\begin{overpic}[height=8cm,tics=10,trim=10 0 20 20,clip]{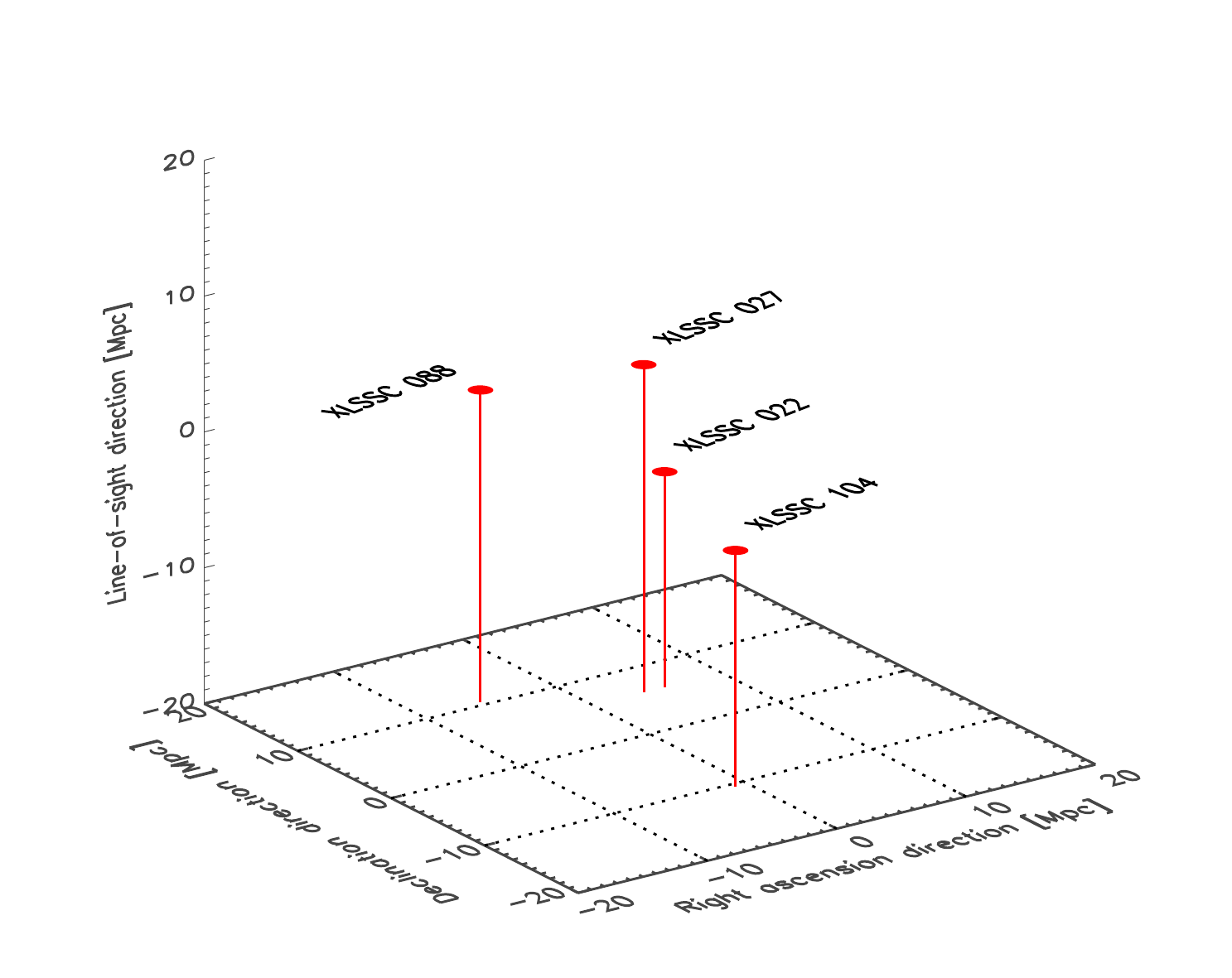}
      \put (78,73) {\large {\bf (XLSSC-d)}}
      \put (81,68.5) {\large {\bf z=0.29}}
    \end{overpic}}}}
   \resizebox{0.5\hsize}{!}{{\fbox{
    \begin{overpic}[height=8cm,tics=10,trim=-25 0 -25 20,clip]{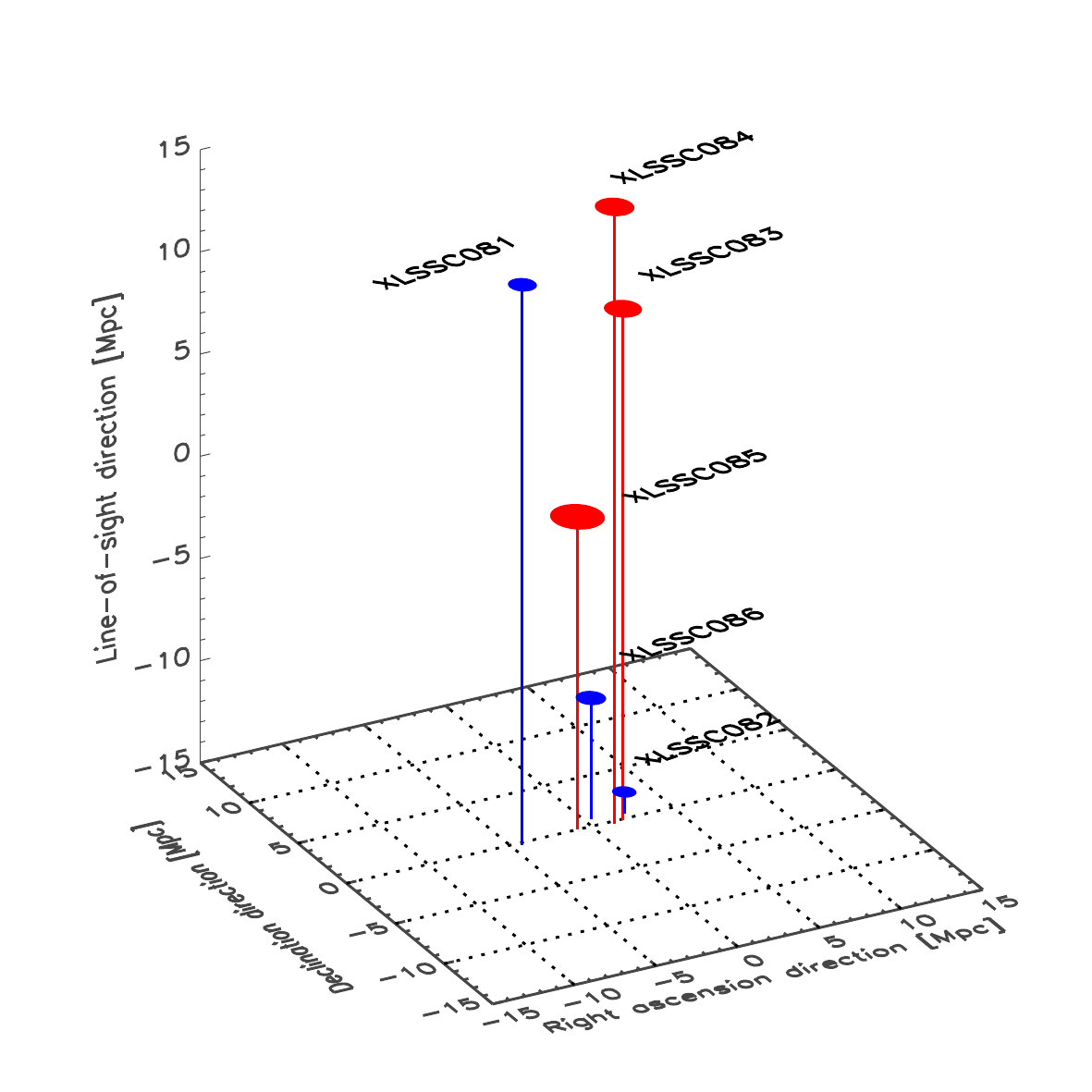}
      \put (78,73) {\large {\bf (XLSSC-e)}}
      \put (81,68.5) {\large {\bf z=0.43}}
    \end{overpic}}}}
    \caption{Three-dimensional configuration of five superstructures identified in the \BXC.
        {\it XLSSC-a}: located in the northern field with an average redshift of $z\sim0.05$.
	{\it XLSSC-b}: a double structure in the northern field at $z\sim0.14$.
		The western component is centred on XLSSC~060 (also known as Abell~329).
	{\it XLSSC-c}: located in the southern field at $z\sim0.17$. It is centred
	on the cluster Abell~4005, which we resolve into two subcomponents (XLSSC~535 and 536).
	{\it XLSSC-d}: located in the northern field at $z\sim0.29$. The central
	pair of clusters has already been identified in the XMM-LSS in \cite{Pacaud2007}.
	{\it XLSSC-e}: This $z=0.43$ supercluster in the northern field is studied in depth
	in \citetalias{Pompei2016}. Clusters published in \citetalias{Pompei2016} that do not
	pertain to the \Chundred\  are in blue. The symbol size reflects the cluster
	masses inferred in \citetalias{Pompei2016}.
	\label{fig:Structures}}
    \end{center}
\end{figure*}

Their characteristics are provided in Table~\ref{tab:SuperProb} and the 3D
configuration of each structure is shown in Fig.~\ref{fig:Structures}.
Interestingly, they can explain the previous measurement of the ACF very well.
Indeed, for separations lower than 12\arcmin, structures XLSSC-c, XLSSC-d and XLSSC-e
account for almost all the excess of pairs above the random distribution.

In the process, we also found two nearby pairs of clusters with distances of less than 10\,Mpc.
They consist of (XLSSC~524, XLSSC~519) at redshift $z=0.270$ and (XLSSC~103, XLSSC~055) at
$z=0.232$.

\begin{table*}
  \begin{center}
   \caption{Properties of the five \ChundredShort\ superstructures.\label{tab:SuperProb}}
   {\small \begin{tabular}{lcccccl}
	\hline\hline
 	  Name       &     RA    &    Dec   &   $z$  &  $N_{cl}$  & $M_{\mathrm{tot,MT}}$ & XLSSC Members   \\
 	    -        &    [deg]  &    [deg]  &    -   &      -     &   $10^{14}\,M_\odot$  &   -   \\
	\hline
	  XLSSC-a    &    36.55  &   -4.06   &   0.05 &     4      &   3.2    & 011, 052, 054, 062 \\
	  XLSSC-b    &    35.39  &   -4.70   &   0.14 &     7      &  10.3    & 041, 050, 060, 087, 090, 095, 112\\
	  XLSSC-c    &   350.67  &  -54.48   &   0.17 &     6      &   6.3    & 514, 518, 520, 530, 535, 536\\
	  XLSSC-d    &    37.22  &   -5.05   &   0.29 &     4      &   8.5    & 022, 027, 088, 104\\
	  XLSSC-e    &    32.87  &   -6.20   &   0.43 &     3      &  11.9    & 083, 084, 085\\
	\hline
   \end{tabular} }
  \tablefoot{RA,Dec: WCS coordinates are in the J2000 system; these were computed from
  the mean of the cluster member positions, except for XLSSC-e where the position of the
  central massive cluster XLSSC~085 was used. $N_{cl}$: number of
  \ChundredShort\ clusters pertaining to the structure. $M_{\mathrm{tot,MT}}$: sum of the
  $\ClMassMT$ masses of all members.}
  \end{center}
  \vspace{-0.1cm}
\end{table*}

\section{Summary and discussion}
\label{sect:Discuss}
%______________________________________________________________

We have defined and presented the \ChundredShort\ galaxy cluster sample, a complete subsample
of the full XXL extended source catalogues, which consists of 100 clusters above a flux
cut of $3\times10^{-14}~\mathrm{erg\,s^{-1}cm^{-2}}$ in a 60\arcsec\ aperture. The selection
function of the sample was carefully estimated, for $\beta$-model clusters, including the
initial pipeline selection, the additional flux cut, as well as the layout and depth of the
XXL observations. We have obtained spectroscopic redshifts for 97 of the clusters and
reliable photometric redshifts for the 3 remaining ones.
Based on the mass and temperature scaling relations self-consistently measured from the
same sample in \citetalias{Giles2016} and \citetalias{Lieu2016}, we compared the redshift
distribution of the sample with model predictions and found that the XXL cluster population
is better reproduced by low $\sigma_8$ models such as the WMAP9 cosmology than models with
higher values of $\sigma_8$ like the one obtained recently from the {\it Planck} satellite.
We then studied the luminosity function of the sample and again obtained results that are
overall consistent with the WMAP9 model, the only discrepancy being a lack of low luminosity,
low redshift clusters. The luminosity function shows no sign of evolution at any redshift,
confirming some earlier findings but contrasting with claims of a negative evolution at
high redshift for more luminous clusters \citep{Mullis2004,Koens2013}.
An attempt to fit the $\Lxxl - \Txxl$ scaling relations indirectly to the
luminosity and redshift distribution of the sample proved consistent with the results
of \citetalias{Giles2016}, which were based on the measured cluster temperatures.
Finally, we reported a significant clustering of sources on scales smaller than 6\arcmin,
which we entirely resolved into five large-scale superclusters identified using FoF
techniques.

In this section, we examine the impact of the main assumptions on the derived results
and consider the implications of the present work for the next steps of the XXL project.

\subsection{The cluster number density}
\label{sect:DiscussDens}

\begin{figure}
   \begin{center}
    \includegraphics[width=7.8cm,viewport=15 0 415 325,clip]{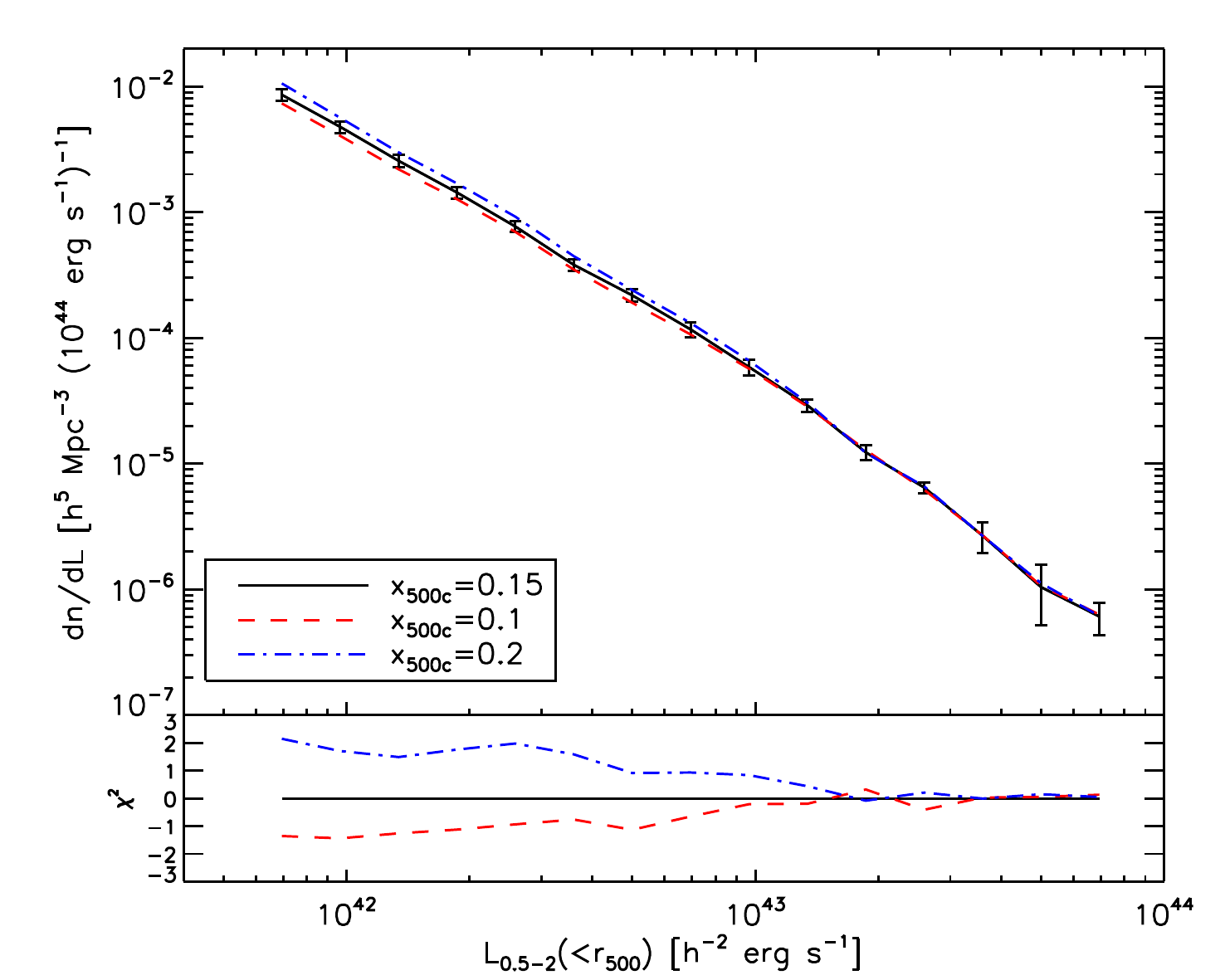}
    \caption{Impact of different choices of $x_{500} = r_c/\ClRad$ on the \ChundredShort\ luminosity function.\label{fig:LfXc}}
   \end{center}
\end{figure}

The deficit of clusters at low luminosity compared to the WMAP9 expectations and the absence
of evolution in the cluster luminosity function are certainly the most significant
results obtained by the analysis of the \Chundred\ number density presented in Sect.~\ref{sect:Cosmo}.
They all rely on the central assumption that the cluster emission profile is represented
well by a simple $\beta$-model with fixed  $\beta=2/3$ and $x_{500}=r_c/\ClRad$ of 0.15.
To check the robustness of these findings, we reproduced the whole analysis for
different values of $x_{500}$. Luminosities in $\ClRadMT$ were re-evaluated from the
measurements in 300\,kpc apertures based on the new $\beta$-model profiles. Then, the
effective volume probed by the survey was updated with the corresponding selection functions
and the altered $\Lxxl-\Txxl$ scaling relations provided in \citetalias{Giles2016}.
The resulting \ChundredShort\ luminosity functions are shown in Fig.~\ref{fig:LfXc} for
$x_{500}=0.1$ and $0.2$, which provide conservative boundaries for this parameter.
The change is greatest for luminosities below $6\times10^{43}\,\mathrm{h^{-2} erg\,s^{-1}}$,
but never exceeds $\sim1\sigma$. Consequently, although the uncertainty on $x_{500}$ would
make a significant contribution to the error budget, it does not yield changes large enough
to affect our conclusions.

\begin{figure}
  \begin{center}
     \includegraphics[width=7.8cm,viewport=15 2 415 325,clip]{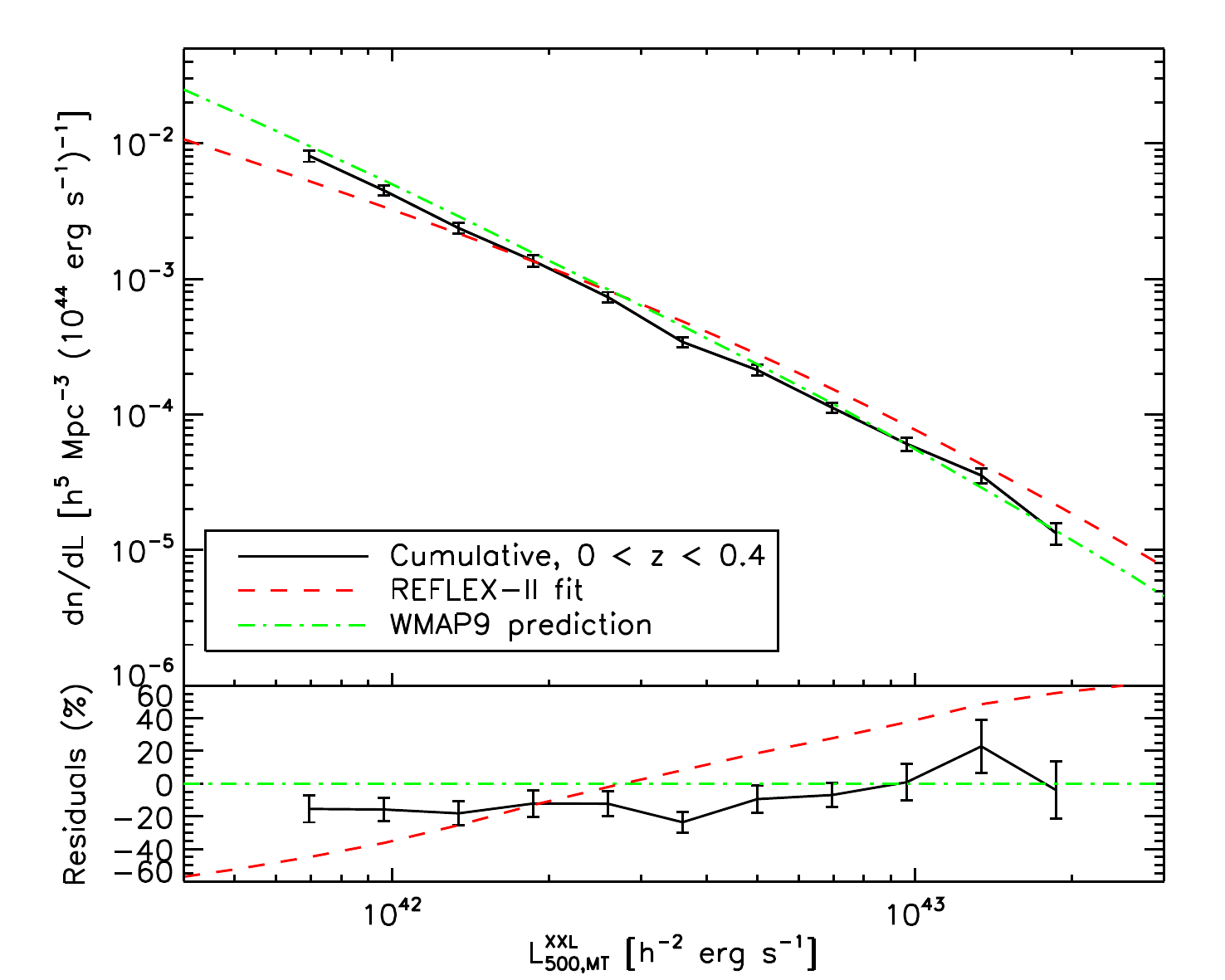}
     \caption{
    Differential luminosity funcion averaged in the redshift range $[$0-0.4$]$, computed in the
     same way as for Fig.~\ref{fig:LumFunc}, in  comparison with measurements from RELEX-II in
     the same redshift range and predictions for the WMAP9 model. The residual plot shows the
     fractional deviation with respect to the model predictions.
     \label{fig:LfuncREFLEX}}
  \end{center}
\end{figure}

To further check that the observed lack of clusters is real, we also compared the \ChundredShort\
measurements at low redshift ($z<0.4$) with the best fit luminosity function obtained by
the REFLEX-II collaboration \citep{Boehringer2014}, the largest complete sample of low redshift
X-ray clusters available to date.
This was rescaled by a global factor of $0.59$ to account for different spectral
bands\footnote{For temperatures between 0.5 and 5~keV, and redshifts between 0 and 0.4, the
K-correction between the $[0.5-2]$ and $[0.1-2.4]$ keV bands mostly shows variations lower than 5\%.}.
As is evident from Fig.~\ref{fig:LfuncREFLEX}, the REFLEX-II luminosity function is significantly
flatter than both the \ChundredShort\ measurements and the WMAP9 prediction. More precisely, in the \Chundred,
we observe 30\% fewer bright objects and 40\% more low luminosity clusters.
Such deviations are well above the uncertainties arising from the assumed spectal band correction,
and also larger than the expected $\sim$15\% fluctuations from Poisson and cosmic variance
over the whole sample.
As the two samples are not drawn from surveys of the same depth, the average redshifts probed
by any given luminosity bin differ. However, the absence of redshift evolution in the \ChundredShort\
luminosity function analysis rules out this possible explanation.
More likely, most of the discrepancy originates in the methods used to estimate the cluster
luminosities. Indeed, the $\ClRad$ apertures used by the REFLEX-II collaboration are not
based on the same set of scaling relations and rely on the cluster luminosities through
an iterative procedure. Then, the initial extraction radius for the luminosity estimates and
the recipe for the aperture correction differ. In addition, the K-correction is applied in
a different way and does not account for the scatter in the $L_X-T$ scaling relation.
Despite these differences, it is possible that some of the mismatch could come from the
properties of the XXL fields themselves. The two fields follow the optical footprint
of optical surveys that purposely avoided the presence of very local bright clusters.
More importantly, on the low-luminosity end, most of the information in the \Chundred\ comes
from the northern field, as a result of the presence of the superstructure XLSSC-a.
Given the density of this structure, which covers the whole northern field, it is likely
that the \ChundredShort\ luminosity function estimates lie on the upper end of the real space
density of clusters in this luminosity range, which would bring the REFLEX-II measurements
into better agreement, but would make the deviation from the WMAP9 model even larger.


From these considerations, we conclude that the lack of observed \Chundred\ clusters at low
luminosity is probably genuine and reflects the fact that the WMAP9 and scaling relation
model used in this paper only provides a first-order description of the cluster density
in the XXL fields.

In the 11\dd\ XMM-LSS field, \citet{Clerc2014} also observed a deficit of clusters in the
$0.5<z<0.7$ range. There is no sign of such a deficit in the rest of the XXL area, so the
earlier results should be ascribed to a cosmic void.

\subsection{Large-scale structures}
Historically, the searches for superstructures in the distribution of galaxy clusters
started with the first Abell catalogues (see e.g. \citealt{Batuski1985} and
\citealt{Kalinkov1995} for the largest samples). Since then, superclusters of lower total
mass have been identified from comprehensive spectroscopic surveys such as the  2dF galaxy
redshift survey \citep{Einasto2007} or the Sloan Digital Sky Survey \citep{Einasto2014},
but remained limited to the local universe ($z\la0.1$). Recently, \citet{Chon2013} extended
the volume used for such studies to $z\la0.3$ and presented the first supercluster
catalogue identified from a complete X-ray selected cluster sample. This new selection
method has the advantage of relying only on galaxy structures that show clear evidence of
a deep potential well.
Although a few isolated very high redshift superclusters are known \citep[e.g.][]{Gal2004},
our work is the first attempt to systematically unveil superstructures up to $z\sim0.5$
in a homogeneous X-ray sample. In addition, the detected structures do not correspond to
the traditional picture of superclusters as very massive systems
($\sim10^{15}-10^{16}\,\mathrm{M_\odot}$) currently starting to collapse, but rather exhibit
masses similar to low redshift massive clusters. These clusters can tell us about the
large-scale structure of the Universe, but they can also provide a unique opportunity to
witness the formation process of current galaxy clusters and understand the origin of their
observational properties. With an XXL sample that is four times larger, we shall therefore
be able to build the first representative sample of more than 10 clusters in the early days
of their formation and use the associated XXL multiwavelength data set to study the galaxy
population in these systems as well as their ICM properties.

The 1.5$\sigma$ evidence of correlation on scales of less than 6 arcmin is encouraging for the XXL goal
of measuring the 3D correlation function. With more than 450 clusters expected for the full sample,
the relative uncertainty on the angular correlation function will decrease by a factor of
$\sim4$, enabling a significant measurement in a few angular bins.
As shown in \citet{Valageas2011}, the signal-to-noise ratio for the 3D correlation function
averaged over the survey volume will be even higher. Their figure~12 demonstrates that a
significant detection of positive correlation should already be obtained for about six bins
between 5 and 30~$\mathrm{h^{-1}Mpc}$ for a 50\dd\ survey with a single mass limits of
$10^{14}~\mathrm{M_\odot}$, a selection that is similar to the \Chundred\
in terms of number density and average mass. However, random catalogs for the 3D cluster distribution
cannot be easily obtained in a model-independent way as we did for the projected sky
distribution. They depend very much on the choice of underlying cosmology and scaling relations.
We therefore defer the measurement of the correlation in 3D to a later work where the
use of the 3D correlation for cosmological constraints will be explored in detail.

\subsection{Cosmological leverage of the \BXC}
\label{sec:CosLeverage}
Obtaining competitive constraints on the dark energy equation of state is a major science
goal of the XXL Survey. To assess the constraining power already achieved with the \Chundred,
we ran an illustrative cosmological analysis based on the luminosity and redshift distribution
of the sample, using the formalism described in Sect.~\ref{sec:LumConst} and
Appendix~\ref{append:Like}. This is very similar to the analysis performed in
\citet{Mantz2008}, except that we use an unbinned likelihood that includes cosmic variance.

As a first attempt, we fix all scaling relation parameters to the fiducial XXL values and
fitted $\sigma_8$ while fixing all other cosmological parameters to their fiducial values.
This resulted in a lower value than the WMAP9 model with $\sigma_8=0.807\pm0.018$, as expected
from the observed deficit of low redshift clusters.
Even with this perfect mass calibration, the achieved uncertainties are of nearly the same
magnitude as the mismatch between the values of $\sigma_8$ obtained by Planck and those
determined by WMAP9 or from late-time tracers of matter fluctuations.
Therefore, the \Chundred\ alone is not large enough to precisely investigate this issue.
We also allowed the parameters of the  $\Lxxl-\Txxl$ scaling relation to
be fitted, including priors according to the measurements of \citetalias{Giles2016}.
Owing to the XXL likelihood model, these measurements are essentially independent of the normalisation
of the mass function (here governed by $\sigma_8$) and mostly result from the temperature
information (which we neglect), so they can be considered as independent constraints. This second fit yields
$\sigma_8=0.814^{+0.037}_{-0.034}$ and implies that the ability to constrain cosmology with
XXL is currently limited by the cluster mass calibration, even with the \ChundredShort\
subsample.
Finally, fitting also for $\Omega_m$, we observe a strong degeneracy and the marginalised
error on $\sigma_8$ becomes $\Delta\sigma_8\sim0.05$ even when the scaling relation
parameters are held fixed. This does not come as a surprise. Even with the final XXL
cluster sample and using the cluster 3D correlation function, which together improve the
constraining power of the survey by a factor $\sim4$, it was already noted in
\citet{Pierre2011} that the combination with other cosmological probes (e.g. CMB
measurements from {\it Planck}) would be necessary to break the cosmological degeneracy and
isolate constraints on the evolution of dark energy.

As already mentioned, the assumed model of the cluster surface brightness profile (as summarised by
the $x_{500}$ parameter) can also have a significant impact on the above results.
The modified models with $x_{500}$=0.1 and 0.2 discussed in Sect.~\ref{sect:DiscussDens}
respectively predict a total of 134 and 100 clusters, i.e. changes comparable with the Poisson noise
standard deviation.
Repeating the $\sigma_8$ analysis for these models, we observe that the statistical errors remain
unchanged, but the best fit value varies (by, respectively,  +0.003 and -0.010).
The shift in $\sigma_8$ is smaller than the most optimistic error bars obtained in the previous
paragraph (i.e. neglecting both uncertainties on scaling relation and cosmological
degeneracies); therefore, a reasonable change in $x_{500}$ has little impact on the cosmological constraints
achievable from the \Chundred. However, this additional uncertainty will become very important for the
modelling of the complete XXL sample. In a later work (D\'emocl\`es et al., in prep.) we
will self-consistently address the determination of $x_{500}$ using the \Chundred\
to reduce this source of modelling uncertainty.
Another concern would be the existence of a subpopulation of clusters with centrally peaked
surface brightness profiles due to either cool-cores or central AGN contamination, since such
clusters might fail to pass the pipeline C1+2 extension criteria.
To evaluate the possible impact of such a subpopulation of clusters on the \Chundred,
we adopted a data-oriented method.
First, we inspected visually a random subsample of 100 XXL point-like sources with a measured pipeline
count rate above the \ChundredShort\ aperture flux limit.
None of those sources was a likely cluster candidate. Assuming a binomial likelihood, it
implies a $1\sigma$ upper limit of 1.1\% on the proportion of genuine clusters among the
sources above the aperture flux classified as point-like by the pipeline. Since the XXL
AGN sample in good pointings contains fewer than 1000 such sources, we estimate an upper limit
of 10 missing clusters due to centrally concentrated profiles. This number is again lower
than the Poisson noise. In addition, we screened all the soft band point-like sources that were
detected as extended in the hard [2-10]~keV band. This enabled us to recover two bright and
very compact low redshift clusters (XLSSC~118 and XLSSC~552) whose properties are detailed
in Table~\ref{SuplTab} and Appendix~\ref{append:ClusterProp}. Although they show traces
of point-source contamination, they seem to be genuinely extended but very centrally peaked
systems. As the impact of cluster morphology on the XXL pipeline selection function does
not seem to strongly affect the results of the present work, we defer a more detailed analysis
to forthcoming XXL papers.

\section{Conclusions}
%______________________________________________________________
In this work, we introduced the \BXC, a subsample of the hundred brightest XXL clusters
detected in the available 46.6\,\dd. The source selection process consists of
a $3\times10^{-14}\,\rm{erg\,s^{-1}cm^{-2}}$ aperture flux cut within a 1\arcmin\ radius in
addition to the initial XXL pipeline.
All but four sources were confirmed with optical spectroscopy and the selection function of
the sample was carefully modelled. We were able to determine well-constrained X-ray temperatures,
and luminosities for most of them and even weak lensing masses for a subsample of 38
clusters. This allowed for a self consistent analysis of the cluster number density
(in this article) together with the cluster scaling relations (in \citetalias{Giles2016} and
\citetalias{Lieu2016}).

We studied the luminosity function of the sample and concluded that it does not evolve
over the considered redshift range. The observed luminosity and redshift distributions match
low $\sigma_8$ models more closely than the currently favoured {\it Planck} cosmology. However, the
alternate WMAP9 model is not perfect either and overpredicts the amount of faint clusters.
We detected a positive angular correlation between the \ChundredShort\ clusters,
which we were entirely able to resolve into five new superclusters detected in the XXL field.
These structures are likely progenitors of local massive clusters and will shed light on
a new phase of the build-up of galaxy clusters across cosmic times.

Basic attempts to assess the cosmological leverage of the sample revealed that,
even with the current subsample, the error budget is already dominated by uncertainties in
the cluster mass calibration, which therefore will be one of the priorities of the XXL
cluster science program over the next few years. The situation will improve significantly
in the near future with the deeper weak-lensing observations currently being performed
over the whole XXL field (see \citetalias{Pierre2016} for details), the introduction of
additional cluster mass indicators, and further cluster follow-up studies.

Extrapolating from the observed \ChundredShort\ cluster population, we expect the whole
XXL X-ray analysis to yield some 450 cluster detections, which is consistent with the current
list of XXL cluster candidates, but significantly less than the initial predictions from
\citet{Pierre2011}. Cluster count predictions in this forecast article were done on the
basis of ad hoc scaling relations and a very rough approximation for the cluster shape
($\beta$-models with a constant core radius of 180\,kpc).
A thorough analysis of the cluster density will be given when the full cluster catalogue
is published. In this future paper, we shall explore the selection function in greater
detail and consider the effect of cluster morphology, cool cores, baryon physics, and AGN
contamination by means of high-resolution hydrodynamical simulations and semi-analytic models
\citepalias[cf.][]{Pierre2016}.

The detailed properties of the \Chundred\  will be distributed through the dedicated XXL
cluster database\footnote{{\tt http://xmm-lss.in2p3.fr:8080/xxldb/}}, together with the seven
complementary clusters detected in contaminated pointings but bright enough to exceed the
\ChundredShort\ aperture flux cut.
A master XXL-100-GC cluster catalogue will also be available in electronic form in the XXL
Master Catalogue Database in Milan\footnote{{\tt http://cosmosdb.iasf-milano.inaf.it/XXL/}}.
Public access to both databases will occur after the paper is accepted.

\begin{acknowledgements}
%______________________________________________________________
 XXL is an international project based around an XMM Very Large Programme
surveying two 25 \dd extragalactic fields at a depth of
$\sim5\times10^{-15}\,$\flux in $[$0.5-2$]$ keV.
The XXL website is {\tt http://irfu.cea.fr/xxl}.
Multiband information and spectroscopic follow-up of the X-ray sources are
obtained through a number of survey programmes, summarised at
{\tt http://xxlmultiwave.pbworks.com/}.\\
This research has made use of the NASA/IPAC Extragalactic Database (NED) which
is operated by the Jet Propulsion Laboratory, California Institute of Technology,
under contract with the National Aeronautics and Space Administration.
This research has made use of {\sc Aladin sky atlas} and {\sc SIMBAD} database,
both developed at CDS, Strasbourg Observatory, France. In particular, we would like
to thank Thomas Boch for his help with {\sc Aladin Lite}.\\
FP acknowledges support from the BMWi/DLR grant 50~OR~1117, the DFG grant RE~1462-6 and
the DfG Transregio Programme TR33.
TR acknowledges support from the DFG Heisenberg grant RE~1462-5.
The Saclay group acknowledges long-term support by the Centre National d'Etudes Spatiales (CNES).
BJM, PAG and MB acknowledge support from STFC grants ST/J001414/1 and ST/M000907/1.
The Dark Cosmology Centre is funded by the Danish National Research Foundation.
\\
The XXL team would also like to thank the VLT support astronomers C.~Hummel and N.~Neumayer
for their dedicated help in the implementation of the XXL ESO Large Program and with the
data reduction.
\end{acknowledgements}

\begin{appendix}

\section{Notes on individual clusters}
\label{append:ClusterProp}

\begin{itemize}

\item \underline{XLSSC~041}: at redshift $z=0.14$, overlaps with a fainter background
galaxy cluster, XLSSC~045, at $z=0.556$ \citep{Adami2011}.

\item \underline{XLSSC~050}: at redshift $z=0.14$, is a very elliptical cluster showing
clear signs of merging activity. In particular, all of its bright galaxy members are
located on opposite sides of the ellipsoid, suggesting a significant offset between the
gas and collisionless material.

\item \underline{XLSSC~052}: a large fraction of the X-ray emission overlaps a bright
saturated star, BD-03~381 from the Tycho catalogs \citep{Hog1998}.

\item \underline{XLSSC~057}: is the X-ray counterpart of Abell~334 \citep{Abell1989}.
The offset between the optical and X-ray detection is 2.9\arcmin.

\item \underline{XLSSC~060}: is the X-ray counterpart of Abell~329 \citep{Abell1989}.
The offset between the optical and X-ray detection is 0.7\arcmin. This cluster is also part
of the {\it Planck} Sunyaev-Zel'dovich cluster catalogue \citep{Planck2015XXVII} as
PSZ2~G167.98-59.95. The infered XXL $\ClMassMT$ falls within 2\% of that reported by the Planck team.

\item \underline{XLSSC~062}: the X-ray emission is possibly dominated by three bright
central galaxies.

\item \underline{XLSSC~083/XLSSC~084}: part of the supercluster studied in
\citetalias{Pompei2016}. Their X-ray emission significantly overlap.

\item \underline{XLSSC~085}: the cluster is surrounded by two X-ray AGNs.
Although they are correctly identified and masked by the pipeline, the cluster remains
an outlier in the $\ClMassWL-\Txxl$ scaling relation presented in \citetalias{Lieu2016}, possibly
indicating a residual contamination.

\item \underline{XLSSC~089}: extremely compact and rich system at $z=0.609$ where most of
the X-ray emission and all of the seven spectroscopically confirmed members are found within
a 30\arcsec\ aperture.

\item \underline{XLSSC~090}: fossil group at $z=0.141$.

\item \underline{XLSSC~091}: is the X-ray counterpart of Abell~362 \citep{Abell1989}.
The offset between the optical and X-ray detection is 2.0\arcmin. This cluster is also part
of the {\it Planck} Sunyaev-Zel'dovich cluster catalogue \citep{Planck2015XXVII} as
PSZ2~G174.40-57.33. The infered XXL  $\ClMassMT$ falls within 14\% of the value reported by the Planck team.

\item \underline{XLSSC~094}: the X-ray emission is probably contaminated by a background
AGN ($z=1.11$) located precisely at the peak of the X-ray emission.

\item \underline{XLSSC~096}: the main galaxy cluster at $z=0.520$ is superimposed on another
redshift overdensity at $z=0.203$. The bright galaxies at $z=0.520$ follow the central
X-ray overdensity, while the low redshift system is much more sparse.

\item \underline{XLSSC~104}: a secondary peak is apparent in the X-ray emission and probably
indicative of some AGN contamination. The cluster is an outlier in the $\ClMassWL-\Txxl$
scaling relation presented in \citetalias{Lieu2016}.

\item \underline{XLSSC~110}: at $z=0.445$ show strong lensing arcs around the probable
brightest cluster galaxy.

\item \underline{XLSSC~113}: nearby group at $z=0.050$ with signs of emission from individual
member galaxies.

\item \underline{XLSSC~118}: bright and surprisingly compact ($x_{500}\sim0.02$) cluster at
$z=0.140$, classified as a point source by the XXL detection pipeline in the [0.5-2] keV band
but recovered as extended in the [2-10] keV band. The surface brightness shows traces of
point-source contamination.

\item \underline{XLSSC~504}: one of the faintest clusters in the \Chundred. The emission is
very compact and peaked with signs of extension to the north-west, suggesting significant
contamination by a central AGN. This is confirmd by the presence of a bright flat-spectrum
radio source detected in the PARKES-MIT-NRAO \citep{Gregory1994} and SPT \citep{Mocanu2013}
surveys.
The temperature measured in \citetalias{Giles2016} without excluding the AGN is very high
but the faint signal does not allow a better analysis.
Consequently, this cluster is not included in the scaling relation and cosmology fits
presented in this paper and in \citetalias{Giles2016} (which we account for with an additional
incompleteness factor of 0.99).

\item \underline{XLSSC~513}: This cluster was detected by the SPT collaboration
\citep{Bleem2015} as SPT-CL~J2316-5453. The inferred XXL $\ClMassMT$ falls within 19\% of
the value reported by the SPT team.

\item \underline{XLSSC~514/XLSSC~515}: their separation on the sky is less than 2\arcmin.
Two clear redshift overdensities are visible at $z=0.101$ and $0.169$ , associated with
two distinct peaks in the X-ray emission. However, the X-ray analysis might be affected
by the superposition.

\item \underline{XLSSC~516}: was assigned a photometric redshift of $z=0.87$.
The X-ray emission, although surrounded by two AGNs, seems genuinely extended.
Despite a first attempt to observe it with VLT/FORS2, the cluster has not yet been confirmed.
However, VLT/HAWK-I near infrared imaging suggests a highly compact cluster, not visible
in the optical images. The photometric redshift is therefore probably underestimated.

\item \underline{XLSSC~523}: is the X-ray counterpart of Abell~S1115 \citep{Abell1989}.
The offset between the optical and X-ray detection is 1.5\arcmin.

\item \underline{XLSSC~526}: is the X-ray counterpart of Abell~S1112 \citep{Abell1989}.
The offset between the optical and X-ray detection is 1.3\arcmin.

\item \underline{XLSSC~535/XLSSC~536}: are the X-ray counterparts of Abell~4005 \citep{Abell1989},
with the optical centre located right in between the two X-ray detections. The two clusters
are aligned in the north-south direction which calls for a redefinition of Abell~4005
as Abell~4005N (XLSSC~535) and Abell~4005S (XLSSC~536).
Taken at face value, the coordinates and redshift of the two subcomponents indicate
a separation of $\sim9$~Mpc. However, their angular separation only corresponds to 700~kpc
and most of the distance could be accounted for by a Doppler shift of
$\sim700~\mathrm{km\,s^{-1}}$, a plausible value for two clusters heading toward each other.
Therefore we cannot exclude that the two clusters are already in an early merging phase.
This pair of clusters forms the centre of a larger superstructure, structure~2, described in
Sect.~\ref{sec:SpatDist}.

\item \underline{XLSSC~539}: shows two reshift structures at $z=0.169$ (3 concordant redshift)
and $z=0.182$ (2 redshifts). In the absence of a better criterion, we opted for the richest
structure.

\item \underline{XLSSC~541}: is the X-ray counterpart of Abell~4027 \citep{Abell1989}.
The offset between the optical and X-ray detection is 0.5\arcmin.

\item \underline{XLSSC~542}: is part of the Planck Sunyaev-Zel'dovich cluster catalogue
ß\citep{Planck2015XXVII} as PSZ2~G325.99-59.48. The inferred XXL $\ClMassMT$ falls within
10\% of the value reported by the Planck team.
The cluster was also detected by the SPT collaboration \citep{Bleem2015} as SPT-CL~J2332-5358.
Their mass estimate for the cluster deviates by only 16\%.

\item \underline{XLSSC~549}: was assigned a redshift of $z=0.808$ based on three,
well-centered, concordant redshifts. However there seems to be a second structure along
the same line of sight at $z=0.568$ with two redshifts.

\item \underline{XLSSC~552}: is a bright and surprisingly compact ($x_{500}\sim0.03$) cluster at
$z=0.152$, classified as a point source by the detection pipeline in the [0.5-2] keV band
but recovered as extended in the [2-10] keV band. Possible but relatively small point-source
contamination.

\end{itemize}

\section{Comparison of luminosity function estimators}
\label{append:LFcomp}

The luminosity function is most often measured (e.g. \citealt{Mullis2004} or
\citealt{Boehringer2014}) by dividing the cluster sample into luminosity bins of width
$\Delta L_i$ and using the estimator
\begin{equation}
 \frac{\mathrm{d}n}{\mathrm{d}L}\left(<L_i>\right) = \frac{1}{\Delta L_i} \sum_{j=0}^{N_i} \frac{1}{V_{eff}(L_j)},
 \label{eq:PunctCorr}
\end{equation}
where $<L_i>$ is the centre of luminosity bin $i$, the summation runs over the $N_i$ detected
clusters in that bin, and $V_{eff}$ is the luminosity dependent effective volume from which
each cluster was selected (accounting for the sky coverage).
In this appendix, we refer to this estimator as the `point correction' method.
It is easy to implement and has the advantage that the error bars of the different bins
are uncorrelated, which renders subsequent statistical analyses easier.

To compute the effective volume, $V_{eff}$, the survey sky coverage first needs to be
recast as a function of the cluster luminosity and redshift, namely
\begin{equation}
   \Omega_S\left(L,T,z\right) = \Omega_S\left(CR_\infty\left[L,T,z\right],x_{500}\ClRad\left[\hat{M}(T,z)\right]\right)
\label{eq:OmegaLTz}
\end{equation}
and
\begin{equation}
   \Omega_S\left(L,z\right) = \frac{\int_T \frac{\mathrm{d}n(L,T,z)}{\mathrm{d}L\,\mathrm{d}T}  \Omega_S\left(L,T,z\right) \mathrm{d}T }{\int_T \frac{\mathrm{d}n(L,T,z)}{\mathrm{d}L\,\mathrm{d}T} \mathrm{d}T},
\end{equation}
where $\hat{M}(T,z)$ is the average mass-temperature scaling relation and  $\mathrm{d}n/\mathrm{d}L\mathrm{d}T$ is the
volumic density of clusters defined in Eq.~(\ref{eq:dnOverdLdT}). This
accounts for the distribution in $\ClMass$ and $T$ at given $\Lxxl$ and $z$ in the
assumed cosmological model to compute an average sky coverage. We note that the mass
function in the chosen cosmology, as well as cluster scaling relations, implicitely enters
the calculation of the sky coverage. This is necessary since the detection efficiency
depends on the mass (through its link to the core radius) and temperature (to compute the
spectral K-correction).

The effective survey volume for given $\Lxxl$ and $z$ is then readily obtained as:
\begin{equation}
    V_{eff}\left(L\right) = \int_z \frac{\mathrm{d}^2V}{\mathrm{d}\Omega\,\mathrm{d}z} \Omega_S(L,z) dz.
\end{equation}

For sparsely populated samples, where the density of points is too low to properly weight
the effective volume between the different luminosities within a bin, the correction for
the sky coverage can become noisy.
In this case one can directly compute the average sky coverage within the luminosity bin
from the same cosmological model that was used to determine the luminosity dependent sky
coverage,
\begin{equation}
   \Omega_{S,i}\left(z\right) = \frac{ \int_{L_{min}^i}^{L_{max}^i} \int_T  \frac{\mathrm{d}n(L,T,z)}{\mathrm{d}L\,\mathrm{d}T}  \Omega_S\left(L,T,z\right) \mathrm{d}T \mathrm{d}L}{\int_{L_{min}^i}^{L_{max}^i} \int_T \frac{\mathrm{d}n(L,T,z)}{\mathrm{d}L\,\mathrm{d}T} \mathrm{d}T \mathrm{d}L} ,
   \label{eq:BinnedSkyCov}
\end{equation}
where $L_{min}^i$ and $L_{max}^i$ are the boundaries of the luminosity bin $i$.

The luminosity function then reads
\begin{equation}
 \frac{\mathrm{d}n}{\mathrm{d}L}(<L_i>) = \frac{1}{\Delta L_i} \frac{N_i}{V_i},
   \label{eq:BinnedCorr}
\end{equation}
where $N_i$ is the total number of clusters in bin $i$ and $V_i$ the effective volume of the
bin obtained from the sky coverage $\Omega_{S,i}$. We term this estimator the `binned
correction' method. The reduced noise in the correction of the sky coverage, compared to the
point estimator, comes at the price of being slightly more model-dependent. However, the
estimator only uses the shape of the luminosity function in the assumed model as a weight
function, so it does not impose the measured luminosity function to match that of the
model. In addition, since this estimator only becomes useful when the shot noise is high,
the error on the underlying model would have to be very large to significantly bias the
final result. Variants of this estimator are fairly common in the analysis of the luminosity
function of X-ray AGNs (although expressed solely in terms of the AGN luminosity and redshift).
For instance, \citet{Page2000} assumed $\mathrm{d}N/\mathrm{d}L\mathrm{d}z$ to be constant
using narrow luminosity bins and showed the resulting estimator to be more robust
than the simple point correction. For larger bins, \citet{Miyaji2001} used precisely the
estimator of Eq.~(\ref{eq:BinnedCorr}), having first derived
$\mathrm{d}N/\mathrm{d}L\mathrm{d}z$ from a binned parametric fit to their data.
Such recipes are meant to reduce the model dependence of the estimator. We did not try
to implement them here as the modelled luminosity function already enters the determination
of $\Omega_S\left(L,z\right)$.

When few clusters per bin are available, the shot noise can remain high, even with the
binned correction, unless the luminosity function is smoothed over very wide luminosity bins.
A possible alternative consists in directly measuring the cumulative luminosity function
with a slight modification of Eq.~(\ref{eq:BinnedCorr}):
\begin{equation}
    n(>L) = \frac{N_{>L}}{V_{>L}},
   \label{eq:CumulFunc}
\end{equation}
where $N_{>L}$ is the number of observed clusters above luminoity $L$, and $V_{>L}$ the bin
averaged effective volume obtained by integrating the sky coverage of
Eq.~(\ref{eq:BinnedSkyCov}) to infinite luminosity.
The differential luminosity function follows from differentiating Eq.~(\ref{eq:CumulFunc}).
This is the `cumulative correction' method that we use for the \ChundredShort\ analysis.
It is affected much less by noise than the point and binned estimates, because the derivative
uses the cluster density in several bins, implying that wider bins are effectively considered.
However, this introduces correlations between the values of the luminosity function in
adjacent bins. This would mostly be a problem while using the luminosity function directly to
constrain physical cluster models or cosmological parameters.
However, this is not desirable since the luminosity function estimate already relies on
the assumed cosmology and cluster model to estimate the effective volume.
For such analyses, one should instead adopt a likelihood formalism similar to the one
presented in Appendix~\ref{append:Like} and used in Sect.~\ref{sec:LumConst}.

Naively, the sensitivity to the shape of the luminosity function in the assumed cosmological
model should become even larger since the effective volume is averaged over a very wide
luminosity bin.
This would be true for the cumulative function of Eq.~(\ref{eq:CumulFunc}), but the effect
is reduced in the case of the differential function because the derivative only results
from finite differences of the cumulative function. This is confirmed by Monte Carlo
computations of the luminosity function errors, which includes the uncertainty on the
$\Lxxl-\Txxl$ scaling relation (see Sect.~\ref{sec:LumFunc}) and yet produces smaller
error bars for the cumulative correction method compared to others.

\begin{figure}
  \begin{center}
    \includegraphics[width=9cm]{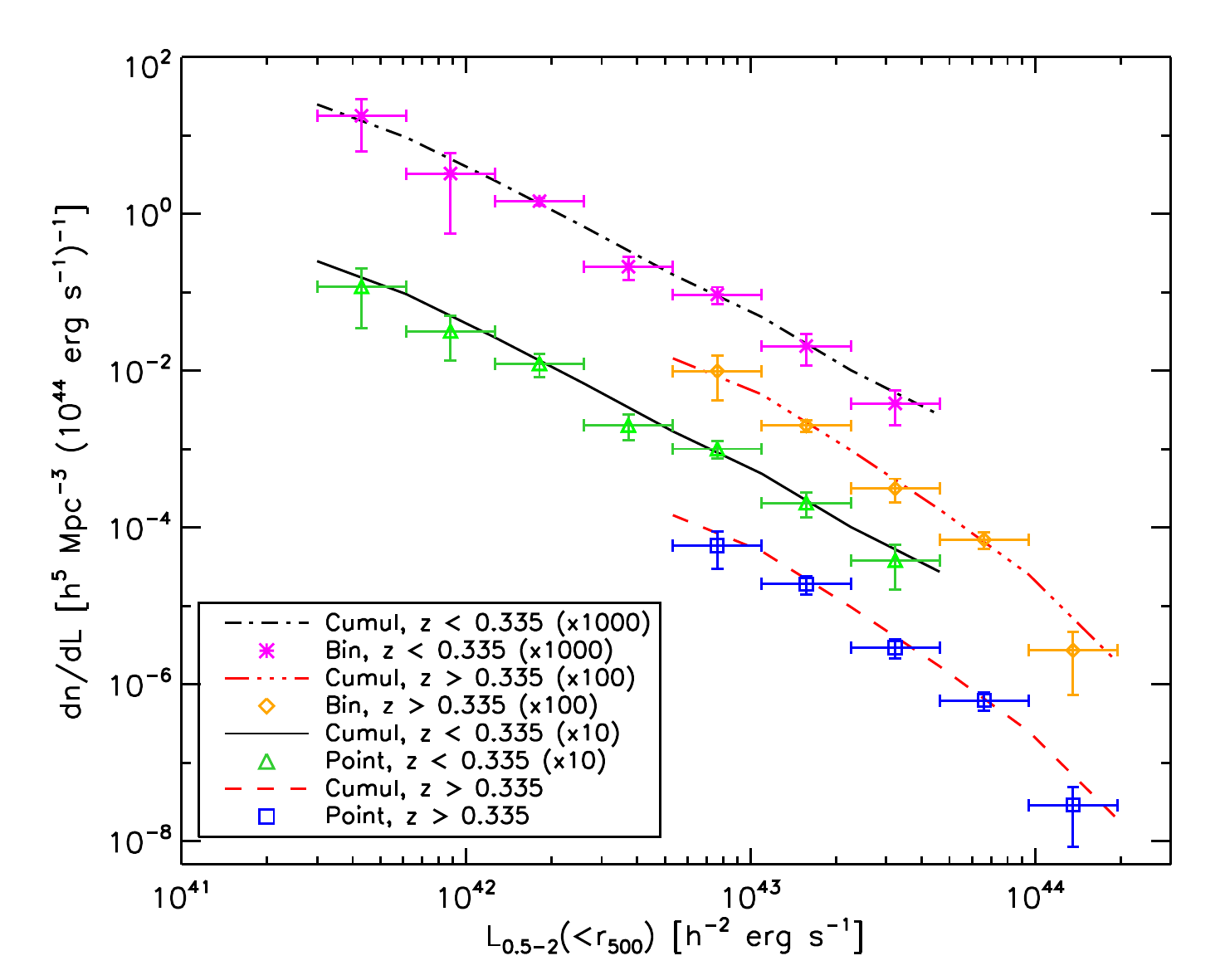}\vspace{-0.1cm}
   \caption{Comparison of the luminosity function obtained using the three
     estimators presented in Appendix~\ref{append:LFcomp}, in the fiducial WMAP9 cosmology,
     and scaling relation model.
     Multiplicative factors were added to separate the different estimates and to
     clarify the plot.\label{fig:LfuncMethods}}
  \end{center}
\end{figure}

A comparison of the \ChundredShort\ luminosity function obtained with these three estimators
is provided in Fig.~\ref{fig:LfuncMethods}, using large luminosity bins to reduce the shot
noise in the point and binned corrrection method. All methods provide very similar values,
justifying the use of the cumulative method which results in tighter constraints on the
differential luminosity function.

\section{Likelihood model}
\label{append:Like}
To infer model parameters ($\mathcal{P}$) from the properties of the \ChundredShort\ clusters,
we make use of a very generic unbinned likelihood model, in which we separate the information
on the number of detected clusters, $N_{det}$, from their distribution in the space of observables $(O)$,
\begin{equation}
	\mathcal{L}(\mathcal{P}|N_{det},\tilde{O}) = P(N_{det}|\mathcal{P}) \prod_{i=1}^{N_{det}} \left[ \int_O P(\tilde{O}|O)  P(O|\mathcal{P}) \mathrm{d}O \right],
    \label{eq:likeForm}
\end{equation}
where $\tilde{O}$ are noisy measurements of the cluster observables (e.g. known redshifts,
luminosities, and masses); $P(\tilde{O}|O)$ therefore describes the measurement process,
while $P(O|\mathcal{P})$ is the expected distribution of the true observables $O$ in the
model described by parameters $\mathcal{P}$.

To proceed further, let us define the cluster density in the observable space, their
redshift distribution, and the number of expected detections in model $\mathcal{P}$ as
\begin{equation}
    \frac{\mathrm{d}N\left(L,T,z\right)}{\mathrm{d}z\,\mathrm{d}L\,\mathrm{d}T} = \frac{\mathrm{d}n\left(L,T,z\right)}{ \mathrm{d}L\,\mathrm{d}T} \frac{\mathrm{d}^2V}{\mathrm{d}\Omega\,\mathrm{d}z} \Omega_S(L,T,z),
\end{equation}
\begin{equation}
    \frac{\mathrm{d}N}{\mathrm{d}z} =  \int_L\int_T \frac{\mathrm{d}N\left(L,T,z\right)}{ \mathrm{d}L\,\mathrm{d}T}\,\mathrm{d}L\,\mathrm{d}T,
\end{equation}
\begin{equation}
    {\langle}N_{det}{\rangle}\, = \int_z \frac{\mathrm{d}N}{\mathrm{d}z}\,\mathrm{d}z,
\end{equation}
where  $\mathrm{d}n/\mathrm{d}L\mathrm{d}T$ is the volumic cluster density defined by Eq.~(\ref{eq:dnOverdLdT}) and
$\Omega_S\left(L,T,z\right)$ is the sky coverage of the survey as expressed in Eq.~(\ref{eq:OmegaLTz}).

If only cluster temperatures, luminosities, and redshifts are considered, the predicted distribution
of clusters in true observable space, present in Eq.~(\ref{eq:likeForm}), becomes
\begin{equation}
    P(O|\mathcal{P}) = \frac{1}{{\langle}N_{det}{\rangle}} \frac{\mathrm{d}N\left(L,T,z\right)}{\mathrm{d}z\,\mathrm{d}L\,\mathrm{d}T}.
    \label{eq:PofO}
\end{equation}
If, in addition, one assumes a Poisson distribution of mean ${\langle}N_{det}{\rangle}$ for $P(N_{det}|\mathcal{P})$,
the likelihood becomes formally the same as the one defined by \cite{Mantz2010a}.

However, this new derivation makes explicit the contribution of the model normalisation, which
allows for simple extensions to the basis model.
For instance, for the fits to the cluster luminosity distribution performed in sections~\ref{sec:LumConst}
and \ref{sec:CosLeverage}, we modified $P(N_{det}|\mathcal{P})$ to account for cosmic
variance as
\begin{equation}
    P(N_{det}|\mathcal{P}) = \int Poi(N_{det}|N_{loc}) \mathcal{LN}\left[N_{loc}|{\langle}N_{det}{\rangle},\sigma_v\right]\,\mathrm{d}N_{loc}
    \label{eq:CosVar}
\end{equation}
where the local density $N_{loc}$ is generated from a log-normal distribution $\mathcal{LN}$ of
mean ${\langle}N_{det}{\rangle}$ and sample variance $\sigma_v^2$, and $N_{loc}$ is then subjected to additional
shot noise through the Poisson law $Poi$. Such log-normal models for the distribution of matter
in the universe have already been used in several studies, starting with \cite{Coles1991}.
We use the analytical model of \cite{Valageas2011} to compute the variance $\sigma_v^2$.
In practice, we combined equations~\ref{eq:PofO} and \ref{eq:CosVar} with $P(\tilde{O}|O)$
arising from log-normal errors on the luminosity, unconstrained temperatures, and perfect redshift
determinations for all the results presented in this article.

A second application of the present likelihood formalism is simply to ignore the information contained
in the model normalisation by setting $P(N_{det}|\mathcal{P})=1$, as assumed in \citetalias{Giles2016}
for the analysis of the scaling relations between luminosity and temperature. In this case, since
the goal was to constrain the scaling relation without including any information from the
spatial density of clusters, the redshift information was also ignored to consider only the
raw distribution of $\Lxxl$ and $\Txxl$ at given z. This implied
\begin{equation}
    P(O|\mathcal{P}) = P(L,T|\mathcal{P},z) = \frac{1}{\mathrm{d}N/\mathrm{d}z} \frac{dN\left(L,T,z\right)}{dzdLdT}.
\end{equation}
\onecolumn

\section{Additional tables}
\vspace{-0.2cm}
\begin{table}[h!]
\begin{center}
\caption{The Bright XXL cluster catalogue.\label{BXCTab}}
{\small \begin{tabular}{C{2.1cm}ccccC{1.0cm}C{1.8cm}cC{1.8cm}cc}
 \hline\hline
 Name &    RA  &   Dec   & z & $N_{gal}$ & $C_{60}$ & $F_{60}$ & $\ClRadMT$ & $\Lxxl$ & $\ClMassMT$  & Ref.\rule{0pt}{2.6ex}\\[1pt]
\hline
     \object{XLSSC 094}  &    30.648 &   -6.732 &         0.886 &         3    &   199 &   4.82\,$\pm$\,0.44 &   0.74 &    25.9\,$\pm$\,2.2 &                  $\star\star$ &  {\footnotesize      ...} \\
     \object{XLSSC 096}  &    30.973 &   -5.027 &         0.520 &         6    &   161 &   3.64\,$\pm$\,0.39 &   1.00 &   5.80\,$\pm$\,0.62 &                 48\,$\pm$\,31 &  {\footnotesize      ...} \\
     \object{XLSSC 102}  &    31.322 &   -4.652 &         0.969 &         3    &   199 &   4.20\,$\pm$\,0.36 &   0.57 &    16.1\,$\pm$\,1.7 &                 19\,$\pm$\,11 &  {\footnotesize      ...} \\
     \object{XLSSC 106}  &    31.351 &   -5.732 &         0.300 &        14    &   681 &   9.13\,$\pm$\,0.39 &   0.86 &   4.44\,$\pm$\,0.21 &                 24\,$\pm$\,11 &  {\footnotesize      ...} \\
     \object{XLSSC 107}  &    31.354 &   -7.594 &         0.436 &         3    &   263 &   5.58\,$\pm$\,0.41 &   0.71 &   4.89\,$\pm$\,0.41 &              15.9\,$\pm$\,7.6 &  {\footnotesize      ...} \\
     \object{XLSSC 100}  &    31.549 &   -6.193 &         0.915 &         6    &   124 &   3.55\,$\pm$\,0.52 &   0.69 &    14.1\,$\pm$\,3.2 &                 26\,$\pm$\,18 &  {\footnotesize      ...} \\
     \object{XLSSC 093}  &    31.699 &   -6.948 &         0.429 &         6    &   418 &   7.23\,$\pm$\,0.41 &   0.81 &   6.47\,$\pm$\,0.42 &                 23\,$\pm$\,11 &  {\footnotesize      ...} \\
     \object{XLSSC 108}  &    31.832 &   -4.827 &         0.254 &         4    &   451 &   6.16\,$\pm$\,0.34 &   0.70 &   1.90\,$\pm$\,0.13 &              12.7\,$\pm$\,5.6 &  {\footnotesize      ...} \\
     \object{XLSSC 095}  &    31.962 &   -5.206 &         0.138 &        12    &   141 &   3.09\,$\pm$\,0.32 &   0.45 &   0.17\,$\pm$\,0.03 &               2.9\,$\pm$\,1.3 &  {\footnotesize      ...} \\
     \object{XLSSC 092}  &    32.071 &   -7.276 &         0.432 &         3    &   166 &   3.14\,$\pm$\,0.33 &   0.77 &   2.81\,$\pm$\,0.31 &                 20\,$\pm$\,11 &  {\footnotesize      ...} \\
     \object{XLSSC 101}  &    32.193 &   -4.436 &         0.756 &         9    &   332 &   5.49\,$\pm$\,0.36 &   0.79 &    16.5\,$\pm$\,1.3 &                 31\,$\pm$\,16 &  {\footnotesize      ...} \\
     \object{XLSSC 109}  &    32.296 &   -6.346 &         0.491 &         2    &   146 &   5.11\,$\pm$\,0.61 &   0.79 &     6.3\,$\pm$\,1.0 &                 23\,$\pm$\,15 &  {\footnotesize      ...} \\
     \object{XLSSC 112}  &    32.514 &   -5.462 &         0.139 &        14    &   178 &   5.89\,$\pm$\,0.62 &   0.65 &   0.61\,$\pm$\,0.08 &               9.0\,$\pm$\,4.1 &  {\footnotesize      ...} \\
     \object{XLSSC 083}  &    32.735 &   -6.200 &         0.430 &         5    &   293 &   4.68\,$\pm$\,0.32 &   0.94 &   4.67\,$\pm$\,0.38 &                 37\,$\pm$\,20 &  {\footnotesize      ...} \\
     \object{XLSSC 084}  &    32.767 &   -6.211 &         0.430 &         6    &   173 &   3.28\,$\pm$\,0.32 &   0.94 &   2.04\,$\pm$\,0.31 &                 36\,$\pm$\,25 &  {\footnotesize      ...} \\
     \object{XLSSC 085}  &    32.870 &   -6.196 &         0.428 &         3    &   206 &   4.09\,$\pm$\,0.36 &   0.98 &   3.88\,$\pm$\,0.40 &    41\,$\pm$\,27$^{\ast\ast}$ &  {\footnotesize      ...} \\
     \object{XLSSC 111}  &    33.111 &   -5.627 &         0.299 &        12    &   707 &  14.00\,$\pm$\,0.57 &   1.02 &   6.65\,$\pm$\,0.32 &                 40\,$\pm$\,18 &  {\footnotesize      ...} \\
     \object{XLSSC 098}  &    33.115 &   -6.076 &         0.297 &         5    &   133 &   3.48\,$\pm$\,0.42 &   0.80 &   1.73\,$\pm$\,0.21 &                 20\,$\pm$\,12 &  {\footnotesize      ...} \\
     \object{XLSSC 099}  &    33.220 &   -6.202 &         0.391 &         1    &   118 &   3.23\,$\pm$\,0.39 &   1.03 &   2.26\,$\pm$\,0.41 &                 46\,$\pm$\,40 &  {\footnotesize      ...} \\
     \object{XLSSC 097}  &    33.342 &   -6.098 &  {\it  0.76 } &   {\it 0 }   &   124 &   4.22\,$\pm$\,0.48 &   0.79 &    13.4\,$\pm$\,1.7 &                 32\,$\pm$\,19 &  {\footnotesize      ...} \\
     \object{XLSSC 110}  &    33.537 &   -5.585 &         0.445 &         4    &   228 &   3.19\,$\pm$\,0.28 &   0.53 &   1.63\,$\pm$\,0.25 &               6.5\,$\pm$\,2.8 &  {\footnotesize       14} \\
     \object{XLSSC 060}  &    33.668 &   -4.553 &         0.139 &        26    &  3553 &  23.03\,$\pm$\,0.43 &   1.14 &   6.31\,$\pm$\,0.08 &                 47\,$\pm$\,20 &  {\footnotesize    7, 11} \\
     \object{XLSSC 072}  &    33.850 &   -3.726 &         1.002 &         5    &   231 &   4.06\,$\pm$\,0.36 &   0.61 &    14.9\,$\pm$\,1.8 &                 19\,$\pm$\,11 &  {\footnotesize    6, 13} \\
     \object{XLSSC 056}  &    33.871 &   -4.682 &         0.348 &         6    &   532 &   7.55\,$\pm$\,0.38 &   0.82 &   4.16\,$\pm$\,0.25 &                 22\,$\pm$\,11 &  {\footnotesize        7} \\
     \object{XLSSC 057}  &    34.051 &   -4.242 &         0.153 &        18    &   463 &   8.01\,$\pm$\,0.43 &   0.73 &   1.18\,$\pm$\,0.07 &              12.9\,$\pm$\,5.8 &  {\footnotesize    7, 11} \\
     \object{XLSSC 023}  &    35.188 &   -3.433 &         0.328 &         4    &   272 &   3.39\,$\pm$\,0.28 &   0.66 &   1.63\,$\pm$\,0.18 &              11.0\,$\pm$\,5.2 &  {\footnotesize        4} \\
     \object{XLSSC 006}  &    35.439 &   -3.772 &         0.429 &        16    &   903 &  18.64\,$\pm$\,0.71 &   0.98 &  17.42\,$\pm$\,0.83 &                 41\,$\pm$\,18 &  {\footnotesize        2} \\
     \object{XLSSC 061}  &    35.485 &   -5.758 &         0.259 &        10    &   167 &   3.66\,$\pm$\,0.40 &   0.68 &   1.09\,$\pm$\,0.14 &              11.3\,$\pm$\,5.8 &  {\footnotesize        7} \\
     \object{XLSSC 036}  &    35.527 &   -3.054 &         0.492 &         3    &   464 &   9.35\,$\pm$\,0.49 &   0.80 &  11.14\,$\pm$\,0.72 &                 24\,$\pm$\,11 &  {\footnotesize        4} \\
     \object{XLSSC 029}  &    36.017 &   -4.225 &         1.050 &         6    &   323 &   3.22\,$\pm$\,0.23 &   0.63 &    19.5\,$\pm$\,1.7 &                 22\,$\pm$\,12 &  {\footnotesize        3} \\
     \object{XLSSC 062}  &    36.061 &   -2.721 &         0.059 &         4    &   103 &   5.04\,$\pm$\,0.69 &   0.42 &   0.11\,$\pm$\,0.02 &  2.3\,$\pm$\,1.0$^{\ast\ast}$ &  {\footnotesize        7} \\
     \object{XLSSC 001}  &    36.238 &   -3.817 &         0.614 &        17    &   522 &   6.43\,$\pm$\,0.33 &   0.78 &  10.10\,$\pm$\,0.72 &                 25\,$\pm$\,12 &  {\footnotesize        1} \\
     \object{XLSSC 054}  &    36.319 &   -5.887 &         0.054 &        28    &   421 &   7.76\,$\pm$\,0.70 &   0.72 &   0.28\,$\pm$\,0.02 &              11.3\,$\pm$\,4.9 &  {\footnotesize        7} \\
     \object{XLSSC 025}  &    36.353 &   -4.680 &         0.265 &        13    &   680 &   6.55\,$\pm$\,0.29 &   0.75 &   2.21\,$\pm$\,0.11 &              15.5\,$\pm$\,6.8 &  {\footnotesize        3} \\
     \object{XLSSC 041}  &    36.378 &   -4.239 &         0.142 &        16    &   454 &  12.54\,$\pm$\,0.64 &   0.67 &   1.19\,$\pm$\,0.07 &               9.7\,$\pm$\,4.3 &  {\footnotesize        4} \\
     \object{XLSSC 050}  &    36.421 &   -3.189 &         0.140 &        16    &   782 &   9.79\,$\pm$\,0.39 &   0.90 &   2.78\,$\pm$\,0.07 &              23.3\,$\pm$\,9.9 &  {\footnotesize        4} \\
     \object{XLSSC 055}  &    36.454 &   -5.896 &         0.232 &        14    &   464 &   7.81\,$\pm$\,0.43 &   0.84 &   2.61\,$\pm$\,0.15 &              21.2\,$\pm$\,9.6 &  {\footnotesize        7} \\
     \object{XLSSC 011}  &    36.540 &   -4.969 &         0.054 &        42    &   349 &   3.23\,$\pm$\,0.22 &   0.83 &   0.15\,$\pm$\,0.01 &              17.1\,$\pm$\,8.7 &  {\footnotesize        3} \\
     \object{XLSSC 052}  &    36.567 &   -2.666 &         0.056 &         3    &   599 &   9.03\,$\pm$\,0.42 &   0.39 &   0.09\,$\pm$\,0.01 &             1.70\,$\pm$\,0.70 &  {\footnotesize        4} \\
     \object{XLSSC 010}  &    36.843 &   -3.362 &         0.330 &         6    &   308 &   4.60\,$\pm$\,0.33 &   0.75 &   2.58\,$\pm$\,0.21 &              16.6\,$\pm$\,8.0 &  {\footnotesize        2} \\
     \object{XLSSC 103}  &    36.886 &   -5.961 &         0.233 &         8    &   164 &   4.27\,$\pm$\,0.41 &   0.91 &   1.30\,$\pm$\,0.14 &                 27\,$\pm$\,17 &  {\footnotesize      ...} \\
     \object{XLSSC 003}  &    36.909 &   -3.300 &         0.836 &         6    &   163 &   3.40\,$\pm$\,0.35 &   0.64 &    12.3\,$\pm$\,1.5 &                 19\,$\pm$\,11 &  {\footnotesize        1} \\
     \object{XLSSC 022}  &    36.917 &   -4.858 &         0.293 &        18    &  1295 &   7.44\,$\pm$\,0.23 &   0.67 &   3.06\,$\pm$\,0.11 &              11.4\,$\pm$\,4.8 &  {\footnotesize        3} \\
     \object{XLSSC 027}  &    37.012 &   -4.851 &         0.295 &        10    &   376 &   3.70\,$\pm$\,0.23 &   0.77 &   1.48\,$\pm$\,0.11 &              17.1\,$\pm$\,8.1 &  {\footnotesize        4} \\
     \object{XLSSC 090}  &    37.121 &   -4.857 &         0.141 &        11    &   217 &   4.53\,$\pm$\,0.37 &   0.51 &   0.43\,$\pm$\,0.05 &               4.2\,$\pm$\,1.8 &  {\footnotesize      ...} \\
     \object{XLSSC 089}  &    37.127 &   -4.733 &         0.609 &         7    &   101 &   3.18\,$\pm$\,0.40 &   0.77 &   6.44\,$\pm$\,0.87 &                 24\,$\pm$\,18 &  {\footnotesize      ...} \\
     \object{XLSSC 104}  &    37.324 &   -5.895 &         0.294 &         6    &   266 &   4.57\,$\pm$\,0.34 &   1.04 &   1.36\,$\pm$\,0.15 &                  $\star\star$ &  {\footnotesize      ...} \\
     \object{XLSSC 088}  &    37.611 &   -4.581 &         0.295 &         7    &   204 &   3.53\,$\pm$\,0.31 &   0.73 &   1.57\,$\pm$\,0.15 &              14.5\,$\pm$\,7.7 &  {\footnotesize      ...} \\
     \object{XLSSC 087}  &    37.720 &   -4.348 &         0.141 &         5    &   530 &  12.21\,$\pm$\,0.59 &   0.62 &   0.92\,$\pm$\,0.09 &               7.7\,$\pm$\,3.2 &  {\footnotesize      ...} \\
\hline
\end{tabular} }
\tablefoot{
Columns description:
Name: Official IAU designation of the cluster.
$z$: Cluster redshift. Photometric redshifts are only reported with two digit accuracy.
RA/Dec: J2000 coordinates in degrees.
$N_{gal}$: number of spectroscopic redshifts for cluster members - 0 means that only photometric redshits are available
$C_{60}$: Net XMM counts in the [0.5-2] keV band within the 60$^{\prime\prime}$ aperture used for the sample selection.
$F_{60}$: Flux in the [0.5-2] keV band within the 60$^{\prime\prime}$ aperture, in units of $10^{-14}\,$\flux.
$\ClRadMT$: Overdensity radius with respect to the critical density in Mpc, obtained from the $\ClMassWL-\Txxl$
scaling relation of \citetalias{Lieu2016} (see Table~\ref{tab:ScalPar}) and the temperatures measured in
\citetalias{Giles2016}.
$\Lxxl$: Rest-frame  [0.5-2] keV luminosity in $\ClRadMT$ in units of $10^{43}\,$\lum.
$\ClMassMT$: mass within $\ClRadMT$ in units of $10^{13}\,M_\odot$, obtained from the
$\ClMassWL-\Txxl$ scaling relation of \citetalias{Lieu2016} (see Table~\ref{tab:ScalPar})
and the temperatures measured in \citetalias{Giles2016}. A $^{\ast\ast}$ sign after the
mass indicates a possible AGN contamination and thus a likely overestimated mass.
We do not provide a mass for clusters that show firm evidence of AGN contaminations (although
$r_{500,MT}$ was still derived from $T_{300kpc}$ using the scaling relation of \citetalias{Lieu2016}).
These clusters are indicated by a $\star\star$ sign.
Ref.: reference to first X-ray detection (as a cluster, when possible).
A reference to the first optical/IR detection is also provided when it is prior to the X-ray detection.
\vspace{-0.05cm}}
\tablebib{(1) \cite{Valtchanov2004}; (2) \cite{Willis2005}; (3) \cite{Pierre2006}; (4) \cite{Pacaud2007};
(5) \cite{Adami2011}; (6) \cite{Willis2013}; (7) \cite{Clerc2014}; (8) \cite{Suhada2012};
(9) \cite{Boehringer2004}; (10) \cite{Voges2000}; (11) \cite{Abell1989}; (12) \cite{Menanteau2010};
(13) \cite{Muzzin2012}; (14) \cite{Limousin2009}; (15) \cite{Jones2009}.}

\end{center}
\end{table}

\addtocounter{table}{-1}
\begin{table}[h!]
\begin{center}
\vspace{-0.1cm}\caption{continued.}
{\small \begin{tabular}{C{2.1cm}ccccC{1.0cm}C{1.8cm}cC{1.8cm}cc}
 \hline\hline
 Name &    RA  &   Dec   & z & $N_{gal}$ & $C_{60}$ & $F_{60}$ & $\ClRadMT$ & $\Lxxl$ & $\ClMassMT$  & Ref.\rule{0pt}{2.6ex}\\[1pt]
\hline
     \object{XLSSC 091}  &    37.926 &   -4.881 &         0.186 &        41    &  3114 &  38.65\,$\pm$\,0.72 &   1.15 &  13.12\,$\pm$\,0.19 &                 51\,$\pm$\,22 &  {\footnotesize       11} \\
     \object{XLSSC 105}  &    38.411 &   -5.506 &         0.429 &         5    &   543 &  12.87\,$\pm$\,0.64 &   1.02 &  12.39\,$\pm$\,0.89 &                 47\,$\pm$\,24 &  {\footnotesize      ...} \\
     \object{XLSSC 502}  &   348.442 &  -53.438 &         0.141 &         4    &   442 &   6.52\,$\pm$\,0.37 &   0.53 &   0.63\,$\pm$\,0.05 &               4.9\,$\pm$\,2.0 &  {\footnotesize      ...} \\
     \object{XLSSC 530}  &   348.833 &  -54.345 &  {\it  0.18 } &   {\it 0 }   &   332 &   4.47\,$\pm$\,0.32 &   0.69 &   0.75\,$\pm$\,0.06 &              10.9\,$\pm$\,4.9 &  {\footnotesize      ...} \\
     \object{XLSSC 501}  &   348.873 &  -53.063 &         0.333 &         7    &   265 &   6.01\,$\pm$\,0.49 &   0.77 &   2.44\,$\pm$\,0.29 &              17.8\,$\pm$\,9.1 &  {\footnotesize      ...} \\
     \object{XLSSC 513}  &   349.221 &  -54.902 &         0.378 &         5    &   409 &  11.19\,$\pm$\,0.65 &   0.94 &   6.50\,$\pm$\,0.44 &                 34\,$\pm$\,17 &  {\footnotesize    8, 12} \\
     \object{XLSSC 525}  &   349.339 &  -53.962 &         0.379 &         2    &   904 &  10.53\,$\pm$\,0.39 &   0.83 &   6.68\,$\pm$\,0.33 &                 24\,$\pm$\,10 &  {\footnotesize      ...} \\
     \object{XLSSC 527}  &   349.570 &  -55.984 &         0.076 &         3    &   100 &   4.80\,$\pm$\,0.62 &   0.93 &   0.20\,$\pm$\,0.05 &                 24\,$\pm$\,27 &  {\footnotesize      ...} \\
     \object{XLSSC 528}  &   349.682 &  -56.204 &         0.302 &         5    &   427 &   4.28\,$\pm$\,0.25 &   0.84 &   2.07\,$\pm$\,0.16 &                 23\,$\pm$\,12 &  {\footnotesize      ...} \\
     \object{XLSSC 529}  &   349.699 &  -56.287 &         0.547 &         6    &   398 &   5.02\,$\pm$\,0.30 &   0.77 &   6.91\,$\pm$\,0.58 &                 23\,$\pm$\,11 &  {\footnotesize       12} \\
     \object{XLSSC 526}  &   349.802 &  -54.087 &         0.273 &         8    &    55 &  15.33\,$\pm$\,3.06 &   0.79 &   5.27\,$\pm$\,0.27 &              18.5\,$\pm$\,8.3 &  {\footnotesize   10, 11} \\
     \object{XLSSC 544}  &   349.816 &  -53.534 &         0.095 &        11    &   927 &  10.52\,$\pm$\,0.40 &   0.79 &   0.77\,$\pm$\,0.03 &              15.2\,$\pm$\,6.6 &  {\footnotesize      ...} \\
     \object{XLSSC 518}  &   349.822 &  -55.325 &         0.177 &         4    &   453 &   6.34\,$\pm$\,0.39 &   0.53 &   0.58\,$\pm$\,0.05 &               5.1\,$\pm$\,2.1 &  {\footnotesize        8} \\
     \object{XLSSC 531}  &   349.876 &  -56.649 &         0.391 &         9    &   190 &   3.65\,$\pm$\,0.36 &   0.97 &   2.73\,$\pm$\,0.38 &                 38\,$\pm$\,30 &  {\footnotesize       12} \\
     \object{XLSSC 534}  &   350.105 &  -53.359 &         0.853 &         4    &   154 &   3.53\,$\pm$\,0.40 &   0.73 &    16.1\,$\pm$\,2.4 &                 27\,$\pm$\,18 &  {\footnotesize      ...} \\
     \object{XLSSC 517}  &   350.449 &  -55.971 &         0.699 &         3    &   155 &   4.07\,$\pm$\,0.46 &   0.70 &     7.3\,$\pm$\,1.1 &                 20\,$\pm$\,12 &  {\footnotesize      ...} \\
     \object{XLSSC 523}  &   350.503 &  -54.750 &         0.343 &         7    &   223 &   5.03\,$\pm$\,0.41 &   0.78 &   2.90\,$\pm$\,0.23 &              18.8\,$\pm$\,9.5 &  {\footnotesize    8, 11} \\
     \object{XLSSC 503}  &   350.646 &  -52.747 &         0.336 &         3    &   230 &   4.92\,$\pm$\,0.43 &   0.64 &   2.47\,$\pm$\,0.24 &              10.5\,$\pm$\,4.9 &  {\footnotesize      ...} \\
     \object{XLSSC 545}  &   350.692 &  -53.388 &         0.353 &         7    &   100 &   3.09\,$\pm$\,0.46 &   0.67 &   1.41\,$\pm$\,0.41 &                 12\,$\pm$\,11 &  {\footnotesize      ...} \\
     \object{XLSSC 514}  &   351.396 &  -54.722 &         0.169 &         9    &   229 &   3.99\,$\pm$\,0.37 &   0.58 &   0.47\,$\pm$\,0.07 &               6.6\,$\pm$\,3.0 &  {\footnotesize    8, 12} \\
     \object{XLSSC 515}  &   351.416 &  -54.741 &         0.101 &        11    &   350 &   5.75\,$\pm$\,0.40 &   0.54 &   0.37\,$\pm$\,0.04 &               4.9\,$\pm$\,2.1 &  {\footnotesize    8, 12} \\
     \object{XLSSC 547}  &   351.427 &  -53.277 &         0.371 &         6    &   164 &   5.32\,$\pm$\,0.52 &   0.92 &   4.09\,$\pm$\,0.40 &                 32\,$\pm$\,18 &  {\footnotesize      ...} \\
     \object{XLSSC 535}  &   351.554 &  -53.317 &         0.172 &        11    &   607 &  11.61\,$\pm$\,0.54 &   0.76 &   2.41\,$\pm$\,0.13 &              14.4\,$\pm$\,6.5 &  {\footnotesize       12} \\
     \object{XLSSC 536}  &   351.557 &  -53.374 &         0.170 &         7    &   282 &   6.28\,$\pm$\,0.47 &   0.66 &   0.47\,$\pm$\,0.08 &               9.5\,$\pm$\,4.5 &  {\footnotesize      ...} \\
     \object{XLSSC 522}  &   351.638 &  -55.022 &         0.395 &         3    &   819 &   3.63\,$\pm$\,0.17 &   0.71 &   2.71\,$\pm$\,0.19 &              15.2\,$\pm$\,7.0 &  {\footnotesize    8, 12} \\
     \object{XLSSC 533}  &   351.712 &  -52.694 &         0.107 &         6    &  1315 &  18.10\,$\pm$\,0.55 &   0.79 &   2.21\,$\pm$\,0.05 &              15.4\,$\pm$\,6.5 &  {\footnotesize        9} \\
     \object{XLSSC 504}  &   351.930 &  -52.425 &         0.243 &         1    &   150 &   5.13\,$\pm$\,0.67 &   1.95 &   1.35\,$\pm$\,0.48 &                  $\star\star$ &  {\footnotesize      ...} \\
     \object{XLSSC 521}  &   352.179 &  -55.567 &         0.807 &         1    &   281 &   6.66\,$\pm$\,0.50 &   0.78 &    17.3\,$\pm$\,1.9 &                 31\,$\pm$\,18 &  {\footnotesize        8} \\
     \object{XLSSC 505}  &   352.250 &  -52.238 &         0.055 &         3    &   545 &  19.22\,$\pm$\,0.95 &   0.66 &   0.47\,$\pm$\,0.03 &               8.6\,$\pm$\,3.7 &  {\footnotesize      ...} \\
     \object{XLSSC 506}  &   352.315 &  -52.497 &         0.717 &         6    &    95 &   3.05\,$\pm$\,0.54 &   0.80 &     8.5\,$\pm$\,1.7 &                 31\,$\pm$\,24 &  {\footnotesize       12} \\
     \object{XLSSC 546}  &   352.416 &  -53.249 &         0.792 &         2    &   256 &   3.67\,$\pm$\,0.30 &   0.67 &    13.1\,$\pm$\,1.4 &                 20\,$\pm$\,10 &  {\footnotesize      ...} \\
     \object{XLSSC 512}  &   352.484 &  -56.136 &         0.402 &         3    &   458 &   3.25\,$\pm$\,0.20 &   0.85 &   2.99\,$\pm$\,0.19 &                 26\,$\pm$\,12 &  {\footnotesize        8} \\
     \object{XLSSC 520}  &   352.502 &  -54.619 &         0.175 &         7    &  1338 &  12.66\,$\pm$\,0.38 &   0.81 &   2.34\,$\pm$\,0.07 &              17.4\,$\pm$\,7.4 &  {\footnotesize    8, 12} \\
     \object{XLSSC 532}  &   352.948 &  -52.669 &         0.392 &         5    &   375 &   4.92\,$\pm$\,0.30 &   0.77 &   3.43\,$\pm$\,0.31 &              19.4\,$\pm$\,9.8 &  {\footnotesize      ...} \\
     \object{XLSSC 519}  &   353.019 &  -55.212 &         0.270 &         3    &   155 &   4.11\,$\pm$\,0.46 &   0.56 &   0.94\,$\pm$\,0.18 &               6.3\,$\pm$\,2.9 &  {\footnotesize        8} \\
     \object{XLSSC 524}  &   353.067 &  -54.702 &         0.270 &         8    &   264 &   3.38\,$\pm$\,0.28 &   0.75 &   1.21\,$\pm$\,0.12 &              15.8\,$\pm$\,8.0 &  {\footnotesize    8, 12} \\
     \object{XLSSC 542}  &   353.113 &  -53.976 &         0.402 &         2    &  3038 &  46.76\,$\pm$\,0.88 &   1.20 &    50.4\,$\pm$\,1.1 &                 74\,$\pm$\,32 &  {\footnotesize   10, 12} \\
     \object{XLSSC 507}  &   353.374 &  -52.252 &         0.566 &         6    &   144 &   3.93\,$\pm$\,0.47 &   0.61 &   4.43\,$\pm$\,0.76 &              11.7\,$\pm$\,6.5 &  {\footnotesize       12} \\
     \object{XLSSC 549}  &   353.515 &  -53.141 &         0.808 &         3    &   201 &   3.33\,$\pm$\,0.31 &   0.71 &    11.3\,$\pm$\,1.9 &                 24\,$\pm$\,20 &  {\footnotesize      ...} \\
     \object{XLSSC 516}  &   353.881 &  -54.586 &  {\it  0.87 } &   {\it 0 }   &   189 &   3.13\,$\pm$\,0.33 &   0.70 &    23.3\,$\pm$\,2.5 &                 29\,$\pm$\,15 &  {\footnotesize        8} \\
     \object{XLSSC 537}  &   354.029 &  -53.876 &         0.515 &        12    &   348 &   5.54\,$\pm$\,0.36 &   0.93 &   8.07\,$\pm$\,0.67 &                 39\,$\pm$\,21 &  {\footnotesize       12} \\
     \object{XLSSC 548}  &   354.193 &  -53.793 &         0.321 &         3    &   175 &   3.09\,$\pm$\,0.33 &   0.43 &   0.51\,$\pm$\,0.13 &               3.0\,$\pm$\,1.3 &  {\footnotesize      ...} \\
     \object{XLSSC 538}  &   354.646 &  -54.623 &         0.332 &         4    &   276 &   3.39\,$\pm$\,0.26 &   0.80 &   1.83\,$\pm$\,0.19 &                 20\,$\pm$\,12 &  {\footnotesize      ...} \\
     \object{XLSSC 543}  &   354.863 &  -55.843 &         0.381 &         2    &   199 &   3.35\,$\pm$\,0.32 &   0.69 &   1.33\,$\pm$\,0.18 &              13.6\,$\pm$\,7.0 &  {\footnotesize      ...} \\
     \object{XLSSC 541}  &   355.431 &  -55.965 &         0.188 &         6    &   415 &   6.77\,$\pm$\,0.40 &   0.81 &   1.42\,$\pm$\,0.09 &              17.7\,$\pm$\,8.0 &  {\footnotesize      ...} \\
     \object{XLSSC 508}  &   355.465 &  -53.145 &         0.539 &         2    &   498 &   9.80\,$\pm$\,0.50 &   0.74 &   4.55\,$\pm$\,0.43 &                 20\,$\pm$\,11 &  {\footnotesize       12} \\
     \object{XLSSC 540}  &   355.632 &  -56.353 &         0.414 &         9    &   483 &   6.53\,$\pm$\,0.36 &   0.78 &   5.52\,$\pm$\,0.34 &              20.1\,$\pm$\,9.3 &  {\footnotesize      ...} \\
     \object{XLSSC 539}  &   355.797 &  -55.881 &         0.184 &         2    &   195 &   3.78\,$\pm$\,0.39 &   0.52 &   0.44\,$\pm$\,0.08 &               4.7\,$\pm$\,2.2 &  {\footnotesize      ...} \\
     \object{XLSSC 509}  &   356.461 &  -54.044 &         0.633 &        12    &   186 &   3.70\,$\pm$\,0.36 &   0.81 &   8.99\,$\pm$\,0.86 &                 29\,$\pm$\,17 &  {\footnotesize      ...} \\
     \object{XLSSC 510}  &   357.539 &  -55.334 &         0.394 &         1    &   380 &   3.93\,$\pm$\,0.26 &   0.71 &   2.96\,$\pm$\,0.20 &              15.2\,$\pm$\,7.2 &  {\footnotesize      ...} \\
     \object{XLSSC 511}  &   357.753 &  -55.371 &         0.130 &         3    &   247 &   3.50\,$\pm$\,0.28 &   0.54 &   0.29\,$\pm$\,0.04 &               5.2\,$\pm$\,2.2 &  {\footnotesize      ...} \\
\hline
\end{tabular} }
\vspace{-0.8cm}
\end{center}
\end{table}

\begin{table}[h!]
\begin{center}
\caption{Supplementary clusters to the \Chundred\ catalogue.\label{SuplTab}}
{\small \begin{tabular}{C{2.1cm}ccccC{1.0cm}C{1.8cm}cC{1.8cm}cc}
 \hline\hline
 Name &    RA  &   Dec   & z & $N_{gal}$ & $C_{60}$ & $F_{60}$ & $\ClRadMT$ & $\Lxxl$ & $\ClMassMT$ & Ref.\rule{0pt}{2.6ex}\\[3pt]
 \hline
     \object{XLSSC 114}  &    30.425 &   -5.031 &         0.234 &         6    &    91 &   3.51\,$\pm$\,0.76 &   1.07 &   1.59\,$\pm$\,0.30 &                 44\,$\pm$\,51 &  {\footnotesize      ...} \\
     \object{XLSSC 113}  &    30.561 &   -7.009 &         0.050 &         9    &   340 &  11.47\,$\pm$\,0.80 &   0.56 &   0.37\,$\pm$\,0.03 &  5.2\,$\pm$\,2.2$^{\ast\ast}$ &  {\footnotesize      ...} \\
     \object{XLSSC 115}  &    32.681 &   -6.588 &         0.043 &        22    &   156 &  12.21\,$\pm$\,1.31 &   0.74 &   0.68\,$\pm$\,0.05 &              12.1\,$\pm$\,6.8 &  {\footnotesize      ...} \\
     \object{XLSSC 550}  &   352.206 &  -52.577 &         0.109 &         3    &   156 &   8.77\,$\pm$\,1.06 &   0.48 &   0.32\,$\pm$\,0.06 &               3.4\,$\pm$\,1.5 &  {\footnotesize      ...} \\
     \object{XLSSC 551}  &   355.444 &  -56.675 &         0.475 &         4    &   189 &   4.66\,$\pm$\,0.54 &   0.67 &   3.96\,$\pm$\,0.67 &              13.7\,$\pm$\,7.9 &  {\footnotesize      ...} \\
\hline
     \object{XLSSC 118}  &    33.692 &   -3.941 &         0.140 &         1    &  5144 &  78.77\,$\pm$\,1.17 &   0.63 &   6.92\,$\pm$\,0.15 &                  $\star\star$ &  {\footnotesize       15} \\
     \object{XLSSC 552}  &   350.629 &  -54.269 &         0.152 &         1    &  3592 &  57.93\,$\pm$\,0.98 &   0.75 &   6.81\,$\pm$\,0.12 & 13.8\,$\pm$\,5.8$^{\ast\ast}$ &  {\footnotesize    12, 8} \\
\hline
\end{tabular} }
\tablefoot{These clusters are above the flux
limit required for inclusion in the main sample, but failed to comply with other
requirements. The five clusters above the horizontal line were detected in ``bad'' pointings
according to the nominal thresholds on the background level and exposure time, while the two
clusters below the line were not part of the pipeline C1+C2 selection.
The table layout is the same as in Table~\ref{BXCTab}.\vspace{-0.8cm}}
\end{center}
\end{table}

\end{appendix}

\twocolumn

\bibliographystyle{aa}
\bibliography{General,XXL}

\end{document}